\documentclass[11pt,a4paper]{article}
\usepackage[english]{babel}
\usepackage[utf8]{inputenc}
\usepackage{epsf}
\usepackage{cite}
\usepackage{graphicx}
\usepackage{subcaption} 

\usepackage{verbatim} 

\usepackage{amsmath}
\usepackage{amssymb}
\usepackage{mathrsfs}

\usepackage{mathbbol}

\usepackage{amsfonts}
\usepackage{bbding}

\usepackage[
colorlinks=true
,urlcolor=black
,anchorcolor=black
,citecolor=black
,filecolor=black
,linkcolor=black
,menucolor=black
,linktocpage=true
,pdfproducer=medialab
,pdfa=true
]{hyperref}

\usepackage{color} 

\usepackage[left=2cm,right=2cm,top=2cm,bottom=2cm]{geometry}
\usepackage{tikz}
\usetikzlibrary{shapes,arrows,positioning,automata,backgrounds,calc,er,patterns}
\usepackage[compat=1.1.0]{tikz-feynman}

\vspace*{1cm}
\allowdisplaybreaks
\begin{document}

\begin{center}
\vspace*{1mm}
\vspace{1.3cm}
{\Large\bf
On the role of leptonic CPV phases in cLFV observables
}

\vspace*{1.2cm}

{\bf A.~Abada $^{a}$, J.~Kriewald $^{b}$ and A.~M.~Teixeira $^{b}$}

\vspace*{.5cm}
$^{a}$ P\^ole Th\'eorie, Laboratoire de Physique des 2 Infinis Irène Joliot Curie (UMR 9012), \\
CNRS/IN2P3,
15 Rue Georges Clemenceau, 91400 Orsay, France

\vspace*{.2cm}
$^{b}$ Laboratoire de Physique de Clermont (UMR 6533), CNRS/IN2P3,\\
Univ. Clermont Auvergne, 4 Av. Blaise Pascal, 63178 Aubi\`ere Cedex,
France

\end{center}

\vspace*{5mm}
\begin{abstract}
\noindent
In extensions of the Standard Model by Majorana fermions, the presence of additional 
CP violating phases has been shown to play a crucial role in lepton number violating processes.
In this work we show that (Dirac and Majorana) CP violating phases can also lead to important effects in  charged lepton flavour violating (cLFV) transitions and decays, in some cases with a 
significant impact for the predicted rates of cLFV observables.
We conduct a thorough exploration of these effects in several cLFV observables, and discuss the implications for future observation.
We emphasise how the presence of leptonic CP violating phases might lead to modified cLFV rates, 
and to a possible loss of correlation between cLFV observables.
\end{abstract}

\newpage
\section{Introduction}
Neutrino oscillations signal the presence of physics beyond the Standard Model (SM). Moreover, leptonic mixings and massive neutrinos open the door to the violation of charged lepton flavour.
Present in several New Physics (NP) models, heavy neutral Majorana fermions (such as right-handed neutrinos) are among the most appealing extensions of the SM, often in association with a neutrino mass generation mechanism, as for example in the case of a fermionic seesaw~\cite{seesaw:I}. 
Such sterile states (singlets under the SM gauge group) can also be present in more complete NP frameworks  as $\mathrm{SO}(10)$ and its extensions.
Depending on their mass scale and interactions with the SM fields, these new neutral fermions can offer 
a minimal framework to address the SM observational problems:~account for neutrino oscillation data, offer an explanation to the baryon asymmetry of the Universe via leptogenesis, and possibly contribute to explain the dark matter problem. 

Heavy neutral leptons (HNL), with masses ranging from the GeV to the tens of TeV, are among the most interesting minimal extensions of the SM, as they can be at the source of significant contributions  to numerous observables, both at high-intensities and at colliders. 
Interestingly, the most minimal tree-level mechanism for neutrino mass generation - the type-I seesaw - calls upon the introduction of at least two such states to account for oscillation data.  
The type-I seesaw~\cite{seesaw:I}  
and its low-scale variants,  
such as the Inverse 
Seesaw (ISS)~\cite{Schechter:1980gr,Gronau:1984ct,Mohapatra:1986bd},  
the Linear Seesaw (LSS)~\cite{Barr:2003nn,Malinsky:2005bi}  and the 
$\nu$-MSM~\cite{Asaka:2005an,Asaka:2005pn,Shaposhnikov:2008pf}, all call upon extending the SM via additional sterile fermions, allowing for Dirac and Majorana mass terms for the neutral lepton sector. 
Irrespective of the actual mechanism of neutrino mass generation under consideration, the mixings of the new states with the active left-handed neutrinos will lead to modifications in both leptonic charged and neutral currents, with a deep phenomenological impact. 
To study and numerically assess the impact of the heavy states, 
it is often convenient to consider simplified ``ad-hoc'' models, in which one adds $n_s$ sterile fermions to the SM field content. Such an approach allows to identify the most relevant effects and the consequences for the observables under scrutiny, and paves the way to a subsequent thorough study of complete models of neutrino mass generation via sterile fermions. 

The role of heavy neutral leptons in what concerns charged lepton flavour violation (cLFV) (see for instance~\cite{Riemann:1982rq,Riemann:1999ab,Illana:1999ww,Mann:1983dv,Illana:2000ic,Alonso:2012ji,Ilakovac:1994kj,Ma:1979px,Gronau:1984ct,Deppisch:2004fa,Deppisch:2005zm,Dinh:2012bp,Abada:2014kba,Abada:2015oba,Abada:2015zea,Abada:2016vzu,Abada:2018nio,Arganda:2014dta}) and lepton number violation (LNV) - see for instance~\cite{Ali:2001gsa,Atre:2005eb,Atre:2009rg,Chrzaszcz:2013uz,Deppisch:2015qwa,Cai:2017mow,Abada:2017jjx,Drewes:2019byd,Maiezza:2015lza,Helo:2013dla,Blaksley:2011ey,Ibarra:2011wi} -
has been extensively explored in recent years, be it in the context of mechanisms of neutrino mass generation, or then in the framework of the above mentioned minimal ad-hoc extensions by one or more heavy states. 
Several of these studies revealed a promising potential of SM extensions via HNL in what concerns cLFV: depending on the mass regime and mixings with the active states, one could expect significant contributions to several observables, well within the future experimental sensitivity (with particularly interesting prospects in the $\mu-e$ sector). 
Moreover, in given scenarios, distinctive patterns and correlations of observables were identified, which in turn could be explored to probe and test these SM extensions, see for example~\cite{Hambye:2013jsa,Calibbi:2017uvl}. 

In addition to the masses of the new states
and their mixings to the active neutrinos, constructions relying on heavy sterile states also open the door to new sources of CP violation: other than new Dirac CP violating (CPV) phases, should the massive states be of Majorana nature, further phases can be present.
The role of these new CPV phases has been explored in analyses dedicated to CP violating observables, as is the case of electric dipole moments (EDM) of charged leptons~\cite{Abada:2015trh,Abada:2016awd,Novales-Sanchez:2016sng,deGouvea:2005jj}. 
New CPV phases have been also recently shown to play a crucial role in what concerns interference effects in LNV (and cLFV) semileptonic meson and tau decays~\cite{Abada:2019bac}, when more than one HNL is involved. Noticeably, while branching fractions of semileptonic meson or tau decays  into 
same-sign and opposite-sign dileptons are expected to be of the same order in the case of SM extensions by a single heavy  Majorana fermion, this is no longer the case 
when the SM is extended by at least two HNLs, due to the possible
interferences that might arise in the presence of multiple states. 
Depending on the CPV phases, one can have a modification 
(enhancement/suppression) of the rates of LNV  
modes and of the lepton number conserving ones. 
As a consequence, the non-observation of a given mode need not be
interpreted in terms of reduced active-sterile couplings, but it could be instead understood in terms of interference effects due to the presence of several sterile states. 
(This effect is particularly amplified for processes 
with different charged leptons in the  final state.) 
Likewise, an experimental signal of a lepton number conserving process and the non-observation of the corresponding 
lepton number violating one do 
not necessarily rule out that the mediators are Majorana fermions~\cite{Abada:2019bac}. 

Similar studies have explored the role of a second heavy neutrino concerning the possibility of resonant CP violation~\cite{Bray:2007ru}, the effect of CP violation in high-scale seesaw scenarios in the context of renormalisation group running and leptogenesis~\cite{Petcov:2005yh,Petcov:2006pc}, the impact for forward-backward asymmetries at an electron-positron collider~\cite{Dev:2019rxh}, while others have compared the expected number of events associated with same-sign and opposite-sign dileptons at colliders in the framework of Left-Right symmetric models~\cite{Anamiati:2016uxp,Das:2017hmg,Dev:2019rxh}. 

\bigskip
In this study we focus exclusively on the role of CPV  phases (Dirac and Majorana) concerning an extensive array of cLFV observables. 
Although the importance of CPV phases in flavour observables is known, we provide here a dedicated thorough analysis, systematically studying the effects of CPV phases in cLFV observables.
We work under the assumption of a unique source of lepton flavour violation, a generalised leptonic mixing matrix which now incorporates the active-sterile mixings.
Following an analysis of the different observables and the implications for future observation, our results suggest that the presence of leptonic CP violating phases can strongly affect the predictions (either suppressing or enhancing the otherwise expected rates), and possibly  
lead to a loss of correlation between observables (typically present in simple SM extensions via heavy sterile fermions). As we will subsequently argue, the confrontation of unexpected cLFV patterns upon observation of certain channels could be suggestive of the presence of non-vanishing phases.
As an example, one could have sizeable rates 
for $\mu\to eee$, and comparatively suppressed rates for $\mu \to e\gamma$.
As we will also point out, another important point concerns accidental cancellations between distinct contributing terms to a given observable, as in the well-known case of $\mu-e$ conversion in nuclei~\cite{Alonso:2012ji}.
The results of our work can be readily generalised for more complete NP models relying on the inclusion of heavy neutral fermions, provided that all complex degrees of freedom are systematically included (e.g. in Casas-Ibarra parametrisation~\cite{Casas:2001sr}), and predictions for cLFV observables revisited.

\bigskip 
This work is organised as follows: the cLFV observables addressed in the framework of the SM extension via sterile fermions are introduced in Section~\ref{sec:cLFVobs}. In Section~\ref{sec:phases.matter}, a first insight into the role of Dirac and Majorana CPV phases is (analytically) presented, relying on a subset of representative cLFV  observables. A phenomenological analysis of the impact of CPV phases on cLFV transitions is conducted in Section~\ref{sec:num:analysis}, including discussions on expected ranges and possible effects on the 
correlation patterns.
In Section~\ref{sec:general-analysis} we offer a comprehensive view of the model's parameter space, highlighting numerous points that could arise upon a future observation of several cLFV decays, followed by a final discussion and outlook in Section~\ref{sec:concs}.
Details on the model, expressions for the decay rates and other relevant amplitudes, as well as further information on the analysis,  are collected  in the Appendices.

\section {cLFV observables: decays and transitions}
\label{sec:cLFVobs}
In this section we present the cLFV processes in the framework of a ``$3+n_S$'' ad-hoc SM extension (see Appendix~\ref{app:model} for a detailed description), providing the expressions for the decay rates and other relevant amplitudes. As further discussed in Appendix~\ref{app:model}, 
the generalised leptonic mixings are encoded in a $(3+n_S)\times(3+n_S)$ unitary mixing matrix, $\mathcal{U}$, of which the upper left $3\times3$ block corresponds to the left-handed leptonic mixing matrix, the would-be PMNS,  $\tilde{U}_\text{PMNS}$, which is no longer unitary.

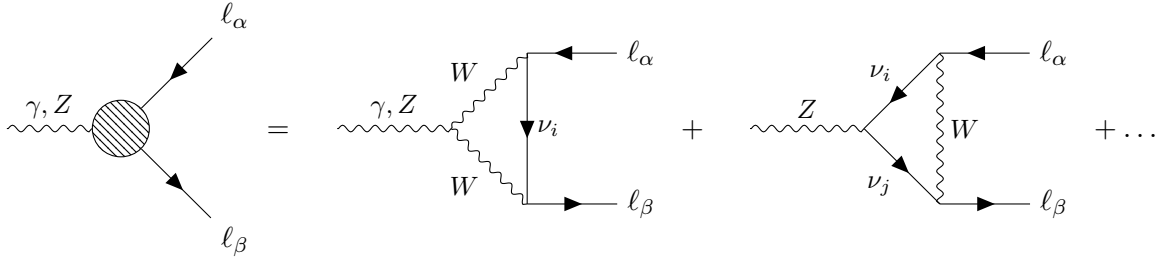
\begin{figure}[h!]
\hspace*{0,5cm}\begin{equation*}
\vcenter{\hbox{
    \begin{tikzpicture}
        \begin{feynman}
            \vertex (m) [blob] at (0,0) {};
            \vertex (B) at (-1.5, 0);
            \vertex (f1) at (1.5, 1.5) {\(\ell_\alpha\)};
            \vertex (f2) at (1.5, -1.5) {\(\ell_\beta\)};
            \diagram* {
            (B) -- [boson, edge label={\small\(\gamma, Z\)}] (m) -- [fermion] (f2),
            (f1) -- [fermion] (m),
            };
        \end{feynman}
    \end{tikzpicture}}}
     =\quad
    \vcenter{\hbox{
    \begin{tikzpicture}
        \begin{feynman}
            \vertex (m) at (0,0);
            \vertex (B) at (-1.5, 0);
            \vertex (f1) at (1, 1);
            \vertex (f2) at (1, -1);
            \vertex (f12) at (2.5, 1) {\(\ell_\alpha\)};
            \vertex (f22) at (2.5, -1) {\(\ell_\beta\)};
            \diagram* {
            (B) -- [boson, edge label={\small\(\gamma, Z\)}] (m) -- [boson, edge label={\small\(W\)}] (f1),
            (m) -- [boson, edge label'={\small\(W\)}] (f2),
            (f1) -- [fermion, edge label={\small\(\nu_i\)}] (f2),
            (f1) -- [anti fermion] (f12),
            (f2) -- [fermion] (f22),
            };
        \end{feynman}
    \end{tikzpicture}
    }}
 +\quad
    \vcenter{\hbox{
    \begin{tikzpicture}
        \begin{feynman}
            \vertex (m) at (0,0);
            \vertex (B) at (-1.5, 0);
            \vertex (f1) at (1, 1);
            \vertex (f2) at (1, -1);
            \vertex (f12) at (2.5, 1) {\(\ell_\alpha\)};
            \vertex (f22) at (2.5, -1) {\(\ell_\beta\)};
            \diagram* {
            (B) -- [boson, edge label={\small\(Z\)}] (m) -- [anti fermion, edge label={\small\(\nu_i\)}] (f1),
            (m) -- [fermion, edge label'={\small\(\nu_j\)}] (f2),
            (f1) -- [boson, edge label={\small\(W\)}] (f2),
            (f1) -- [anti fermion] (f12),
            (f2) -- [fermion] (f22),
            };
        \end{feynman}
    \end{tikzpicture}
    }}
     +\hdots
\end{equation*}

\caption{
Vertex diagrams  contributing the cLFV decays. The flavour of the charged leptons is denoted by $\alpha, \beta, ... = e, \mu,\tau$; in the neutral fermion internal lines, $i,j=1,..., 3+n_S$ denote the neutral fermion mass eigenstates.
 }
 \label{Fig:penguins}
\end{figure}

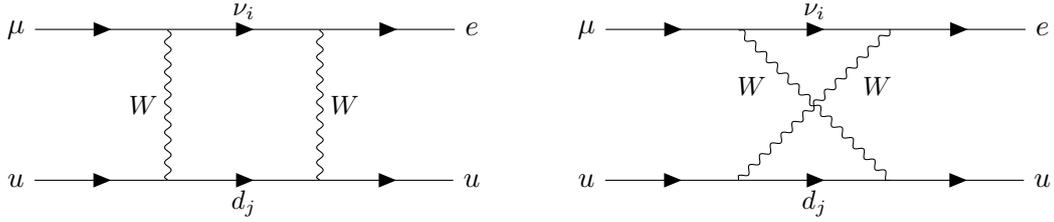
\begin{figure}[h!]
\begin{equation*}
\vcenter{\hbox{
\begin{tikzpicture}
    \begin{feynman}
        \vertex (f1) at (0,1.0) {\(\mu\)};
        \vertex (f11) at (2.0,1.0);
        \vertex (f12) at (4.0, 1.0);
        \vertex (f2) at (6.0, 1.0) {\(e\)};
        \vertex (f3) at (0, -1.0) {\(u\)};
        \vertex (f31) at (2.0, -1.0);
        \vertex (f32) at (4.0, -1.0);
        \vertex (f4) at (6.0, -1.0) {\(u\)};
        \diagram*{
        (f1) -- [fermion] (f11) -- [fermion, edge label={\small\(\nu_i\)}] (f12) -- [fermion] (f2),
        (f3) -- [fermion] (f31) -- [fermion, edge label'={\small\(d_j\)}] (f32) -- [fermion] (f4),
        (f11) -- [boson, edge label'={\small \(W\)}] (f31),
        (f12) -- [boson, edge label={\small \(W\)}] (f32),
        };
    \end{feynman}
\end{tikzpicture}
}}
\quad\quad
\vcenter{\hbox{
\begin{tikzpicture}
    \begin{feynman}
        \vertex (f1) at (0,1.0) {\(\mu\)};
        \vertex (f11) at (2.0,1.0);
        \vertex (f12) at (4.0, 1.0);
        \vertex (f2) at (6.0, 1.0) {\(e\)};
        \vertex (f3) at (0, -1.0) {\(u\)};
        \vertex (f31) at (2.0, -1.0);
        \vertex (f32) at (4.0, -1.0);
        \vertex (f4) at (6.0, -1.0) {\(u\)};
        \diagram*{
        (f1) -- [fermion] (f11) -- [fermion, edge label={\small\(\nu_i\)}] (f12) -- [fermion] (f2),
        (f3) -- [fermion] (f31) -- [fermion, edge label'={\small\(d_j\)}] (f32) -- [fermion] (f4),
        (f11) -- [boson, edge label'={\small \(W\)},near start] (f32),
        (f12) -- [boson, edge label={\small \(W\)},near start] (f31),
        };
    \end{feynman}
\end{tikzpicture}
}}
\end{equation*}
\caption{Example of  box diagrams (depicted for the box contributions to neutrinoless $\mu-e$ conversion). In the quark internal lines, $j=1,...,3$ 
  runs over the quark families; 
  in the neutral fermion ones, $i=1,...,3+n_S$.}  
\label{Fig:boxes}
\end{figure}

The cLFV observables addressed here receive contributions at the one-loop level arising from dipoles ($\gamma$- and $Z$-penguins) and/or from boxes. The vertex diagrams contributing to the cLFV decays are depicted in Fig.~\ref{Fig:penguins} and the box diagrams (adapted for the $\mu-e$ conversion rate for the sake of illustration) are presented in Fig.~\ref{Fig:boxes}.

In the following subsections, we provide analytical expressions for the cLFV observables in the general $3+n_s$ framework (from Section 3 onwards the discussion will be focused on the case $n_s = 2$).

\subsection{Leptonic cLFV decays}
We first provide the expressions for the branching ratios of the ``pure'' leptonic cLFV decays, i.e. the radiative 
decays $\ell_\beta \to \ell_\alpha \gamma$ and the 3-body decays\footnote{In the most general 3-body cLFV decay, $\ell_\beta \to \ell_\alpha \ell_\gamma \ell_\gamma^\prime$ the primed final state denotes the possibility of having equal or opposite charges for both $\ell_\gamma$ and $\ell_\gamma^{(\prime)}$.}
$\ell_\beta \to \ell_\alpha \ell_\gamma \ell_\gamma^\prime$. For the  sake of brevity, we present here the expression for $\ell_\beta \to 3 \ell_\alpha$ decays (the full expression for the most general case 
can be found in Ref.~\cite{Ilakovac:1994kj}). Notice that in our numerical analysis we have considered all possible decays for $\ell_\beta \to \ell_\alpha \ell_\gamma \ell_\gamma^\prime$.

In SM extensions via
$n_s$ heavy sterile fermions, the rates for the radiative and three-body decays are then given by~\cite{Alonso:2012ji}  
\begin{equation}
    \mathrm{BR}(\ell_\beta\to \ell_\alpha \gamma) \,=
    \frac{\alpha_w^3\,
      s_w^2}{256\,\pi^2}\,\frac{m_{\beta}^4}{M_W^4}\,
\frac{m_{\beta}}{\Gamma_{\beta}}\, 
    \left|G_\gamma^{\beta \alpha} \right|^2\:, 
\end{equation}
\begin{eqnarray}
    \mathrm{BR}(\ell_\beta\to 3\ell_\alpha) &=&
    \frac{\alpha_w^4}{24576\,\pi^3}\,\frac{m_{\beta}^4}{M_W^4}\,
\frac{m_{\beta}}{\Gamma_{\beta}}\times\left\{2\left|\frac{1}{2}F_\text{box}^{\beta
      3\alpha} +F_Z^{\beta\alpha} - 2 s_w^2\,(F_Z^{\beta\alpha} -
    F_\gamma^{\beta\alpha})\right|^2 \right.  \nonumber\\ 
     &+& \left. 4 s_w^4\, |F_Z^{\beta\alpha} -
    F_\gamma^{\beta\alpha}|^2 + 16
    s_w^2\,\mathrm{Re}\left[(F_Z^{\beta\alpha} - \frac{1}{2}F_\text{box}^{\beta
        3\alpha})\,G_\gamma^{\beta \alpha
        \ast}\right]\right.\nonumber\\ 
     &-&\left. 48 s_w^4\,\mathrm{Re}\left[(F_Z^{\beta\alpha} -
      F_\gamma^{\beta\alpha})\,G_\gamma^{\beta\alpha \ast}\right] + 32
    s_w^4\,|G_\gamma^{\beta\alpha}|^2\left[\log\frac{m_{\beta}^2}{m_{\alpha}^2}
      - \frac{11}{4}\right] \right\}\,.  
\end{eqnarray}
Here, $M_W$ is the $W$ boson mass, $\alpha_w = g_w^2/4\pi$ denotes the weak coupling, $s_w$ the sine of the weak mixing angle, and $m_{\beta}$ ($\Gamma_\beta$) the mass (total width)
of the decaying charged lepton of flavour $\beta$.  
The form factors present in the above equations are given by~\cite{Alonso:2012ji, Ilakovac:1994kj} 
\begin{eqnarray}
    G_\gamma^{\beta \alpha} &=& \sum_{i =1}^{3 + n_s}
    \mathcal{U}_{\alpha i}^{\phantom{\ast}}\,\mathcal{U}_{\beta i}^\ast\,
    G_\gamma(x_i)\:,\label{eq:cLFV:FF:Ggamma} \\
     F_\gamma^{\beta \alpha} &=& \sum_{i =1}^{3 + n_s}
    \mathcal{U}_{\alpha i}^{\phantom{\ast}}\,\mathcal{U}_{\beta i}^\ast
    \,F_\gamma(x_i)\:,
   \\ 
    F_Z^{\beta \alpha} &=& \sum_{i,j =1}^{3 + n_s}
    \mathcal{U}_{\alpha i}^{\phantom{\ast}}\,\mathcal{U}_{\beta j}^\ast
    \left[\delta_{ij} \,F_Z(x_j) + 
    C_{ij}\, G_Z(x_i, x_j) + C_{ij}^\ast \,H_Z(x_i,
    x_j)\right]\:, 
    \label{eq:cLFV:FF:FZ}
    \\  
    F_\text{box}^{\beta 3 \alpha} &=&\sum_{i,j = 1}^{3+n_s}
    \mathcal{U}_{\alpha i}^{\phantom{\ast}}\,\mathcal{U}_{\beta
      j}^\ast\left[\mathcal{U}_{\alpha i}^{\phantom{\ast}} \,\mathcal{U}_{\alpha
        j}^\ast\, G_\text{box}(x_i, x_j) - 2 \,\mathcal{U}_{\alpha
        i}^\ast \,\mathcal{U}_{\alpha j}^{\phantom{\ast}}\, F_\text{Xbox}(x_i, x_j)
      \right]\:,\label{eq:cLFV:FF:Fbox}
    \end{eqnarray}
in which the sums run over the neutral mass eigenstates ($i,j=1,...,3+n_S$). 
The loop functions are given in Appendix~\ref{app:cLFV.others}, with the corresponding arguments defined as $x_i ={m_{i}^2}/{M_W^2}$. 
Furthermore, $\mathcal{U}$ is the $(3+n_s)\times (3+n_s)$ lepton mixing matrix (unitary) and $C_{ij}$ is defined as
\begin{equation}\label{eq:Cijdef}
 C_{ij} = \sum_{\rho = 1}^3
  \mathcal{U}_{i\rho}^\dagger \,\mathcal{U}_{\rho j}^{\phantom{\dagger}}\:. 
\end{equation}

\subsection{Neutrinoless $\mu-e$ conversion in nuclei}
The coherent conversion rate\footnote{In the present discussion we only consider the coherent conversion; for a general discussion of spin-dependent contributions to the process, we refer to Refs.~\cite{Cirigliano:2017azj,Davidson:2017nrp}.} in the presence of a nuclei (N) can be cast as~\cite{Alonso:2012ji}   
\begin{equation}\label{eq:def:CRfull}
    \mathrm{CR}(\mu - e,\,\mathrm{N}) = \frac{2 G_F^2\,\alpha_w^2\,
      m_\mu^5}{(4\pi)^2\,\Gamma_\text{capt.}}\left|4 V^{(p)}\left(2
    \widetilde F_u^{\mu e} + \widetilde F_d^{\mu e}\right) + 4 V^{(n)}\left(
    \widetilde F_u^{\mu e} + 2\widetilde F_d^{\mu e}\right)  + s_w^2
    \frac{G_\gamma^{\mu e}D}{2 e}\right|^2\,.  
\end{equation}
In the above expression, $\Gamma_\text{capt.}$ denotes the capture rate for the nucleus N; 
$D$, $V^{(p)}$ and $V^{(n)}$ correspond to nuclear form factors whose
values are given in~\cite{Kitano:2002mt}, with $e$ 
being the unit electric charge. 
The form factors present in the above equations are given by~\cite{Alonso:2012ji, Ilakovac:1994kj} 
\begin{eqnarray}
    \widetilde F^{\mu e}_d &=& -\frac{1}{3}s_w^2 F_\gamma^{\mu e} - F_Z^{\mu e}\left(\frac{1}{4} - \frac{1}{3}s_w^2 \right) + \frac{1}{4}F^{\mu e dd}_\text{box}\ ,\\
    \widetilde F^{\mu e}_u &=& \frac{2}{3}s_w^2 F_\gamma^{\mu e} + F_Z^{\mu e}\left(\frac{1}{4} - \frac{2}{3}s_w^2 \right) + \frac{1}{4}F^{\mu e uu}_\text{box}\,.
\end{eqnarray}
In addition to the form factors previously defined in Eqs.~(\ref{eq:cLFV:FF:Ggamma} - \ref{eq:cLFV:FF:Fbox}), $\widetilde{F}_{u,d}^{\mu e}$ call upon the following box contributions, 
\begin{eqnarray}
     F_\text{box}^{\mu e uu} &=& \sum_{i = 1}^{3 + n_s}\sum_{q_d = d, s,
      b} \mathcal{U}_{e i}^{\phantom{\ast}}\,\mathcal{U}_{\mu i}^\ast\, V_{u q_d}^{\phantom{\ast}}\,V_{u
      q_d}^\ast \:F_\text{box}(x_i, x_{q_d})\,, 
    \label{eq:cLFV:FF:mueuu}\\
    F_\text{box}^{\mu e dd} &=& \sum_{i = 1}^{3 + n_s}\sum_{q_u = u, c,
      t} \mathcal{U}_{e i}^{\phantom{\ast}}\,\mathcal{U}_{\mu i}^\ast\, V_{q_u
      d}^{\phantom{\ast}}\,V_{q_u d}^\ast \:F_\text{Xbox}(x_i, x_{q_u})\,, 
    \label{eq:cLFV:FF:muedd}    
\end{eqnarray}
in which $x_{q} ={m_{q}^2}/{M_W^2}$ and $V$ is the Cabibbo-Kobayashi-Maskawa (CKM) quark mixing matrix.  

Other cLFV transitions and decays also occur in the presence of muonic atoms. This is the case of Muonium oscillations and decays ($\text{Mu}-\overline{\text{Mu}}$, $\text{Mu}\to e e $), and of the Coulomb enhanced decay $\mu e \to e e$. The expressions for the cLFV rates of the latter can be found in Appendix~\ref{app:cLFV.others}; here we only emphasise that all depend on combinations of $\mu e$ form factors already introduced for the radiative and 3-body decays (corresponding to $Z$- and photon-penguins, and box diagrams). 

\subsection{cLFV $Z$ decays}
\label{sec:Zdecays}
As the topology of cLFV $Z$ decays is closely related to contributions at the origin of several cLFV leptonic decays and transitions ($Z$-penguins), we thus also include  
$Z \to \ell_\alpha \bar\ell_\beta$ decays in our study\footnote{In view of the very distinct topological contributions, we do not include cLFV Higgs decays in the present work (for a recent discussion see~\cite{Arganda:2014dta,Arganda:2015naa,Arganda:2015uca,Arganda:2017vdb}).}. 
For convenience, we summarise here the analytical expressions needed for the LFV $Z$-decays in the Feynman-t'Hooft gauge, given in Refs.~\cite{DeRomeri:2016gum, Abada:2014cca, Abada:2015zea, Illana:1999ww}, in the convention of {\tt LoopTools}~\cite{Hahn:1998yk}.
The decay width (for $\alpha\neq\beta$) is given by
\begin{equation}
    \Gamma(Z\to \ell_\alpha\bar\ell_\beta) = \frac{\alpha_w^3}{192\pi^2 c_w^2}M_Z\left|\mathcal F_Z^{\beta\alpha}\right|^2\,,
\end{equation}
in which the form factor $\mathcal F_Z^{\beta\alpha} = \sum_{i=1}^{10} \mathcal F_{Z,\beta\alpha}^{(i)}$ receives contributions from 10 different diagrams, as given in~\cite{DeRomeri:2016gum, Abada:2014cca, Illana:1999ww}, and $M_Z$ is the mass of the $Z$-boson.
The contributions of the different diagrams (neglecting the charged lepton masses) are given by
\begin{eqnarray}
    \mathcal F_{Z,\beta\alpha}^{(1)} &=& \frac{1}{2}\sum_{i,j = 1}^{3 + n_s}\mathcal U_{\alpha i}\,\mathcal U_{\beta j}^\ast\left[-C_{ij} x_i x_j M_W^2 C_0 + C_{ij}^\ast \sqrt{x_i x_j}\left(M_Z^2 C_{12} - 2C_{00} + \frac{1}{2}\right)\right]\ ,\\
    \mathcal F_{Z,\beta\alpha}^{(2)} &=& \sum_{i,j = 1}^{3 + n_s}\mathcal U_{\alpha i}\,\mathcal U_{\beta j}^\ast\left[-C_{ij}\left((C_0 + C_1 + C_2 + C_{12}) - 2C_{00} + 1\right) +  C_{ij}^\ast \sqrt{x_i x_j}M_W^2 C_0 \right]\,,
\end{eqnarray}
where $C_{0, 1, 2, 12, 00}\equiv C_{0, 1, 2, 12, 00}(0, M_Z^2, 0, M_W^2, m_{\nu_i}^2, m_{\nu_j}^2)$ are the Passarino-Veltman functions~\cite{Passarino:1978jh} in {\tt LoopTools}~\cite{Hahn:1998yk} notation.
We further have
\begin{eqnarray}
    \mathcal F_{Z,\beta\alpha}^{(3)} &=& 2 c_w^2 \sum_{i=1}^{3+n_s}\mathcal U_{\alpha i}\,\mathcal U_{\beta i}^\ast \left[ M_Z^2 (C_1 + C_2 + C_{12}) - 6C_{00} + 1\right]\,,\\
    \mathcal F_{Z,\beta\alpha}^{(4)} + \mathcal F_{Z,\beta\alpha}^{(5)} &=& -2 s_w^2 \sum_{i=1}^{3+n_s}\mathcal U_{\alpha i}\,\mathcal U_{\beta i}^\ast x_i M_W^2 C_0\,,\\
    \mathcal F_{Z,\beta\alpha}^{(6)} &=& -(1 - 2 s_w^2)\sum_{i=1}^{3+n_s}\mathcal U_{\alpha i}\,\mathcal U_{\beta i}^\ast x_i C_{00}\,,
\end{eqnarray}
with $C_{0, 1, 2, 12, 00}\equiv C_{0, 1, 2, 12, 00}(0, M_Z^2, 0, m_{\nu_i}^2, M_W^2, M_W^2)$, and
\begin{equation}
    \mathcal F_{Z,\beta\alpha}^{(7)} + \mathcal F_{Z,\beta\alpha}^{(8)} + \mathcal F_{Z,\beta\alpha}^{(9)} + \mathcal F_{Z,\beta\alpha}^{(10)} = \frac{1}{2}(1 - 2 c_w^2)\sum_{i=1}^{3+n_s}\mathcal U_{\alpha i}\,\mathcal U_{\beta i}^\ast \left[(2 + x_i)B_1 + 1\right]\,,
\end{equation}
in which we have $B_1 \equiv B_1(0,m_{\nu_i}^2, M_W^2)$.
We evaluate all Passarino-Veltman functions with the public Fortran code {\tt LoopTools}~\cite{Hahn:1998yk} wrapped into our dedicated python code.

Since current searches do not distinguish the charges of the final state leptons, in our numerical analysis we consider the CP averaged decay rate, that is
\begin{equation}
    \Gamma (Z\to \ell_\alpha^\pm \ell_\beta^\mp) = \frac{1}{2}\left[\Gamma(Z\to \ell_\alpha^+\ell_\beta^-) + \Gamma(Z\to \ell_\alpha^-\ell_\beta^+)\right]\,.
\end{equation}

\subsection{cLFV searches: current status and future prospects}
\label{sec:exp_bounds}
As extensively argued, the observation of one (or several) cLFV processes would be a clear signal of physics beyond the SM.
Currently, there is a vast world-wide 
array of dedicated experiments and searches, at different energy scales, aiming at discovering 
such transitions.
In Table~\ref{tab:cLFVdata} we list current experimental bounds and future sensitivities\footnote{Note that for the Mu3e experiment~\cite{Blondel:2013ia} we display a second more optimistic future sensitivity, reflecting the potential of having a very high intensity muon beam available; in our analysis we use the latter (optimal) sensitivity.} for the observables considered in this work.
\renewcommand{\arraystretch}{1.3}
\begin{table}[h!]
    \centering
    \hspace*{-7mm}{\small\begin{tabular}{|c|c|c|}
    \hline
    Observable & Current bound & Future Sensitivity  \\
    \hline\hline
    $\text{BR}(\mu\to e \gamma)$    &
    \quad $<4.2\times 10^{-13}$ \quad (MEG~\cite{TheMEG:2016wtm})   &
    \quad $6\times 10^{-14}$ \quad (MEG II~\cite{Baldini:2018nnn}) \\
    $\text{BR}(\tau \to e \gamma)$  &
    \quad $<3.3\times 10^{-8}$ \quad (BaBar~\cite{Aubert:2009ag})    &
    \quad $3\times10^{-9}$ \quad (Belle II~\cite{Kou:2018nap})      \\
    $\text{BR}(\tau \to \mu \gamma)$    &
     \quad $ <4.4\times 10^{-8}$ \quad (BaBar~\cite{Aubert:2009ag})  &
    \quad $10^{-9}$ \quad (Belle II~\cite{Kou:2018nap})     \\
    \hline
    $\text{BR}(\mu \to 3 e)$    &
     \quad $<1.0\times 10^{-12}$ \quad (SINDRUM~\cite{Bellgardt:1987du})    &
     \quad $10^{-15(-16)}$ \quad (Mu3e~\cite{Blondel:2013ia})   \\
    $\text{BR}(\tau \to 3 e)$   &
    \quad $<2.7\times 10^{-8}$ \quad (Belle~\cite{Hayasaka:2010np})&
    \quad $5\times10^{-10}$ \quad (Belle II~\cite{Kou:2018nap})     \\
    $\text{BR}(\tau \to 3 \mu )$    &
    \quad $<3.3\times 10^{-8}$ \quad (Belle~\cite{Hayasaka:2010np})  &
    \quad $5\times10^{-10}$ \quad (Belle II~\cite{Kou:2018nap})     \\
    & & \quad$5\times 10^{-11}$\quad (FCC-ee~\cite{Abada:2019lih})\\
        $\text{BR}(\tau^- \to e^-\mu^+\mu^-)$   &
    \quad $<2.7\times 10^{-8}$ \quad (Belle~\cite{Hayasaka:2010np})&
    \quad $5\times10^{-10}$ \quad (Belle II~\cite{Kou:2018nap})     \\
    $\text{BR}(\tau^- \to \mu^-e^+e^-)$ &
    \quad $<1.8\times 10^{-8}$ \quad (Belle~\cite{Hayasaka:2010np})&
    \quad $5\times10^{-10}$ \quad (Belle II~\cite{Kou:2018nap})     \\
    $\text{BR}(\tau^- \to e^-\mu^+e^-)$ &
    \quad $<1.5\times 10^{-8}$ \quad (Belle~\cite{Hayasaka:2010np})&
    \quad $3\times10^{-10}$ \quad (Belle II~\cite{Kou:2018nap})     \\
    $\text{BR}(\tau^- \to \mu^-e^+\mu^-)$   &
    \quad $<1.7\times 10^{-8}$ \quad (Belle~\cite{Hayasaka:2010np})&
    \quad $4\times10^{-10}$ \quad (Belle II~\cite{Kou:2018nap})     \\
    \hline
    $\text{CR}(\mu- e, \text{N})$ &
     \quad $<7 \times 10^{-13}$ \quad  (Au, SINDRUM~\cite{Bertl:2006up}) &
    \quad $10^{-14}$  \quad (SiC, DeeMe~\cite{Nguyen:2015vkk})    \\
    & &  \quad $2.6\times 10^{-17}$  \quad (Al, COMET~\cite{Krikler:2015msn,Adamov:2018vin})  \\
    & &  \quad $8 \times 10^{-17}$  \quad (Al, Mu2e~\cite{Bartoszek:2014mya})\\
    \hline
    \hline 
    $\mathrm{BR}(Z\to e^\pm\mu^\mp)$ & \quad$< 4.2\times 10^{-7}$\quad (ATLAS~\cite{Aad:2014bca}) & \quad$\mathcal O (10^{-10})$\quad (FCC-ee~\cite{Abada:2019lih}\\
    $\mathrm{BR}(Z\to e^\pm\tau^\mp)$ & \quad$< 5.2\times 10^{-6}$\quad (OPAL~\cite{Akers:1995gz}) & \quad$\mathcal O (10^{-10})$\quad (FCC-ee~\cite{Abada:2019lih}\\
    $\mathrm{BR}(Z\to \mu^\pm\tau^\mp)$ & \quad$< 5.4\times 10^{-6}$\quad (OPAL~\cite{Akers:1995gz}) & \quad $\mathcal O (10^{-10})$\quad (FCC-ee~\cite{Abada:2019lih}\\
    \hline
    \end{tabular}}
    \caption{Current experimental bounds and future sensitivities on cLFV observables considered in this work. All limits are given at $90\%\:\mathrm{C.L.}$, and the Belle II sensitivities correspond to an integrated luminosity of $50\:\mathrm{ab}^{-1}$.}
    \label{tab:cLFVdata}
\end{table}
\renewcommand{\arraystretch}{1.}

\section{Phases do matter}\label{sec:phases.matter}
In what follows we offer a first insight into the role of CP violating phases regarding a subset of (representative) observables.
All other observables considered  in the full phenomenological analysis of Section~\ref{sec:num:analysis} can be understood from a generalisation of the discussion here carried. 
We work under the hypothesis of having $n_s=2$  massive sterile states; the neutral spectrum thus comprises 5 states, with masses $m_i$ (with $i=1,...,5$), including the 3 light (mostly active) neutrinos and two heavier states, with masses $m_{4,5}$. The leptonic mixings (whose precise origin is not specified) are parametrised by a $5\times5$ unitary mixing matrix, $\mathcal{U}$.
As described in Appendix~\ref{app:model}, the full mixing matrix $\mathcal{U}$ 
can be cast in terms\footnote{For concreteness, we have chosen the six Dirac CP phases $\delta_{\alpha 4}$ and $\delta_{\alpha 5}$ to be non-vanishing.} of 10 real mixing angles $\theta_{ij}$, 6 Dirac phases  $\delta_{ij}$ and 4 Majorana phases, $\varphi_i$. 

\bigskip
The study carried in  this  section is accompanied by a brief analytical discussion, relying on a simplified approach to this ``3+2 toy model''. In particular, and for the purpose of deriving clear (and compact) analytical expressions, in this section we work under the following assumptions:
firstly, the mixings between the active and the sterile states (i.e., $\theta_{\alpha i}$ with $\alpha = e, \mu, \tau$ and $i=4,5$) are assumed to be sufficiently small so that $\cos \theta_{\alpha i} \approx 1$ to a very good approximation. The $3\times 2$ rectangular matrix encoding the active-sterile mixings can then be parametrised as \begin{equation}
\mathcal{U}_{\alpha (4,5)} \approx 
\left (\begin{array}{cc}
s_{14} e^{-i(\delta_{14}-\varphi_4)} &
s_{15} e^{-i(\delta_{15}-\varphi_5)} \\
s_{24} e^{-i(\delta_{24}-\varphi_4)} &
s_{25} e^{-i(\delta_{25}-\varphi_5)} \\
s_{34} e^{-i(\delta_{34}-\varphi_4)} &
s_{35} e^{-i(\delta_{35}-\varphi_5)} 
\end{array}
\right)\,,
\end{equation}
with $s_{\alpha i} = \sin \theta_{\alpha i}$, and where 
$\delta_{\alpha i}$ ($\varphi_i$) denote Dirac (Majorana) phases. We further assume $\theta_{\alpha 4} \approx \theta_{\alpha 5}$, and take the heavy states to be nearly mass-degenerate\footnote{This offers the advantage of further simplifying the expressions, allowing to factor out the loop functions from the sums over the heavy states.}, $m_4 \approx m_5 \gg \Lambda_\text{EW}$ (typically of the order of a few~TeV). 

We emphasise that all numerical results (displayed in the present section and throughout the full manuscript) are obtained without relying on any approximation, taking into account all the contributions present in the most general setup. 

As this first discussion is dedicated to understanding and rendering visible the role of phases, no experimental constraints will be applied (certain observables might thus reach values already in disagreement with current experimental bounds). 

\subsection{cLFV decay rates: sensitivity to CPV 
phases}
In what follows, we focus on $\mu-e$ sector flavour violation, and consider
the following subset of observables:
BR($\mu \to e \gamma$), BR($\mu \to 3 e $) and BR($Z \to e \mu$). We then devote a brief dedicated discussion to $\mu-e$ conversion in nuclei. 

\begin{figure}
    \centering
\mbox{\hspace*{-5mm}    \includegraphics[width=0.51\textwidth]{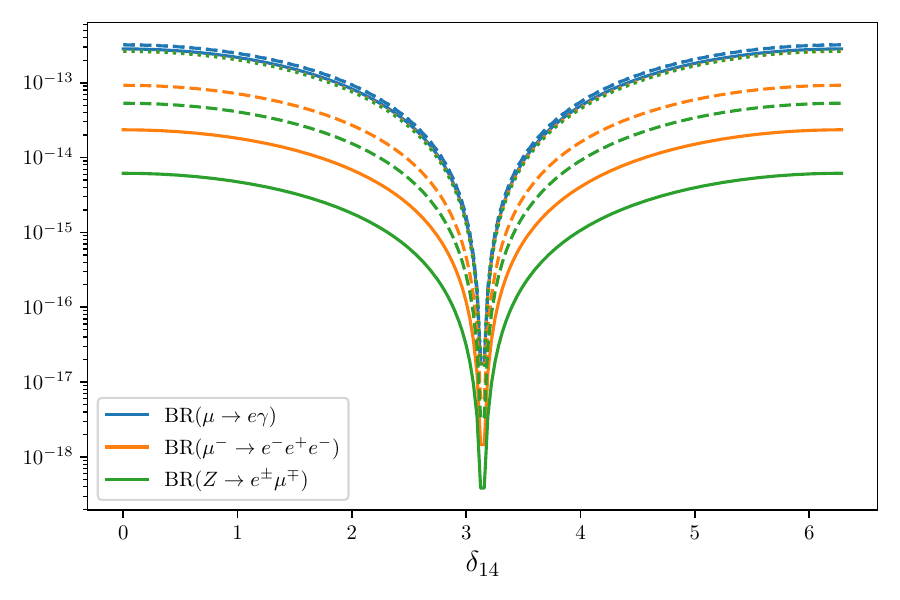} \hspace*{2mm}   
    \includegraphics[width=0.51\textwidth]{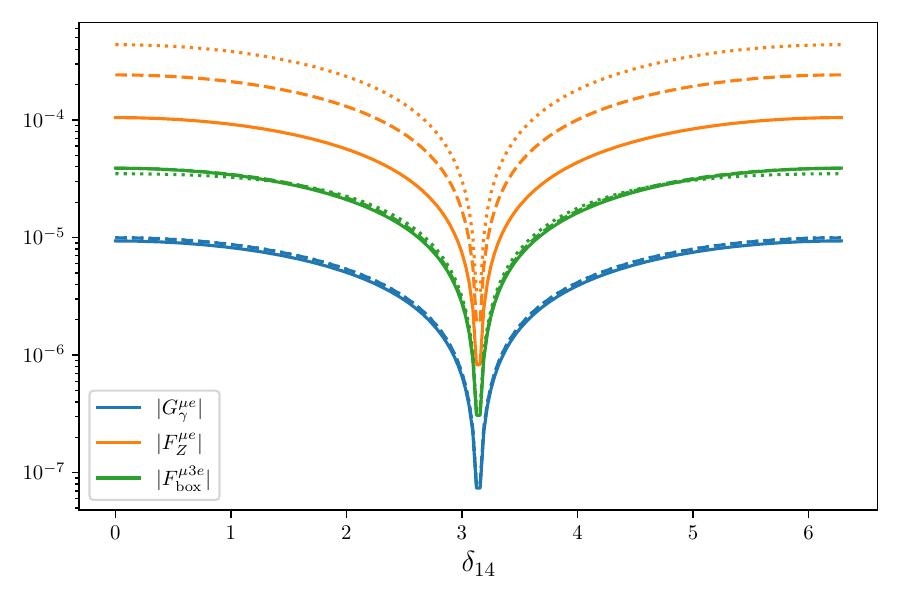}}
    \caption{Dependence of cLFV observables and several form factors (contributing to the different cLFV decay rates) on the CP violating Dirac phase $\delta_{14}$  (all other phases set to  zero).
    On the left panel we present BR($\mu \to e \gamma$) (blue), BR($\mu \to 3 e $) (orange) and BR($Z \to e \mu$) (green); on the right one finds $|G_\gamma^{\beta \alpha}|$ (blue), $|F_Z^{ \beta \alpha}|$ (orange) and $|F_\text{box}^{\beta3\alpha}|$ (green), choosing for illustrative purposes $\alpha=e$ and $\beta=\mu$. 
    In both panels, solid, dashed and dotted lines respectively correspond to the following heavy fermion masses: $m_4=m_5=1, 5, 10~\text{TeV}$.}
\label{fig:cLFV.FF.delta14}
\end{figure}

\paragraph{The role of Dirac phases} 
In Fig.~\ref{fig:cLFV.FF.delta14} we display the dependence of the above mentioned cLFV rates (and their form factors)
on the Dirac phases.
We set as an illustrative (benchmark) choice the following 
values for the mixing angles,
$\theta_{14}=\theta_{15}= 10^{-3}$, 
$\theta_{24}=\theta_{25}= 0.01$ and 
$\theta_{34}=\theta_{35}= 0$. 
Moreover, all phases are set to zero except the Dirac phase $\delta_{14}$. 
We also consider three representative values of the heavy fermion masses $m_4 = m_5 =1, 5, 10$~TeV (associated with solid, dashed and dotted lines).
As can be seen in the left panel, all considered observables have a clear dependence on $\delta_{14}$ (the only non-vanishing phase considered), with the associated rates exhibiting a strong cancellation (typically amounting to around four orders of magnitude) for $\delta_{14} = \pi$, for all considered masses of the heavy sterile states. This behaviour can be understood by considering the pattern shown by the form factors contributing to cLFV radiative and 3-body muon decays, all displaying an (analogous) suppression for $\delta_{14} = \pi$. 

Working in the limits above referred to, 
in Appendix~\ref{app:analytic.phase.observables} we present analytical expressions for the form factors contributing to the purely leptonic decays, including the full dependence on all phases. Regarding the dipole contributions, and in the case in which only 
$\delta_{14}\neq 0$, one has 
\begin{equation}\label{eq:Gmue:delta14}
    G_\gamma^{\mu e} \approx s_{1 4}s_{2 4} e^{-\frac{i}{2}(\delta_{14})} 2 \cos\left(\frac{\delta_{14}}{2}\right) G_\gamma (x_{4,5})\,,
\end{equation}
thus implying that in the simplest case of $\mu \to e \gamma$ decays, the corresponding branching fraction for the radiative decays is given by
\begin{equation}
    \mathrm{BR}(\mu \to e \gamma)\propto |G_\gamma^{\mu e}|^2 \approx 4 s_{14}^2 s_{24}^2 \cos^2\left(\frac{\delta_{14}}{2}\right)  G_\gamma^2(x_{4,5})\,,
\end{equation}
with $x_{4,5}=m_4^2/M_W^2=m_5^2/M_W^2$, thus indeed approximately vanishing for $\delta_{14}=\pi$. Similar results can be obtained for the photon penguin form factor $F_\gamma^{\mu e}$,  
as well as for one of the terms in the form factor
$F_Z^{\mu e}$ (i.e. $F_Z^{(1)}$, see Appendix~\ref{app:analytic.phase.observables}), all contributing to the rate of $\mu \to 3 e$. 
For $F_Z^{(2)}$, after carrying the sum over $\rho=e, \mu, \tau$, one finds\footnote{While one can in general neglect the contribution 
of the light (mostly active) neutrinos to the form factors here considered, that is not the case for $F_Z^{(2)}$, since the associated loop function $G_Z(x,y)$ does not vanish in the limit $x\sim 0, y \gg 1$.
As can be seen in Appendix~\ref{app:analytic.phase.observables}, despite being more complex, the term corresponding to the ``light-heavy'' contribution exhibits a similar dependence on the Dirac phases; here we only consider the dominant ``heavy-heavy'' contribution.}
\begin{equation}
F_Z^{(2)} \approx 4\,   s_{14} s_{24}  e^{-\frac{i}{2}(\delta_{14})} \left(s_{14}^2 + s_{24}^2 
\right)\, \cos\left(\frac{\delta_{14}}{2}\right) \widetilde G_Z (x_{4,5})\,,
\end{equation}
also exhibiting a suppression for $\delta_{14}=\pi$.
In the above equation we introduced $\widetilde G_Z(x_{4,5})\equiv  G_Z(x_{4,5}, x_{4,5})$, which we also use in the following for loop functions that depend on 2 parameters, in the limit of degenerate masses (cf. Appendix~\ref{app:cLFV.others}).
Finally, the remaining contributing term can be approximately given by 
\begin{equation}
F_Z^{(3)} \approx 4\,   s_{14} s_{24}  e^{-\frac{i}{2}(\delta_{14})} \cos\left(\frac{\delta_{14}}{2}\right)
\left[s_{14}^2  \cos(\delta_{14})
+ s_{24}^2  
\right] \widetilde H_Z (x_{4,5})\,,
\end{equation}
again revealing the same behaviour\footnote{Due to the  contributions associated with the combination $C_{ij}$ (sum over all flavours), mixings involving the tau sector also contribute; taking them into account would lead to a similar dependence on 
$\cos \delta_{14}/2$.}, which is also 
present in 
the box diagram contributions to $\mu \to 3 e$. 
The latter lead to 
\begin{align}
& F_\text{box}^{(1)} \approx  4\,   s_{14}^3 s_{24} 
\cos(\delta_{14}) \cos\left(\frac{\delta_{14}}{2}\right) 
\widetilde G_\text{box} (x_{4,5})\,, \nonumber \\
& F_\text{box}^{(2)} \approx  -8\,   s_{14}^3 s_{24} 
\cos\left(\frac{\delta_{14}}{2}\right) 
\widetilde F_\text{Xbox} (x_{4,5})\,.
\end{align}
This brief discussion explains the behaviour of the different form factors, as depicted in the right panel of 
Fig.~\ref{fig:cLFV.FF.delta14}.
Albeit carried in a 
limiting case (degenerate heavy states, identical mixings, etc.), this 
discussion is helpful in understanding the  more complex behaviours which will emerge upon the numerical analysis 
presented in Section~\ref{sec:num:analysis}. 

\bigskip
Concerning cLFV $Z$ decays, we do not explicitly discuss their phase dependence here; let us just notice that the form factors parameterising  $Z$ decays (cf. Section~\ref{sec:Zdecays}) can be understood as $Z$-penguins at non-vanishing momentum transfer, i.e. $q^2 = M_Z^2$, and are therefore expected to have a very similar behaviour in what concerns the impact of the CPV phases. 
This can be quantitatively confirmed in Fig.~\ref{fig:cLFV.FF.delta14} upon comparison of the dependence of the $Z$-penguin form factor $F_Z^{\mu e}$ and the branching fraction of the $Z \to e \mu$ decay on $\delta_{14}$.

\begin{figure}[h!]
    \centering
\includegraphics[width=0.48\textwidth]{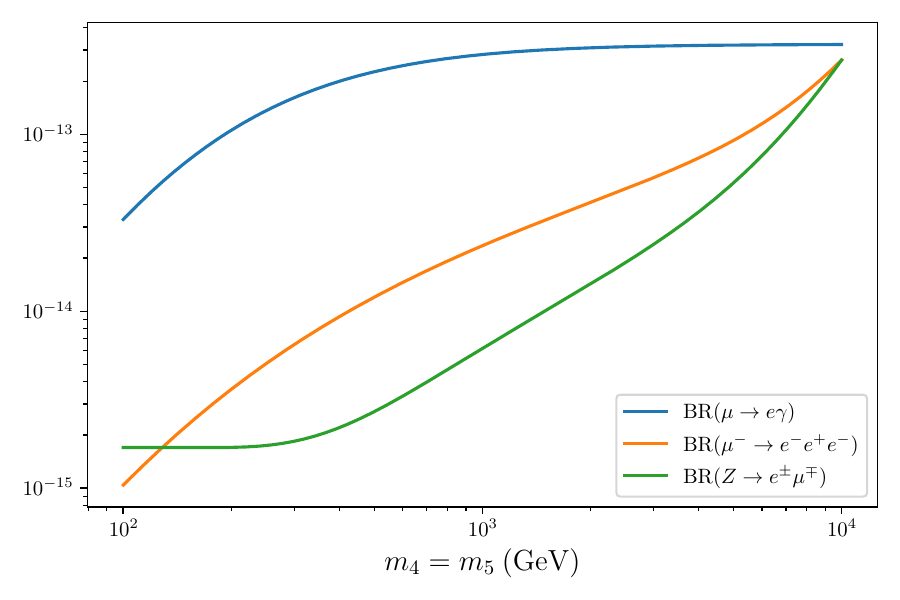}
\includegraphics[width=0.48\textwidth]{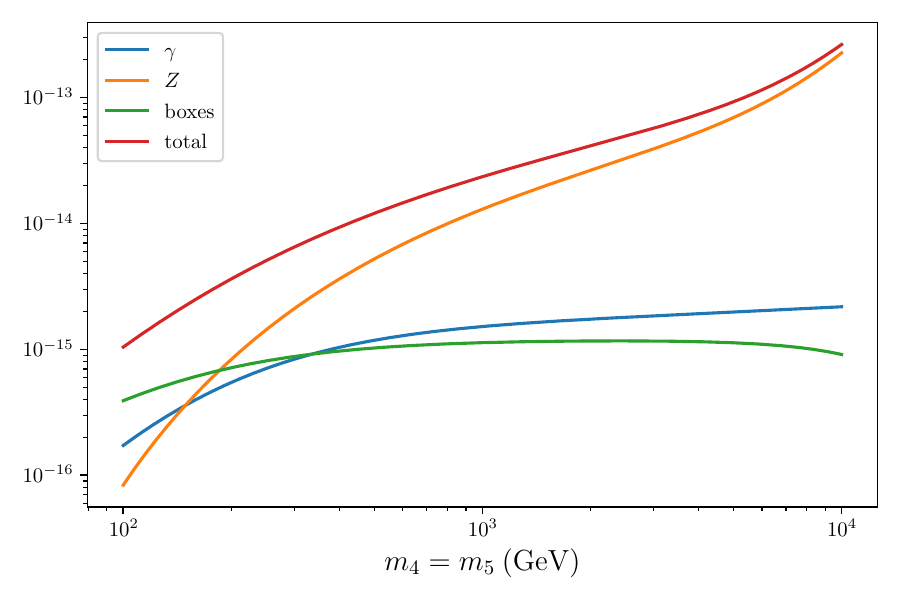}
\caption{cLFV observables (left panel) and choice of contributing form factors to the different rates (right panel), as a function of the degenerate heavy sterile mass, $m_4=m_5$ (in~GeV), for vanishing CPV phases. 
On the left panel we present
BR($\mu \to e \gamma$) (blue), BR($\mu \to 3 e $) (orange) and BR($Z \to e \mu$) (green); on the right, one finds the contributions of the $\gamma$-penguin form factors
$F_\gamma^{\beta \alpha}$ and $G_\gamma^{\beta \alpha}$ (blue), the $Z$-penguin form factor $F_Z^{ \beta \alpha}$ (orange) and the box form factor $F_\text{box}^{\beta3\alpha}$ (green) to the total branching ratio of decays of the form $\ell_\beta\to3\ell_\alpha$ (red), 
choosing for illustrative purposes $\alpha=e$ and $\beta=\mu$. 
}
\label{fig:clfvFF_M}
\end{figure}

\bigskip
For completeness, and before moving to the possible impact of the Majorana phases, notice that the individual form factors - and hence the full rates - depend on the mass of the new fermions (already manifest from the behaviour shown in Fig.~\ref{fig:cLFV.FF.delta14}), as shown in both panels of Fig.~\ref{fig:clfvFF_M} (which were obtained considering vanishing values of all phases).

\paragraph{Impact of  Majorana phases on cLFV decay rates}

\begin{figure}[h!]
    \centering
\mbox{\hspace*{-5mm}   
    \includegraphics[width=0.51\textwidth]{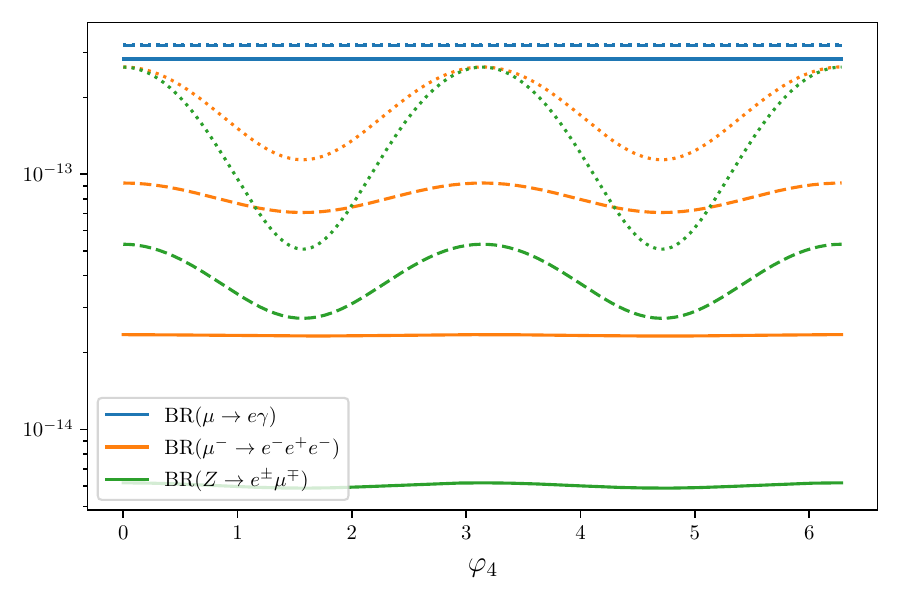} \hspace*{2mm}   
    \includegraphics[width=0.51\textwidth]{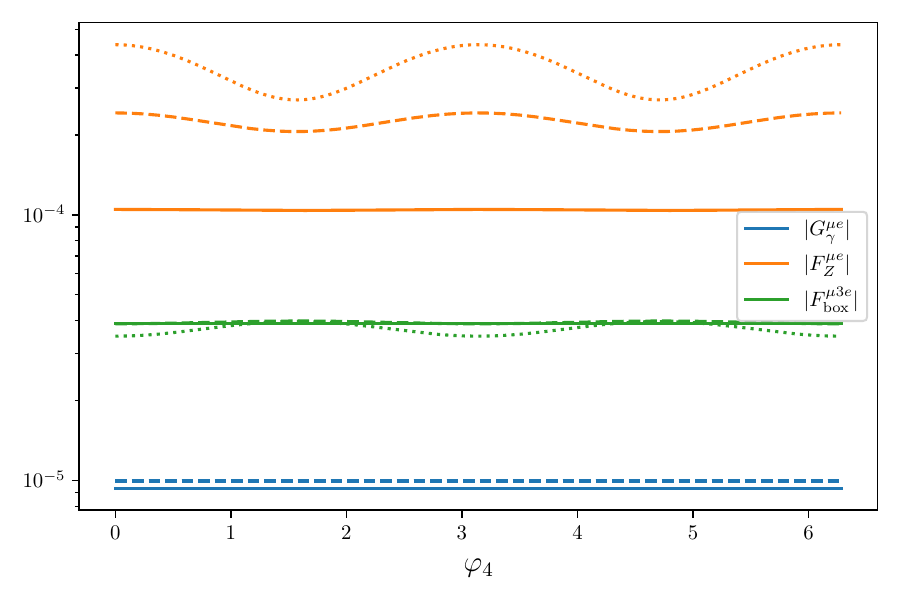}
    }
\caption{
Dependence of cLFV observables and several contributing form factors on the CP violating Majorana phase $\varphi_{4}$  (with all other phases set to  zero).
On the left panel we present
BR($\mu \to e \gamma$) (blue), BR($\mu \to 3 e $) (orange) and BR($Z \to e \mu$) (green); on the right, one has 
$|G_\gamma^{\mu e}|$ (blue), $|F_Z^{ \mu e}|$ (orange) and $|F_\text{box}^{\mu 3e}|$ (green).
In both panels, solid, dashed and dotted lines respectively correspond to $m_4=m_5=1, 5, 10~\text{TeV}$.}
\label{fig:cLFV.FF.phi4}
\end{figure}

In what follows, we set the mixing angles to the same values as before, and choose the same three values of the heavy fermion masses. All phases are set to zero except $\varphi_{4}$.  The results for the observables (the same as studied concerning the Dirac phase) and the contributing form factors are displayed in Fig.~\ref{fig:cLFV.FF.phi4}, as a function of the Majorana phase $\varphi_{4}$. 

As expected, there  is no dependence of the radiative decays on the Majorana phase (cf. the full expression for BR($\ell_\beta \to \ell_\alpha \gamma$) given in Appendix~\ref{app:analytic.phase.observables}). This is also true 
for all dipole and dipole-like contributions.    
In contrast, the three-body decays (and the cLFV $Z$ decays) do exhibit a non-negligible dependence on the Majorana phase, as can be verified from both panels of Fig. \ref{fig:cLFV.FF.phi4}.
This is especially true for heavier mass regimes, in which the relative contribution of the form factors sensitive to $\varphi_4$ 
($Z$-penguins and to a lesser extent box-contributions)
become more important (cf. right-handed panel of Fig.~\ref{fig:clfvFF_M}).
Indeed, in the simplified limits of the form factors (see 
Appendix~\ref{app:analytic.phase.observables}), one verifies that only two contributions in the form factors depend on the Majorana phase, $F_Z^{(3)}$ and $F_\text{box}^{(1)}$. 
In the presence of a single non-vanishing Majorana phase, their expressions are:
\begin{align}\label{eq:Majorana:FF}
    & F_Z^{(3)} \approx 4s_{14}s_{24} \left(s_{14}^2 +
    s_{24}^2 \right) \cos^2(\varphi_4) \widetilde H_Z(x_{4,5}) \,, \nonumber \\
    & F_\text{box}^{(1)} \approx 4 s_{14}^3 s_{24} \cos^2(\varphi_4) \widetilde G_\text{box}(x_{4,5})\,.
\end{align}
The impact of the Majorana phase on the cLFV $Z$ decays can be also understood in analogy from the dependence of the corresponding $Z$ penguin form factor. This is readily visible from inspection of Fig.~\ref{fig:cLFV.FF.phi4}, which reveals a very similar dependence on $\varphi_4$.

\begin{figure}
    \centering
\mbox{\hspace*{-5mm}
\includegraphics[width=0.51\textwidth]{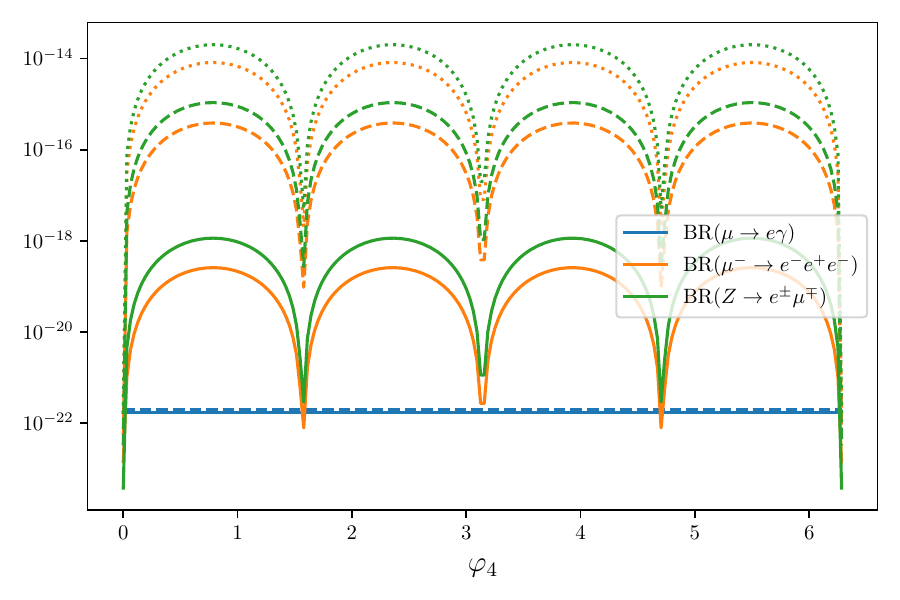} \hspace*{2mm}
    \includegraphics[width=0.51\textwidth]{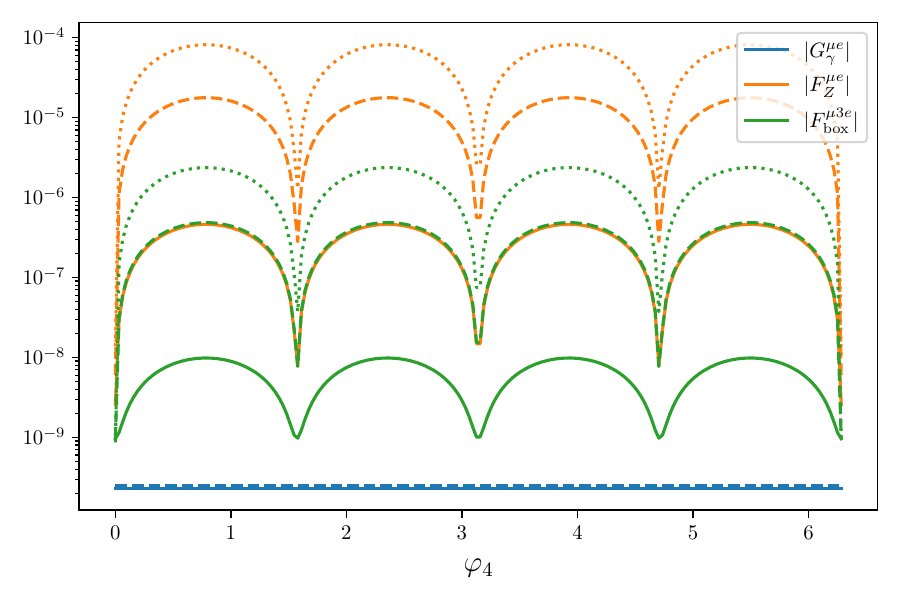}}
\caption{Dependence of cLFV observables (left) and several contributing form factors contributing to the decay rates (right) on the
Majorana CPV phase $\varphi_{4}$, for non-vanishing Dirac CPV phase, $\delta_{14}=\pi$.
We again present on the left
BR($\mu \to e \gamma$) (blue), BR($\mu \to 3 e $) (orange) and BR($Z \to e \mu$) (green), while on the right one finds $|G_\gamma^{\mu e}|$ (blue), $|F_Z^{ \mu e}|$ (orange) and $|F_\text{box}^{\mu 3e}|$ (green). We consider fixed values of the mass of heavy sterile states: 
solid, dashed and dotted lines respectively correspond to $m_4=m_5=1, 5, 10~\text{TeV}$.}
\label{fig:clfvFF_phi_delta_pi}
\end{figure}

\paragraph{Joint Dirac-Majorana phase effects}
A first view of the joint effect of Majorana and Dirac phases can be obtained by setting one to a fixed non-vanishing value, while the other is varied over its full range (i.e. $\in [0, 2\pi ]$). This is shown in Fig.~\ref{fig:clfvFF_phi_delta_pi}, where we re-evaluate the dependence of the cLFV rates, and of a subset of form factors, on the Majorana phase $\varphi_4$ (similar to what was presented in Fig.~\ref{fig:cLFV.FF.phi4}), but now taking $\delta_{14}=\pi$. 

The effects arising from the presence of both phases 
are clearly manifest, especially when compared with the plots of Figs.~\ref{fig:cLFV.FF.delta14} and~\ref{fig:cLFV.FF.phi4}. Recall that $\delta_{14}=\pi$
was found to lead to a strong cancellation of the form factors (see discussion of Fig.~\ref{fig:cLFV.FF.delta14}); thus, the non-vanishing contributions to the observables (3-body and $Z$ decays) are associated with the form factors exhibiting a non-trivial dependence on $\varphi_4$ - as mentioned before.
Relying again on the simple analytical estimates (see Appendix~\ref{app:analytic.phase.observables}), one now verifies that taking $\delta_{14}=\pi$ leads to the following modification of the form factors sensitive to the Majorana phases (and hence of the observables): 
\begin{align}\label{eq:Majorana_fixedDirac:FF}
    & F_Z^{(3)} \propto
    s_{14}s_{24} \left(s_{14}^2 -
    s_{24}^2 \right) \sin(2\varphi_4)\,  \widetilde H_Z(x_{4,5}) \,, \nonumber \\
    & F_\text{box}^{(1)} \propto
    s_{14}^3 s_{24} \sin(2\varphi_4) \, \widetilde G_\text{box}(x_{4,5})\,,
\end{align}
explaining the modified pattern (shift of $\pi/2$ and dependence on $2\varphi_4$) visible in 
Fig.~\ref{fig:clfvFF_phi_delta_pi}.

\subsection{Neutrinoless $\mu-e$ conversion in nuclei and CP violating phases}
\label{sec:muecon_simple}
\begin{figure}[h!]
    \centering
\mbox{\hspace*{-5mm}
\includegraphics[width=0.51\textwidth]{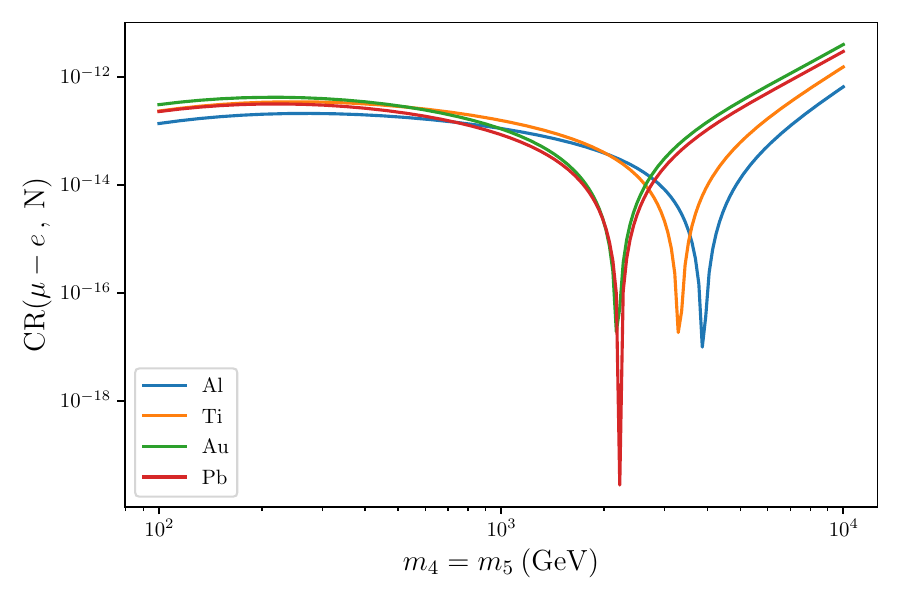} \hspace*{2mm}
\includegraphics[width=0.51\textwidth]{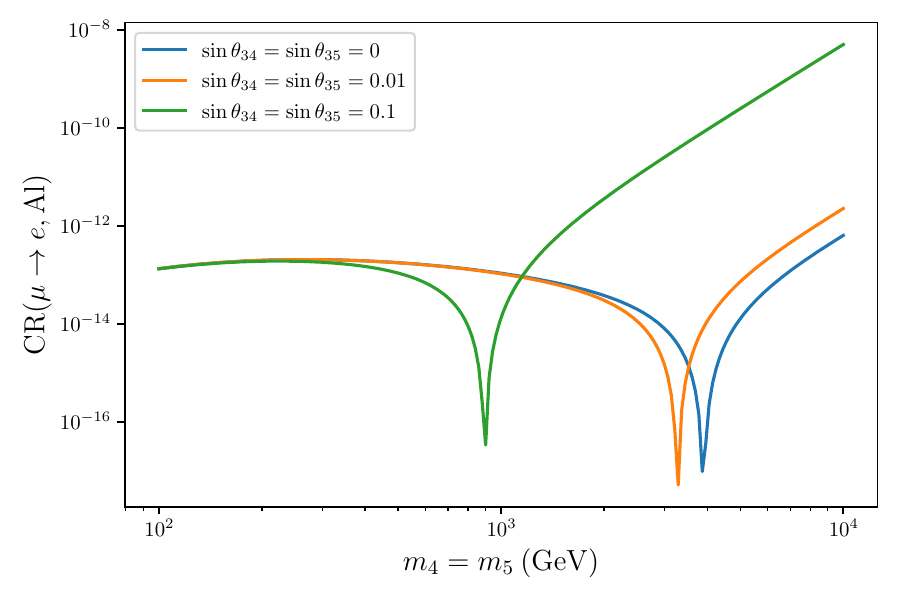}}
\caption{Neutrinoless $\mu-e$ conversion in nuclei as 
a function of the degenerate heavy sterile mass, $m_4=m_5$ (in~GeV). On the left, we set 
$\theta_{1j}=10^{-3}$, $\theta_{2j}=0.01$ and $\theta_{3j}=0$ ($j=4,5$), and consider different muonic atoms: Aluminium (blue), Titanium (orange), Gold (green) and Lead (red). On the right, CR($\mu-e$, Al) for different values of the 
tau-sterile mixing angles: $\theta_{3j}=0$ (blue), $\theta_{3j}=0.01$ (orange) and  $\theta_{3j}=0.1$ (green), again with $\theta_{1j}=10^{-3}$, $\theta_{2j}=0.01$ (with $j=4,5$).}
\label{fig:CR_M:nuclei_tau}
\end{figure}

We begin by illustrating how the predictions for the conversion rate (in particular the dependence on the heavy fermion mass) reflect the nature of the muonic atom, i.e. 
the chosen target nucleus (see Eq.~(\ref{eq:def:CRfull})). This is shown on the left panel of Fig.~\ref{fig:CR_M:nuclei_tau}, where we display CR($\mu-e$, N) as a function of the (degenerate) heavy masses, for Aluminium, Titanium, Gold and Lead nuclei, which have been chosen for past and future experimental searches. We fix the active-sterile mixing angles as before ($\theta_{14}=\theta_{15}=10^{-3}$, $\theta_{24}=\theta_{25}=0.01$ and $\theta_{34}=\theta_{35}=0$), and set all phases to zero. We recover the behaviour originally pointed out in~\cite{Alonso:2012ji}, with the distinct rates all exhibiting a sharp cancellation for a given value  of the heavy fermion mass  - which we denote $m_{4,5}^c$ (the actual value of $m_{4,5}^c$, and the ``depth'' observed in the conversion rate, both depend on the considered nucleus).    

Until now, and for simplicity, we have not considered mixings of the sterile states to the third generation of leptons; however, and even though we are studying cLFV in the $\mu-e$ sector, certain observables are sensitive to tau mixings through the $C_{ij}$ coupling (sum over all flavours) which arises from the $Z$-penguin contribution, cf. Eq.~(\ref{eq:cLFV:FF:FZ}). On the right panel of Fig.~\ref{fig:CR_M:nuclei_tau}, we consider Aluminium nuclei, and again depict CR($\mu-e$, Al) vs. the heavy sterile masses, for different choices of $\theta_{3j}$ ($j=4,5$), keeping the other mixing angles and phases as before.  
As expected, and despite seemingly indirect, the impact is significant, both to the rate itself, and in what concerns the value of $m_{4,5}^c$ associated with the cancellation in the conversion rate. This is all the most important when $\theta_{3j} \gg \theta_{1j},\theta_{2j}$ 
($j=4,5$), as can be inferred from the green line. 

\begin{figure}
    \centering
\mbox{\hspace*{-5mm}\includegraphics[width=0.51\textwidth]{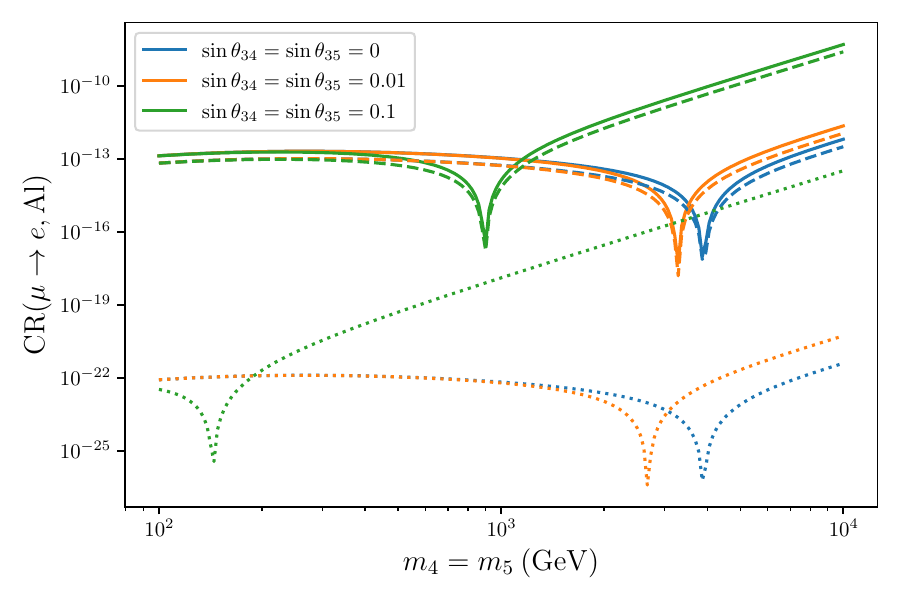}
\hspace*{2mm}
\includegraphics[width=0.51\textwidth]{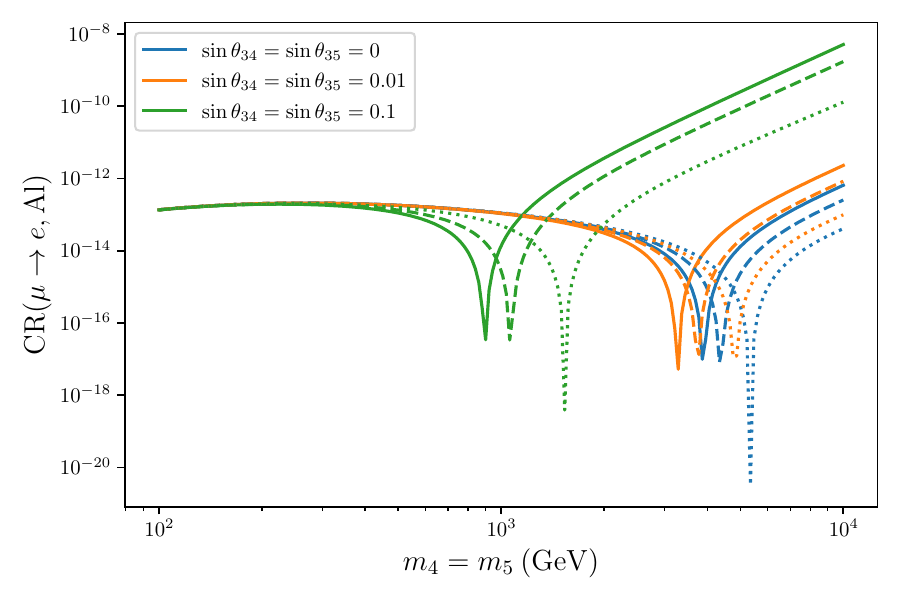}}
\caption{Neutrinoless $\mu - e$ conversion in Aluminium as 
a function of the degenerate heavy sterile mass, $m_4=m_5$ (in~GeV).  We set $\theta_{1j}=10^{-3}$, $\theta_{2j}=0.01$, for different values of the tau-sterile mixing angles: $\theta_{3j}=0$
 (blue), $\theta_{3j}=0.01$ (orange) and  $\theta_{3j}=0.1$ (green), with $j=4,5$.
 On the left, we set all Majorana phases to zero and consider three choices of the Dirac phase, $\delta_{14}=0, \pi/2$ and $\pi$, respectively corresponding to solid, dashed and dotted lines. Conversely, on the right panel
  all Dirac phases are set to zero, and we consider three choices of the Majorana phase, $\varphi_4=0, \pi/4$ and $\pi/2$, corresponding to solid, dashed and dotted lines.
  }
\label{fig:CR_Al_M:cancel_phases}
\end{figure}

The impact of the CPV phases on the conversion rate is studied in both panels of Fig.~\ref{fig:CR_Al_M:cancel_phases}: for three choices of $\theta_{3j}=0, 0.01$ and $0.1$, we consider the impact of Dirac and Majorana phases on 
CR($\mu-e$,~Al), displayed as a function of the masses of the heavy states. 
For the case of vanishing $\varphi_{4}$, the choice of Dirac phases leads to effects which could already be expected from the discussion of the previous subsection\footnote{We notice that although the full expression is considerably more involved, the form factors contributing to the conversion rate include those already presented for the radiative and 3-body cLFV decays. Additional ones (i.e. boxes with an internal quark line) only depend on a single combination of $\mathcal{U}_{e i}^{\phantom{\ast}}\,\mathcal{U}_{\mu i}^\ast$, and exhibit a behaviour similar to the dipole contributions, being only sensitive to Dirac phases.}. 
In particular, and as can be seen on the left panel of Fig.~\ref{fig:CR_Al_M:cancel_phases}, notice the very important suppression for $\delta_{14}=\pi$; for the latter case, and for sizeable 
$\theta_{3j}$ one also observes a significant displacement of $m_{4,5}^c$, lighter by almost an order of magnitude. This is the result of a cancellation of (numerically) very small terms. 

The effect of the Majorana phases (for vanishing Dirac phases) is more interesting: while having already a visible effect on the overall scale of CR($\mu-e$, Al) - a reduction by a factor $\sim 100$ between $\varphi_4=0$ and  
$\varphi_4=\pi/2$ -, they also modify the value of $m_{4,5}^c$
($m_{4,5}^c (\varphi_4=0)\approx 850~\text{GeV}$ while $m_{4,5}^c (\varphi_4=\pi/2)\approx 1.5~\text{TeV}$). 

\begin{figure}[h!]
\mbox{ \hspace*{-5mm}   \centering
\includegraphics[width=0.51\textwidth]{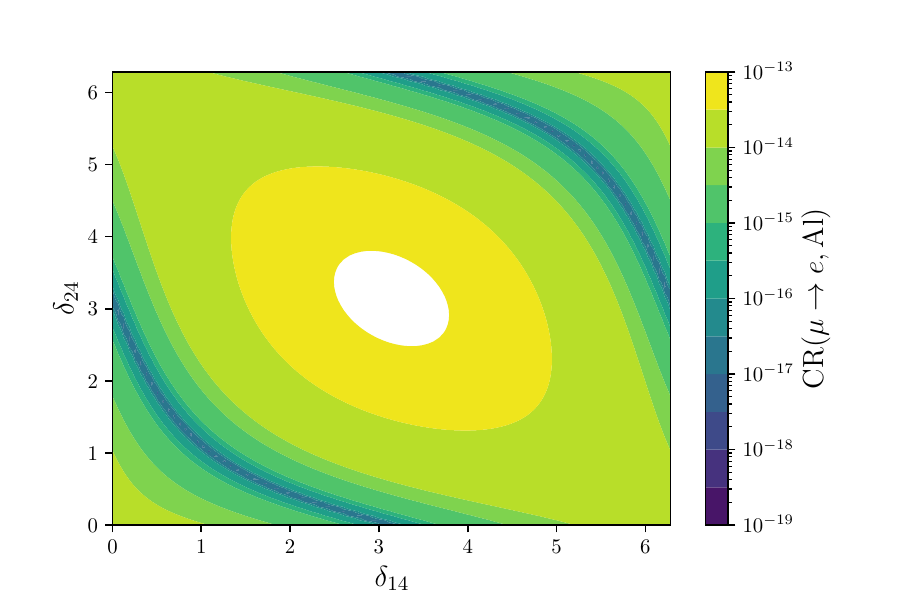}\hspace*{2mm}
\includegraphics[width=0.51\textwidth]{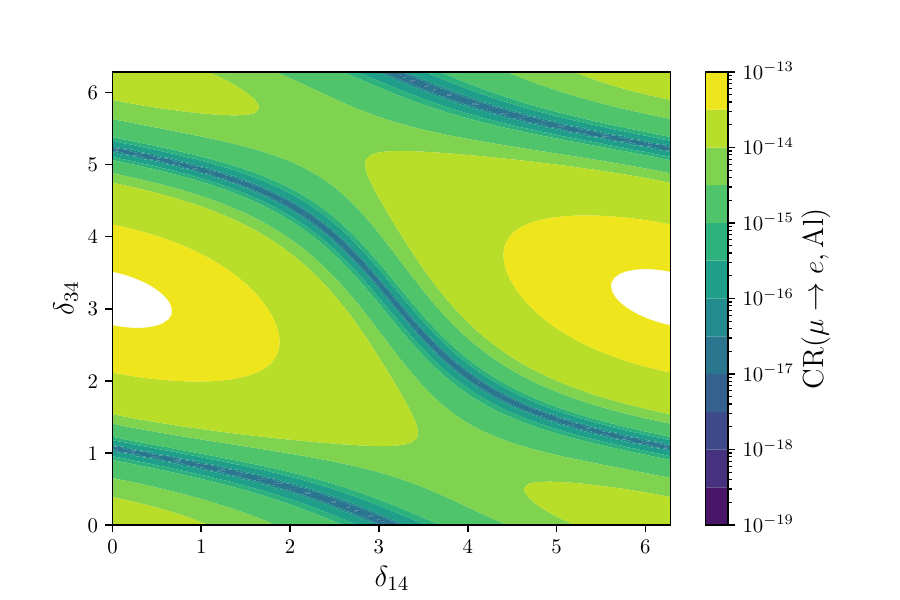}}\\
\mbox{\hspace*{-5mm}
\includegraphics[width=0.51\textwidth]{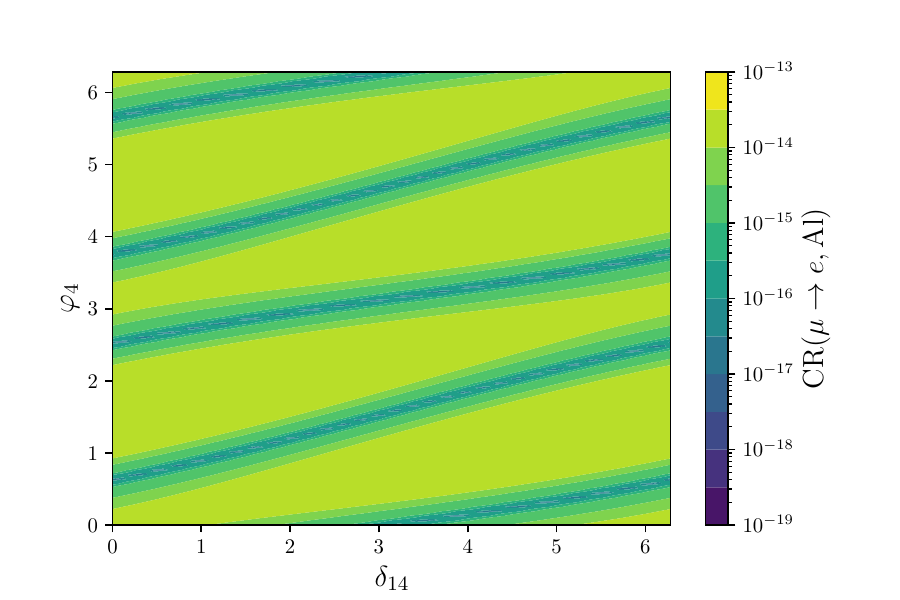}\hspace*{2mm}
\includegraphics[width=0.51\textwidth]{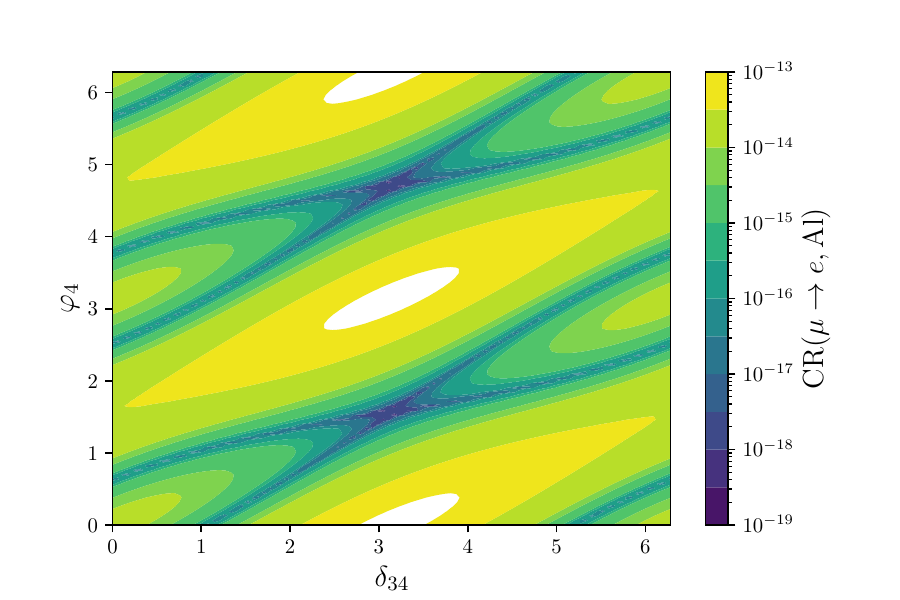}}
\caption{Contour plots for cLFV $\mu-e$ conversion in Aluminium, for fixed values of the degenerate heavy sterile mass, $m_4=m_5=1$~TeV, 
for $\theta_{1j}=10^{-3}$, $\theta_{2j}=0.01$ and $\theta_{3j}=0.1$ ($j=4,5$) and varying CPV phases: 
on the top row, spanned by pairs of Dirac phases, $(\delta_{14}-\delta_{24})$ and 
$(\delta_{14}-\delta_{34})$, respectively left and right panels; bottom row, spanned by Dirac-Majorana phases, 
$(\delta_{14}-\varphi_{4})$ and 
$(\delta_{34}-\varphi_{4})$, respectively left and right panels. 
The colour scheme denotes the associated value of CR($\mu-e$, Al) as indicated by the colour bar to the right of each plot (white regions denote CR($\mu-e$, Al)$> 10^{-13}$).}
\label{fig:CR_Al:contour_phases}
\end{figure}

Finally, we consider for illustrative purposes the simultaneous effects of two phases (Dirac, or combined Dirac and Majorana). This is shown in the contour plots of Fig.~\ref{fig:CR_Al:contour_phases}, for which we again consider $m_4=m_5=1$~TeV, and fix the active-sterile mixings to 
$\theta_{1j}=10^{-3}$, $\theta_{2j}=0.01$ and $\theta_{3j}=0.1$ ($j=4,5$).

The different plots in Fig.~\ref{fig:CR_Al:contour_phases} summarise (and generalise) the previous findings of Figs.~\ref{fig:CR_M:nuclei_tau} and~\ref{fig:CR_Al_M:cancel_phases}: for fixed values of the ``standard'' input parameters (i.e. the mass of the heavy states and the active-sterile mixing angles), a variation of the phases - individually or as a joint effect - can lead to significant
changes in the predictions for the conversion rate.
This is seen in the panels of Fig.~\ref{fig:CR_Al:contour_phases}, with the rates ranging from as low as $10^{-18}$ (dark blue), to values above $10^{-14}$ (bright yellow), or even beyond 
$10^{-13}$ (white).
Here, one can also observe regions of constructive interference, which are of different origin.
For example, in the upper left plot of Fig.~\ref{fig:CR_Al:contour_phases} we show the conversion rate as a function of $\delta_{14}$ and $\delta_{24}$; if both phases are close to $\pi$, the suppression vanishes (as it depends on  $\cos((\delta_{14}-\delta_{24})/2)$), while the complex exponential multiplying the contributions depends on $(\delta_{14}+\delta_{24})/2$, thus leading to a ``sign-flip'' for values of $\delta_{14}\simeq\delta_{24}\simeq\pi$ (cf. App.~\ref{app:analytic.phase.observables}).
This affects the signs of the individual contributions
to the different form factors, and can lead to an overall constructive interference, as visible in the plots.
In the remaining panels in which one observes an enhancement of the rate with respect to vanishing CPV phases (conversion rate as a function of $\delta_{14}$ and $\delta_{34}$, and $\delta_{34}$ and $\varphi_{4}$ respectively), the source of constructive interference solely lies in the $Z$-penguin form factor. 
In this case, the interference occurs between terms that depend on the tau-sterile mixing angles $\theta_{3j}\,,\:j=4,5$ and the remaining terms, in conjunction with the effects of other phases.

\subsection{Muonium anti-Muonium oscillations}
Other cLFV observables rely on combinations of the already discussed form factors, so we will not address them individually. However, a few remarks concerning Muonium oscillations\footnote{The cLFV Muonium decays are also expected to be impacted in the same way, as can be understood from the corresponding form factors collected in Appendix~\ref{app:analytic.phase.observables}.} are in order, as $\text{Mu}-\overline{\text{Mu}}$ is unique in the sense that it only receives contributions from box diagrams, and thus offers a direct access to this topology (and associated form factors).
Notice that the oscillation probability is proportional to a single effective four-fermion coupling $G_{M\overline{M}}$~\cite{Clark:2003tv,Cvetic:2005gx}.
Moreover, as can be seen from the corresponding expressions for $G_{M\overline{M}}$ (see Eq.~(\ref{eq:Gmumu}) in Appendix~\ref{app:cLFV.others}), this observable is only sensitive to $\mu-e$ flavour violation. 

\begin{figure}
    \centering
    \mbox{ \hspace*{-5mm}   \centering
\includegraphics[width=0.51\textwidth]{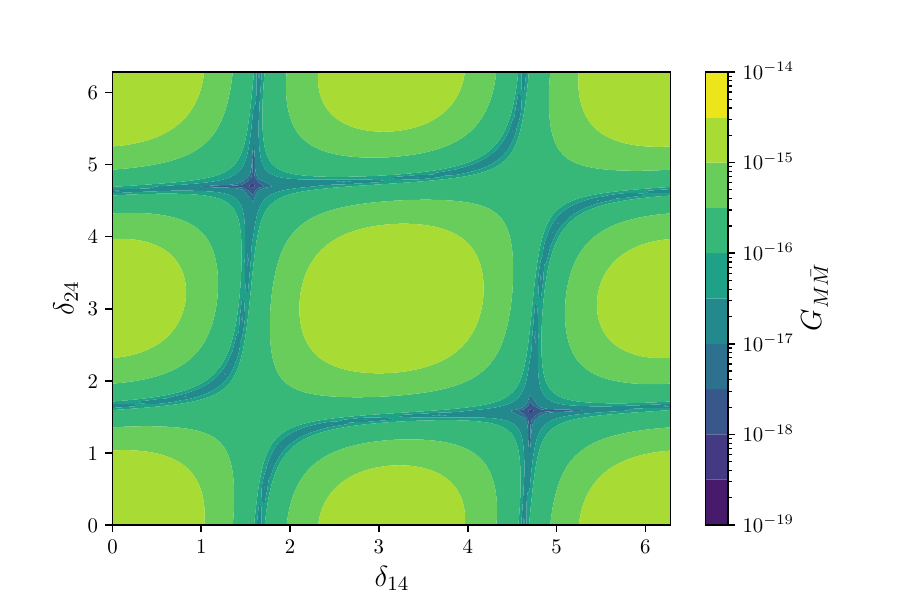}\hspace*{2mm}
\includegraphics[width=0.51\textwidth]{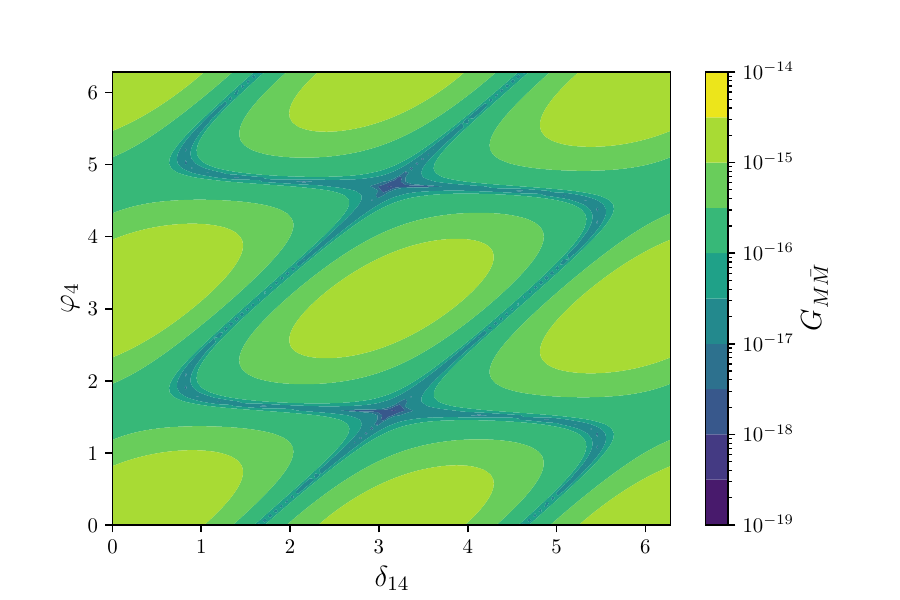}}
    \caption{Contour plots for the effective coupling $G_{M\overline{M}}$ of $\mathrm{Mu}-\overline{\mathrm{Mu}}$ oscillations, for fixed values of the degenerate heavy sterile mass, $m_4=m_5=1$~TeV, with 
$\theta_{1j}=10^{-3}$, $\theta_{2j}=0.01$ 
and varying pairs of CPV phases: 
$(\delta_{14}-\delta_{24})$ and 
$(\delta_{14}-\varphi_{4})$, respectively in the left and right panels.
Colour code as in Fig.~\ref{fig:CR_Al:contour_phases}.}
    \label{fig:GMM:contour_phase}
\end{figure}

In Fig.~\ref{fig:GMM:contour_phase}, we present contour plots for the effective coupling $G_{M\overline{M}}$, spanned by varying pairs of Dirac CP violating phases ($\delta_{14}$ and $\delta_{24}$ on the left panel), and Dirac-Majorana CPV phases 
($\delta_{14}$ and $\varphi_{4})$ on the right panel). As previously done, we set $m_4=m_5=1$~TeV, and 
$\theta_{1j}=10^{-3}$, $\theta_{2j}=0.01$  ($j=4,5$). (Recall that $\text{Mu}-\overline{\text{Mu}}$ oscillations are not sensitive to $\theta_{3j}$.)

\subsection{cLFV and CP violating phases in the 
$\pmb{\tau-\ell}$ sectors}

Leptonic cLFV tau decays (i.e. $\tau \to \ell_\alpha \gamma$ or  $\tau \to 3\ell_\alpha$
with $\alpha=e, \mu$) receive contributions from form factors whose structure is analogous to that of the corresponding muon decays (allowing for the different flavour composition of the final state leptons); likewise, tau leptons can be present as final states of cLFV $Z$ decays. Since the dependence of the observables on the CPV phases is in all similar to what has been discussed  for the $\mu-e$ sector, we refrain from a dedicated analytical study (these observables will be included in the numerical study of Section \ref{sec:num:analysis}).
Notice that due to the large tau mass, one can also have semileptonic cLFV tau decays, with the final state composed of a light lepton and (light) mesons. However, we will not address them in the present study. 

\subsection{Other possible enhancements}
\label{sec:enh_others}
\begin{figure}[t!]
    \centering
\mbox{   \hspace*{-5mm}  \includegraphics[width=0.51\textwidth]{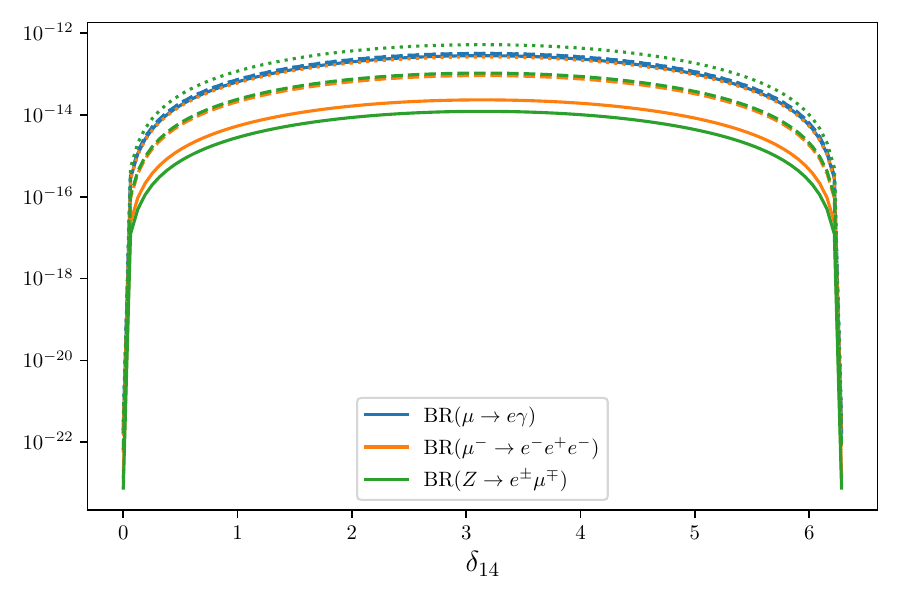}\hspace*{2mm} 
    \includegraphics[width=0.51\textwidth]{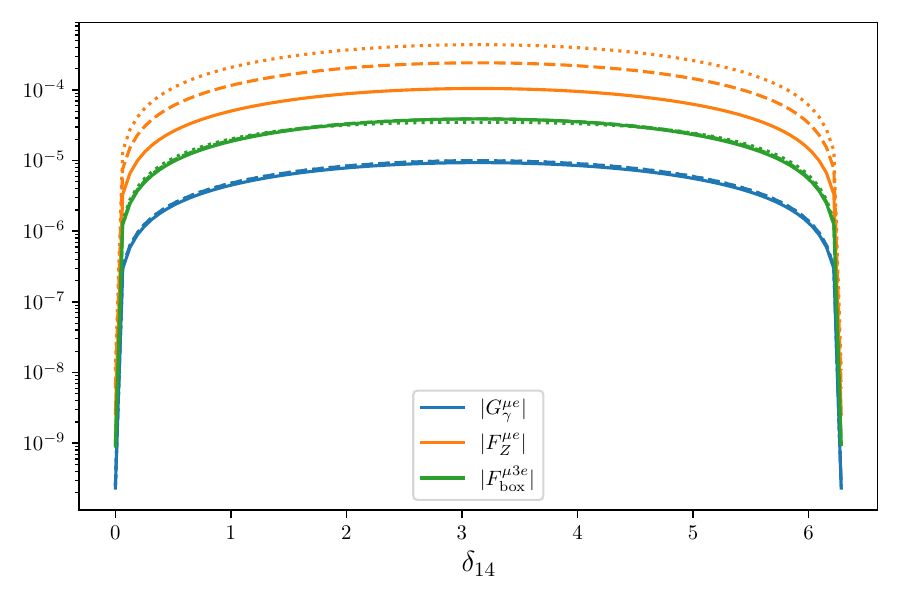}}\\
\mbox{  \hspace*{-5mm}   \includegraphics[width=0.51\textwidth]{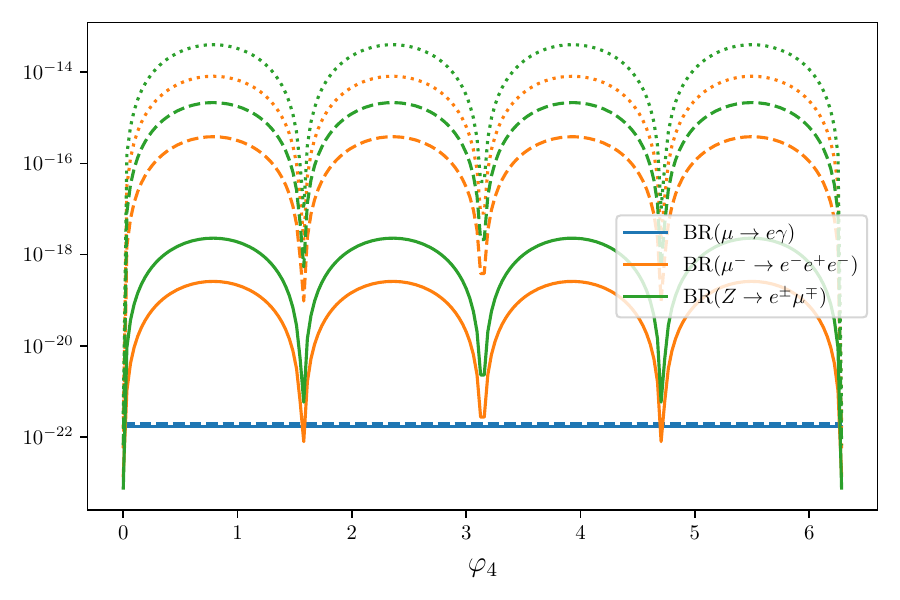}\hspace*{2mm}
    \includegraphics[width=0.51\textwidth]{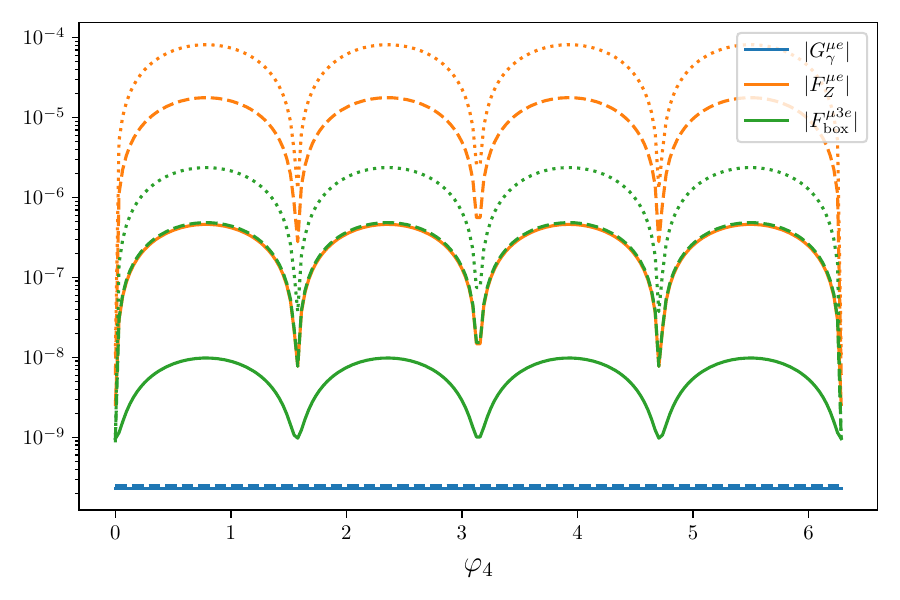}}
    \caption{Dependence of cLFV observables and several form factors (contributing to the different decay rates) on the CPV phases $\delta_{14}$ and $\varphi_4$, 
    as done in Figs.~\ref{fig:cLFV.FF.delta14} and~\ref{fig:cLFV.FF.phi4}, but with $\sin\theta_{15} = - \sin\theta_{14}$. 
    }
    \label{fig:cLFV.FF.delta14.signs}
\end{figure}

So far, and relying on the simplifying approximation $\theta_{\alpha 4} \approx \theta_{\alpha 5}$, we have mostly addressed effects of destructive interference leading to a strong suppression of the cLFV
observables (due to a cosine dependence of the corresponding form factors on the CPV phases, see Eqs. (\ref{eq:Gmue:delta14}-\ref{eq:Majorana_fixedDirac:FF})). 
However, in the most general case 
different behaviours (in particular generic enhancements)
can be encountered upon 
relaxation of $\theta_{\alpha 4} \approx \theta_{\alpha 5}$. 
For example, by considering a simple sign difference in one of the flavours ($\theta_{14} = - \theta_{15}$), we are led to a generic cancellation as in 
this case, one finds a sinus-like dependence of the observables on the phases. 
For instance the photon dipole form factor is now given by 
\begin{equation}\label{eq:Gmue:delta14:minus}
    G_\gamma^{\mu e} \approx - i s_{1 4}s_{2 4} e^{-\frac{i}{2}(\delta_{14})} 2 \sin\left(\frac{\delta_{14}}{2}\right)  G_\gamma (x_{4,5})\,,
\end{equation}
to be compared with Eq.~(\ref{eq:Gmue:delta14}). 
Thus, non-vanishing phases now generically lead to enhancements, which can be quite important.
This is illustrated in Fig.~\ref{fig:cLFV.FF.delta14.signs}, where we show results analogous to those displayed in  Figs.~\ref{fig:cLFV.FF.delta14} and~\ref{fig:cLFV.FF.phi4},  but now taking $\theta_{14} = - \theta_{15}$.
It can be seen that the opposite sign of $\theta_{14}$ and $\theta_{15}$ effectively leads to the same behaviour as a shift in $\delta_{14} \to \delta_{14} + \pi$ (see Figs.~\ref{fig:cLFV.FF.delta14} and~\ref{fig:clfvFF_phi_delta_pi}). 

\section {Phenomenological study: interference effects}\label{sec:num:analysis}
Following the simple analysis of the previous section, which allowed a clear view of the effect of the CPV Dirac and Majorana phases, we now proceed to a more complete numerical study. In what follows, we will survey in a comprehensive way the simple ``3+2 toy model'', carrying a phenomenological analysis of a larger set of cLFV observables, now taking into account the available experimental constraints (see  Appendix~\ref{app:constraints}). The latter include limits on the active-sterile mixings, results of direct and indirect searches for the heavy states, and EW precision tests, among many others. Finally, current bounds on searches for cLFV transitions are taken into account (cf. Table~\ref{tab:cLFVdata}).

In order to further explore the effects of the new non-vanishing CPV phases in a realistic way, we perform a random scan of the active-sterile mixing angles\footnote{Here we relax the assumption of $\theta_{\alpha 4} = \pm \theta_{\alpha 5}$ by means of adding gaussian noise with a relative $1\,\sigma$ deviation of $10\,\%$ to the samples, i.e.
$\theta_{\alpha 5} = \pm\theta_{\alpha 4} \pm 10\,\%\,$.}
for $m_4 = m_5 = 1,\,5\:\mathrm{TeV}$.  
Leading to the numerical results displayed in this section, we first perform a scan with the phases set to zero and (randomly) select $2000$ points consistent with all experimental data at the $3\,\sigma$ level (corresponding to the blue points in the plots presented in this section).
For each of the selected points we then randomly vary the phases $\delta_{14}$, $\delta_{24}$, $\delta_{34}$, and $\varphi_4$ in the interval $(0, \,2\pi)$, drawing 100 samples from a uniform distribution (shown in orange).
Finally, we further add to the data set 
additional points which  correspond to having systematically 
varied the phases on a grid for the ``special'' values $\{0, \frac{\pi}{4}, \frac{\pi}{2}, \frac{3\pi}{4}, \pi\}$ (shown in green).
This procedure allows to exhaustively study the impact of having non-vanishing CPV phases, especially regarding correlations between observables.
Due to computational limitations, in this section we still do not take into account $\varphi_5$ nor $\delta_{i5}$; in the limit of degenerate masses, and nearly degenerate angles, non-vanishing values of the latter can be understood as leading to a phase shift 
(e.g. $\propto \cos(\varphi_4 - \varphi_5)$, see Appendix~\ref{app:analytic.phase.observables}). 

\subsection{Correlation of $\pmb{\mu-e}$ observables} 
\begin{figure}[h!]
    \centering
\mbox{ \hspace*{-5mm}     \includegraphics[width=0.51\textwidth]{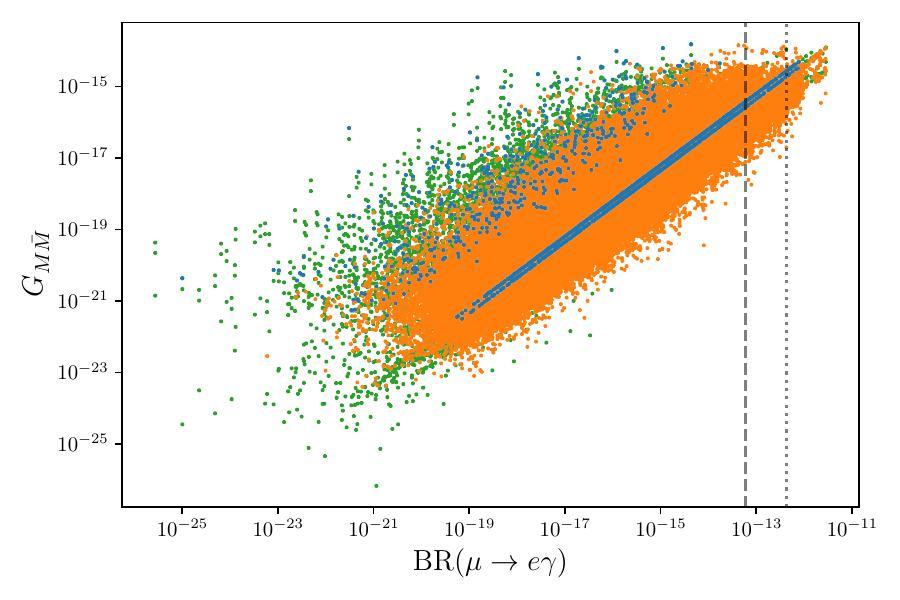}\hspace*{2mm}
    \includegraphics[width=0.51\textwidth]{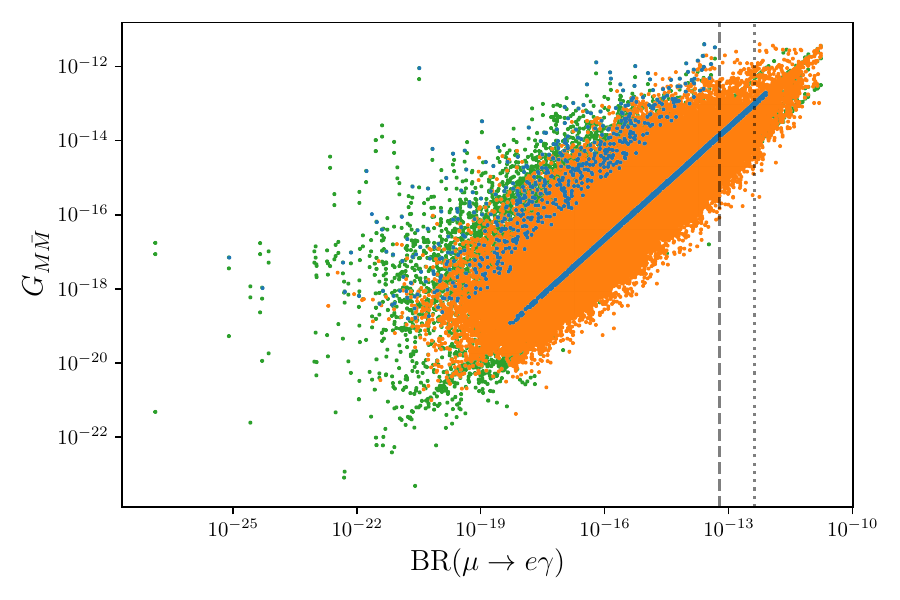}}\\
\mbox{ \hspace*{-5mm}     \includegraphics[width=0.51\textwidth]{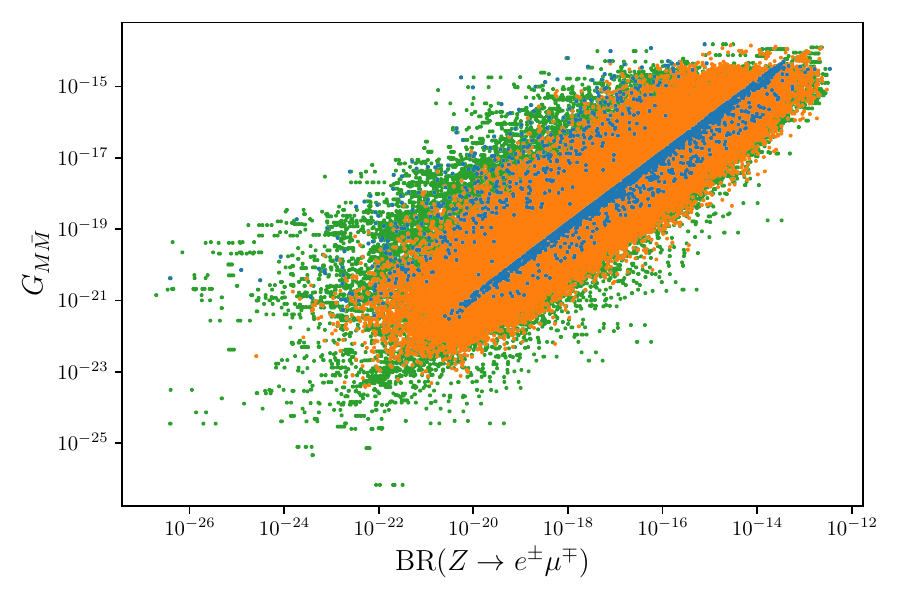}\hspace*{2mm}
    \includegraphics[width=0.51\textwidth]{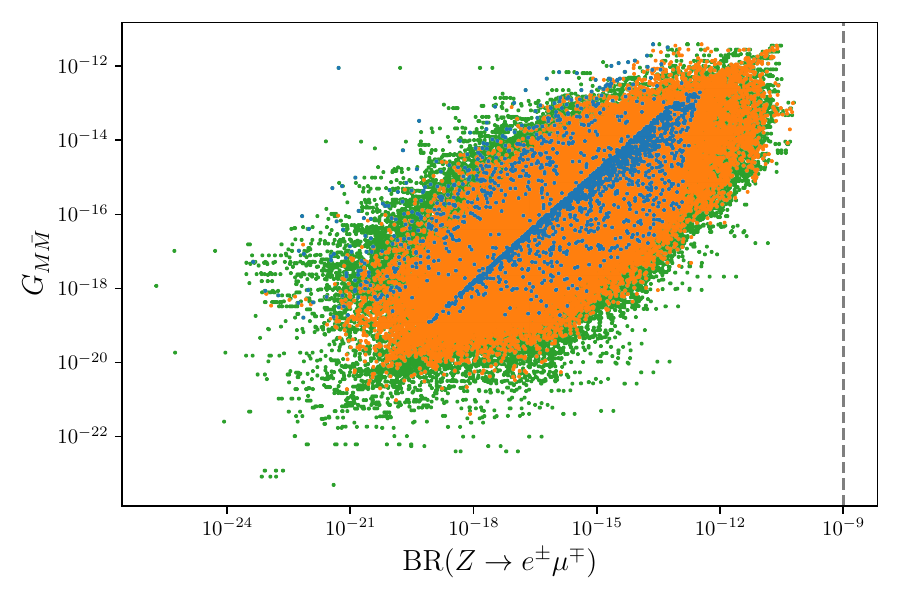}}\\
\mbox{ \hspace*{-5mm}     \includegraphics[width=0.51\textwidth]{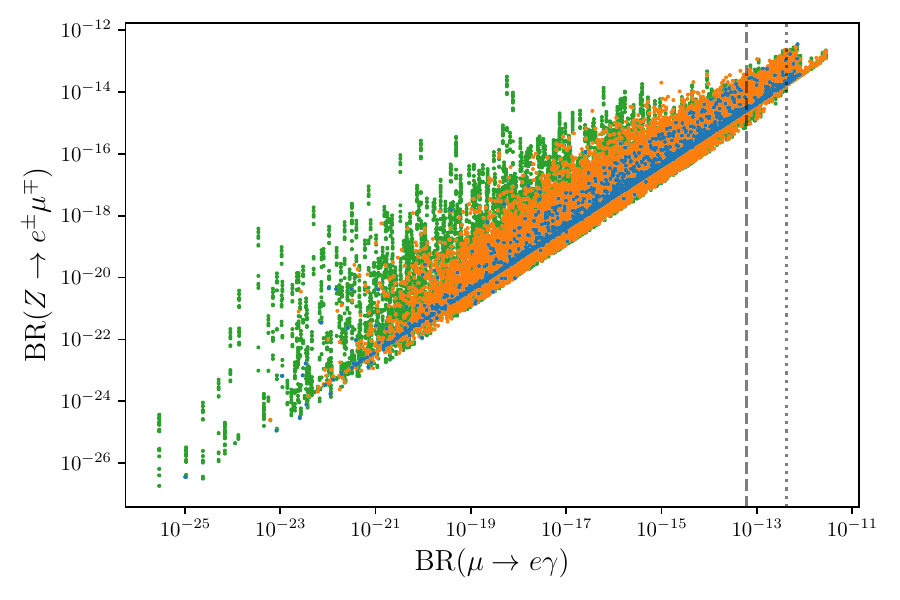}\hspace*{2mm} 
    \includegraphics[width=0.51\textwidth]{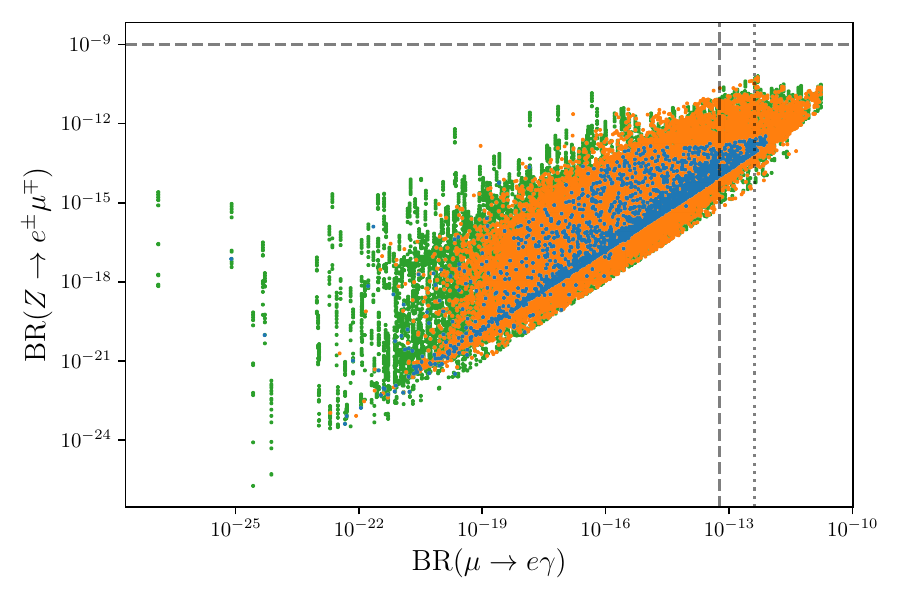}}
    \caption{Correlation of $\mu-e$ flavour violating observables (depending on unique topologies), for varying values of the CPV Dirac and Majorana phases: 
    blue points correspond to vanishing phases, orange denote random values of $\delta_{\alpha 4}$ and $\varphi_4$ in the interval $(0, 2\,\pi)$, and green points refer to $\delta_{\alpha 4}, \varphi_4 =\{0, \frac{\pi}{4}, \frac{\pi}{2}, \frac{3\pi}{4}, \pi\}$ (see text). Dotted  (dashed) lines denote current  bounds (future sensitivity).
    On the left panels, $m_4 = m_5 = 1\:\mathrm{TeV}$, while on the right we set  $m_4 = m_5 = 5\:\mathrm{TeV}$.}
    \label{fig:scatter_GMM_mueg_Zemu}
\end{figure}

We begin by considering cLFV observables in the $\mu-e$ sector which receive contributions from unique topologies (dipole, $Z$-penguins and boxes) at one-loop level, and address the impact of CPV phases on the expected correlations. More specifically, on Fig.~\ref{fig:scatter_GMM_mueg_Zemu}, we consider $\mu \to e \gamma$, $Z \to e \mu$ decays and the probability of muonium-antimuonium oscillations, $\propto G_{M\overline{M}}$, displaying the results for two heavy mass regimes, 1 and 5~TeV.

Focusing first on the blue sets of points\footnote{In several plots one observes that in some
extreme cases the predictions of certain observables already lie above the experimental limit; this merely reflects having taken points 
which are consistent with cLFV bounds at the $3\,\sigma$ level, while the denoted experimental limits correspond to 90\% C.L..} (corresponding to vanishing CP violating phases), one confirms that there is a strong correlation between the 
three considered $\mu -e$ flavour violating observables, as expected. This is particularly manifest in the upper row of 
Fig.~\ref{fig:scatter_GMM_mueg_Zemu}, since both 
BR($\mu \to e \gamma$) and $G_{M\overline{M}}$ do not depend on $\theta_{3j}$; on the other hand, non-vanishing values for $\theta_{3j}$ do contribute to $Z\to e\mu$ decays (through the $C_{ij}$ term), hence one observes a spread of the points along the central straight line. This well-known behaviour (correlated predictions for a given value of the propagator's mass) has been explored in the literature as a means to test the underlying BSM construction, see e.g.~\cite{Hambye:2013jsa,Calibbi:2017uvl}.  
Once CP violating phases are (randomly) taken into account, the correlation between the observables  is strongly affected - if not lost, as can be seen by the significant spread of the corresponding orange points. This is especially visible for the plots in the right  ($m_4 = m_5 = 5\:\mathrm{TeV}$). The situation becomes even more 
degraded once the ``special'' values of the phases are considered (those leading to vanishing values of certain contributions, as discussed in Section~\ref{sec:phases.matter}). As expected, the predictions for the observables are dramatically impacted, as visible from the green points. 
Finally, notice that the effect of phases 
can lead to either an increase or reduction of the cLFV rates with respect to the corresponding ones obtained in the vanishing phase 
limit. In some cases (like for BR($\mu \to e \gamma$), with $m_{4,5}=5$~TeV) the new predictions may even be in conflict with current experimental bounds.

\begin{figure}[h!]
    \centering
 \mbox{ \hspace*{-5mm}    \includegraphics[width=0.51\textwidth]{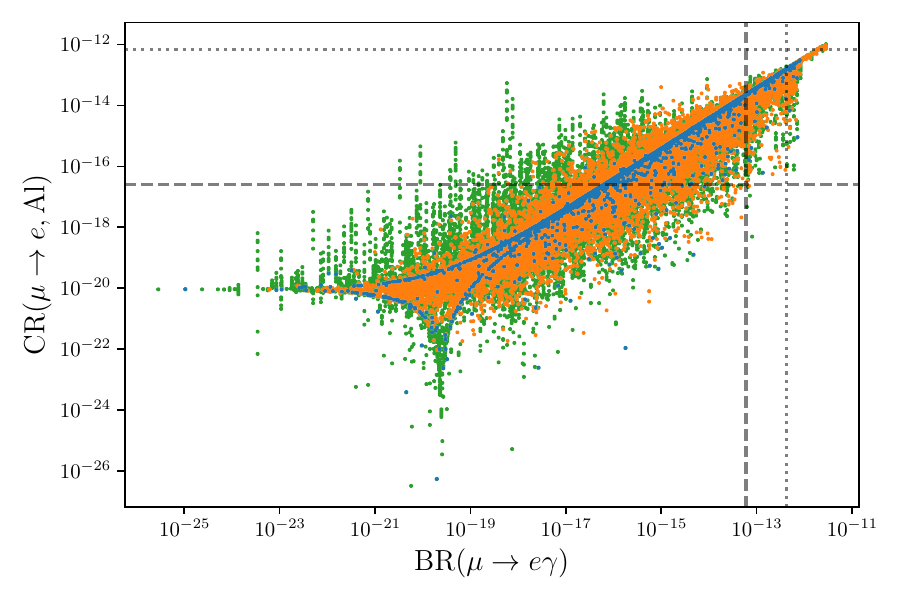}\hspace*{2mm}
    \includegraphics[width=0.51\textwidth]{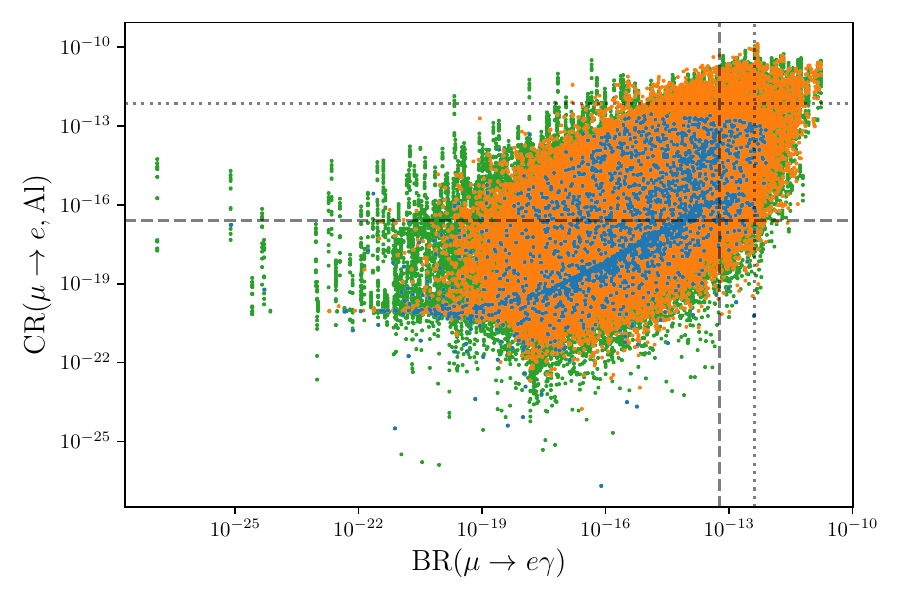}}\\
\mbox{ \hspace*{-5mm}     \includegraphics[width=0.51\textwidth]{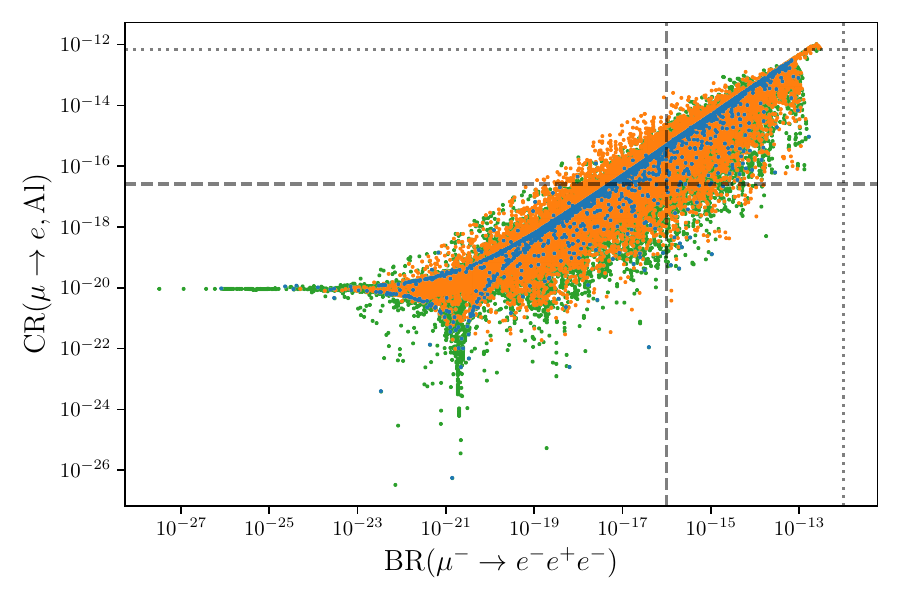}\hspace*{2mm}
    \includegraphics[width=0.51\textwidth]{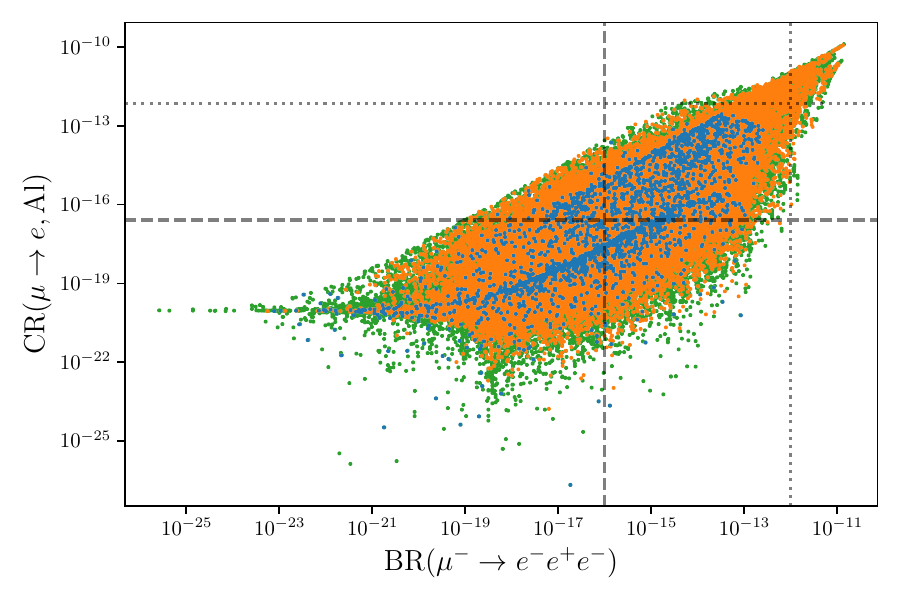}}
    \mbox{ \hspace*{-5mm}     \includegraphics[width=0.51\textwidth]{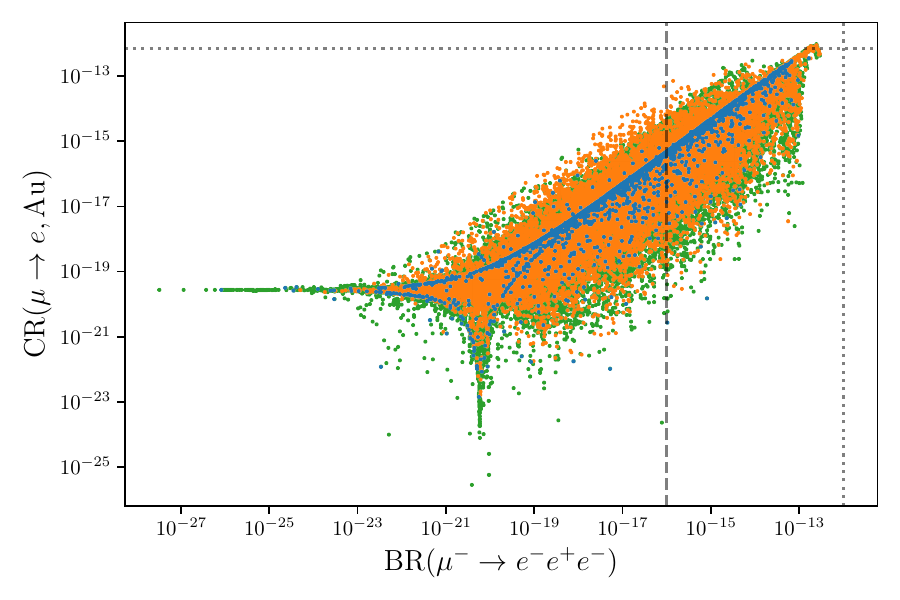}\hspace*{2mm}
    \includegraphics[width=0.51\textwidth]{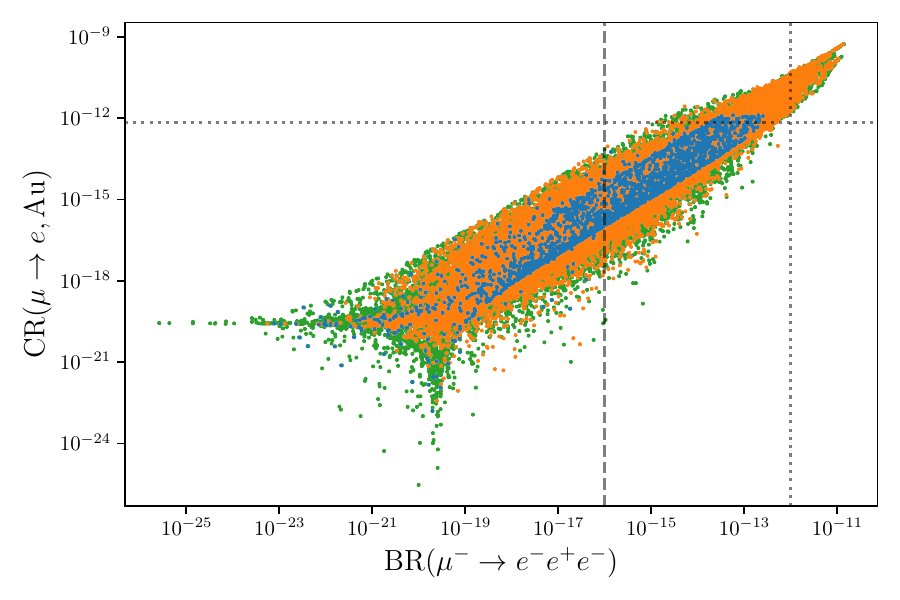}}
    \caption{
    Correlation of general $\mu-e$ flavour violating observables, for varying values of the CPV Dirac and Majorana phases. Line and colour code as in Fig.~\ref{fig:scatter_GMM_mueg_Zemu}.
    On the left panels, $m_4 = m_5 = 1\:\mathrm{TeV}$, while on the right we set  $m_4 = m_5 = 5\:\mathrm{TeV}$.}
    \label{fig:scatter_GMM_mu3e_CRmue}
\end{figure}

\medskip
In Fig.~\ref{fig:scatter_GMM_mu3e_CRmue} we consider the joint behaviour of general $\mu-e$ flavour violating observables, which now receive non-vanishing contributions from various  types of diagrams (dipole, penguins and boxes). In particular, on the first row we present CR($\mu-e$, Al) vs. BR($\mu \to e\gamma$), while on the second CR($\mu-e$,~Al) vs. BR($\mu \to 3e$), for 
$m_4 = m_5 = 1\text{ and }5\:\mathrm{TeV}$ (left and right columns, respectively). 
Despite the mixings between $\nu_\tau$ and the sterile states also leading to a spread in the case of vanishing phases
(for $\mu-e$ conversion and  3-body decays), one still observes a visible correlation between the different sets of observables (see blue points)\footnote{The spread and visible cancellations for vanishing CP-violating phases are a consequence of accidental cancellations in $\mu-e$ conversion as discussed in Section~\ref{sec:muecon_simple} and of accidental cancellations due to opposite-sign mixing angles (e.g. $\theta_{14}\approx - \theta_{15}$) leading to a suppression of dipole operators as discussed in Section~\ref{sec:enh_others}.}.
As already verified in Fig.~\ref{fig:scatter_GMM_mueg_Zemu}, the presence of CP violating phases - especially for certain values of the latter - leads to a strong loss of correlation, as visible from the dispersion of the orange and green points (again more important for $m_4 = m_5 = 5\:\mathrm{TeV}$). 
On the third and final row of Fig.~\ref{fig:scatter_GMM_mu3e_CRmue}, we present the prospects for the correlation between  $\mu \to 3 e$ decays and neutrinoless muon-electron conversion, but now for Gold (Au) nuclei.
Notice that there are significant differences between 
Al and Au, which become particularly manifest for  
$m_4 = m_5 = 5~\mathrm{TeV}$. In this case the correlation between the conversion rate in Gold and the three-body decays is more prominent than for Aluminium nuclei. This is a consequence of milder cancellations of the type discussed in Section~\ref{sec:muecon_simple}: as seen from 
Figs.~\ref{fig:CR_M:nuclei_tau} and~\ref{fig:CR_Al_M:cancel_phases}, in the limit of vanishing phases 
(and with $\theta_{3j}=0$), the values of the heavy propagator mass for which the accidental cancellation occurs ($m_{4,5}^c$) are considerably lower for Gold than Aluminium, and effects of CP phases and/or non-vanishing $\theta_{3j}$ tend to further shift $m_{4,5}^c$ to lower values. 

\bigskip
Consequently, the CP violating phases might thus have an important impact
regarding a future interpretation of data:
let us consider a hypothetical scenario in which collider searches strongly hint for the presence of sterile states with masses close to 1~TeV. Should BR($\mu \to 3 e$)$\approx 10^{-15}$ be measured in the future, one could expect an observation of CR($\mu-e$, Al)~$\approx \mathcal{O}(10^{-14})$, be it at COMET or Mu2e. However, in the presence of CP violating 
phases, the expected range for the muon-electron conversion is vast, with CR($\mu-e$, Al) potentially as low as 
$10^{-18}$.

\subsection{Prospects for other observables}
In view of the diversity of cLFV observables, and of the associated (future) experimental prospects, we have so far focused our phenomenological discussion on the most promising $\mu-e$ cLFV transitions and decays. 
Before concluding this section, we address the impact of the CP violating phases for the $\mu-\tau$ sector (i.e. for $\tau \to 3\mu$ and $Z \to \mu \tau$ decays), as well as for cLFV Muonium decays. 

\begin{figure}
    \centering
    \includegraphics[width=0.48\textwidth]{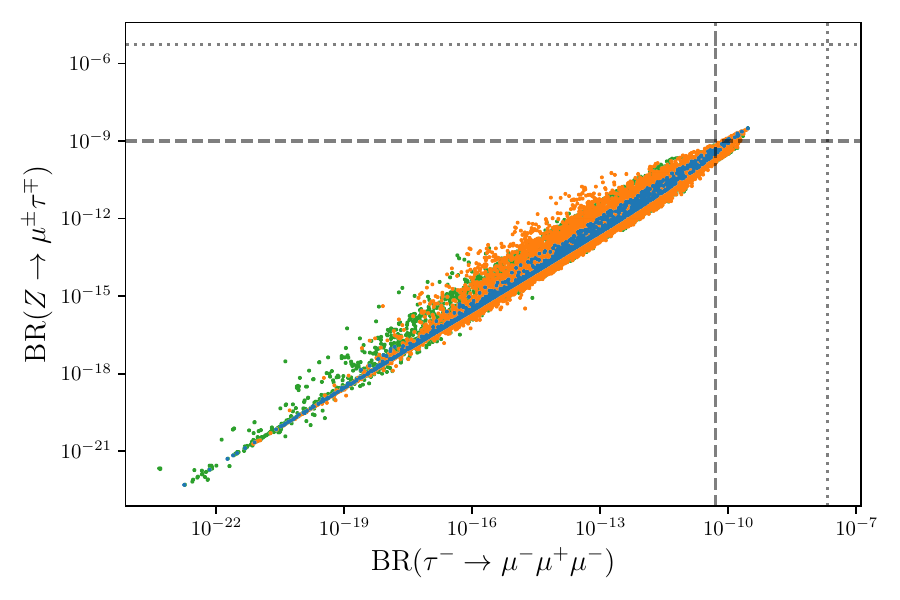}
    \includegraphics[width=0.48\textwidth]{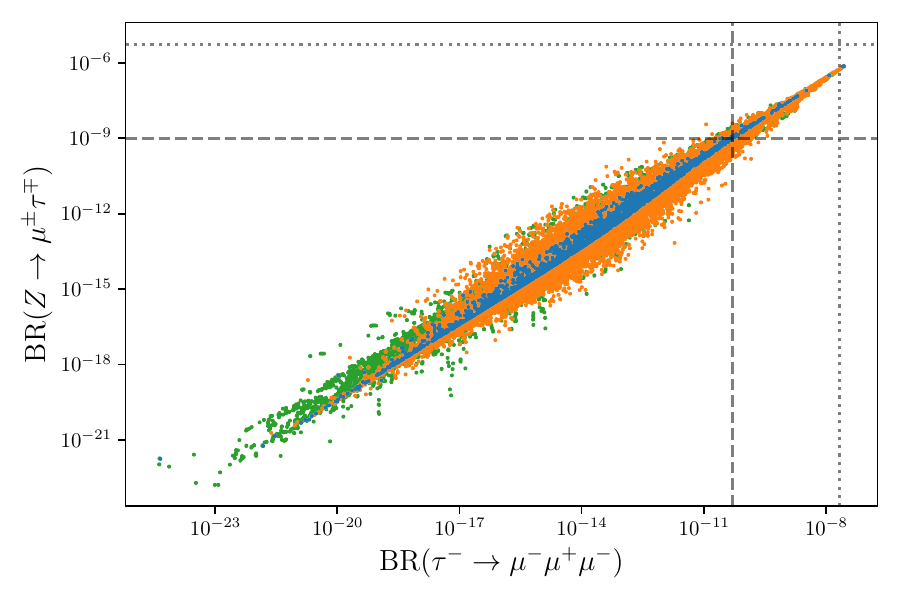}
    \caption{Correlation of $Z \to \mu \tau$ and $\tau \to 3\mu$ decays, for varying values of the CPV Dirac and Majorana phases. Line and colour code as in Fig.~\ref{fig:scatter_GMM_mueg_Zemu}.
    On the left panels, $m_4 = m_5 = 1\:\mathrm{TeV}$, while on the right we set  $m_4 = m_5 = 5\:\mathrm{TeV}$.}
    \label{fig:tau3mu:Ztaumu}
\end{figure}

In Fig.~\ref{fig:tau3mu:Ztaumu} we summarise the results of a study analogous to those presented in  Figs.~\ref{fig:scatter_GMM_mueg_Zemu} and~\ref{fig:scatter_GMM_mu3e_CRmue}, displaying the correlated behaviour of high- and low-energy $\mu-\tau$ sector cLFV observables for (non-) vanishing Dirac and Majorana CPV phases, and for two values of the degenerate heavy neutral leptons' masses. Although the general prospects for observation are comparatively less promising, one nevertheless encounters the same phase-induced distortion of the correlation between observables, which was present in the 
limit of vanishing phases.
Being also dominated by penguin transitions, the tau-lepton decay modes $\tau^-\to \mu^-e^+e^-$ and $\tau^-\to e^-\mu^+\mu^-$ (i.e. only one flavour violating vertex) do not offer any additional insight with respect to the $\tau\to 3\mu$ and $\tau \to 3e$ counterparts, and we find similar predictions for the associated rates.
On the other hand, tau-lepton decays with an additional flavour violating coupling, that is $\tau^-\to e^-\mu^+e^-$ and $\tau^-\to \mu^- e^+ \mu^-$, are transitions which are purely mediated by box diagrams. Thus, these are strongly suppressed when compared to other modes, with typically very small branching ratios, $\mathrm{BR} \lesssim 10^{-17}$, thus clearly beyond any future sensitivity reach.

\begin{figure}
    \centering
\mbox{\hspace*{-5mm}    \includegraphics[width=0.51\textwidth]{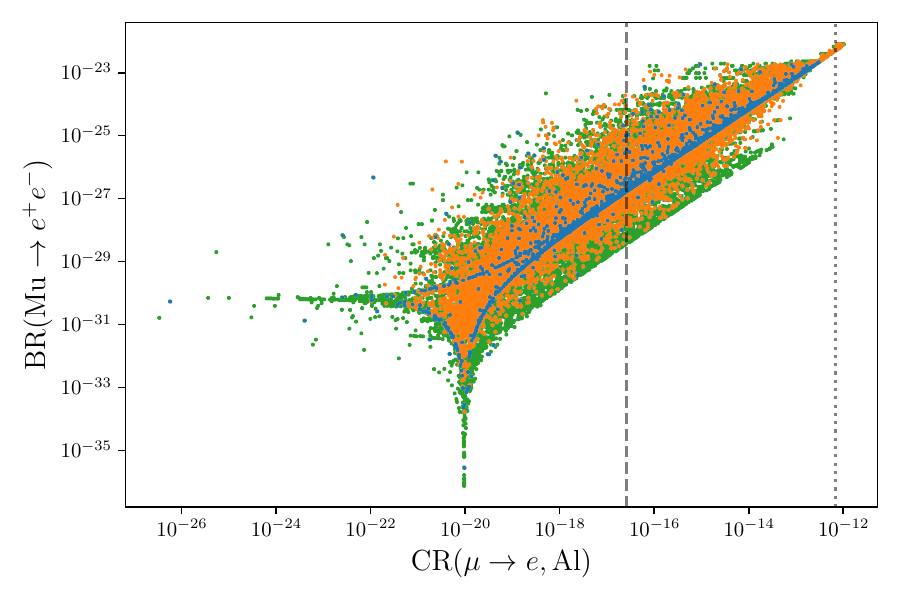}\hspace*{2mm}
    \includegraphics[width=0.51\textwidth]{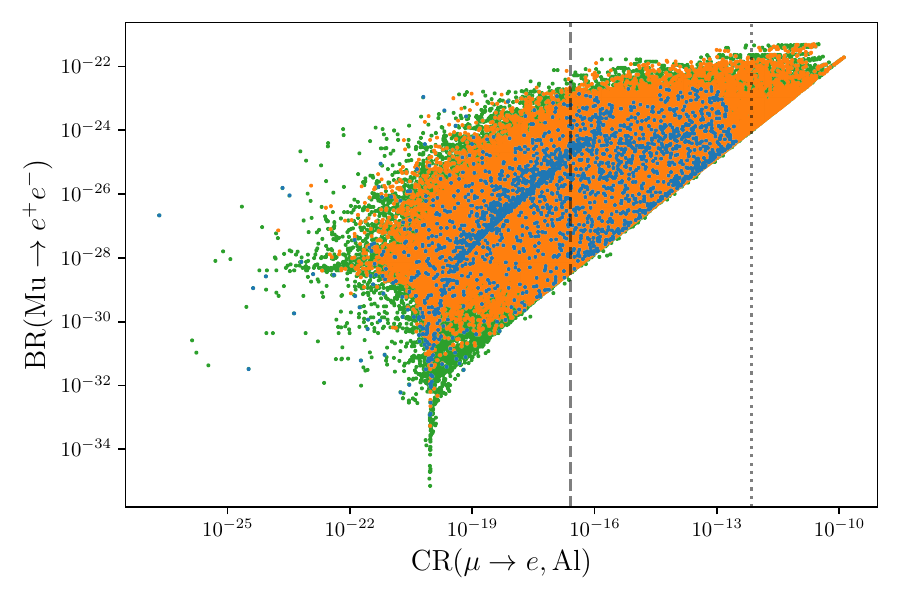}}
    \caption{Correlation of BR($\text{Mu}\to ee$) and CR($\mu-e$, Al), for varying values of the CPV Dirac and Majorana phases. Line and colour code as in Fig.~\ref{fig:scatter_GMM_mueg_Zemu}.
    On the left panel, $m_4 = m_5 = 1\:\mathrm{TeV}$, while on the right we set  $m_4 = m_5 = 5\:\mathrm{TeV}$.}
    \label{fig:Muoniumdecay}
\end{figure}

In Fig.~\ref{fig:Muoniumdecay}, we display for completeness\footnote{Other cLFV observables - such as the Coulomb enhanced decays $\mu e\to ee$(see Refs.~\cite{Koike:2010xr,Uesaka:2016vfy,Uesaka:2017yin,Kuno:2019ttl,Abada:2015oba}), were also studied. Although the associated rate for $\mathrm{Al}$ typically lies some orders of magnitude below $\mu-e$ conversion, effects due to non-vanishing CPV phases can however lead to comparable rates, with a maximum of $\mathrm{BR}(\mu e\to ee,\,\mathrm{Al})\sim10^{-16}$.} the prospects for $\text{Mu}\to ee$ decays, depicting its correlation with CR($\mu-e$, Al).  
In the most optimal scenarios, one can expect BR($\text{Mu}\to ee)\sim \mathcal{O}(10^{-22})$.

\section{Overall view and further discussion}\label{sec:general-analysis}
So far, we have  analysed the implications of non-vanishing CPV phases in ``idealised and simplified'' scenarios, assuming certain relations between the model's parameters. 
We followed this approach to explore and maximise the effect of phases on the predictions for the cLFV observables.
Clearly, broader scenarios must be considered, and in the present section we relax several of the previous assumptions.
This aims at a more comprehensive overview, and allows a better confrontation with (hypothetical) future data.

\subsection{Comprehensive overview of the parameter space}
We conclude our study by a final comprehensive overview of this very simple SM extension. The numerical data presented in this subsection is now obtained as follows:
firstly, and in what concerns the masses of the two heavy  states\footnote{In more general scenarios in which the masses of the sterile states and their mixings with the active neutrinos are a priori unrelated to each other, effects of the CPV phases are expected to be less striking, but nevertheless important (and in general driven by the heaviest state).}, we no longer take them to be degenerate, but rather assume their masses to be sufficiently close to allow for interference effects\footnote{See Ref.~\cite{Abada:2019bac} for a related discussion regarding the mass splitting between $m_4$ and $m_5$.}; in practice, and for fixed 
$m_4$, random values of $m_5$ are obtained from 
half-normal distributions with the scale set to a value representative of the width of the sterile states (in this case $\sim50~\mathrm{GeV}$).
Concerning the active-sterile mixing angles, these are now independently varied: more specifically, we draw 
samples from log-uniform distributions, further randomly varying their signs.  
For $m_4=1$~TeV, the ranges of the parameters to be here explored 
are then
\begin{eqnarray}\label{eqn:6d_scan_ranges}
     && m_5 - m_4 \in\,[0.04, 210]\,\mathrm{GeV}\,,\nonumber \\
    && |\sin\theta_{14, 5}|\in\, [2.0\times 10^{-5}, 3\times10^{-3}]\,,\nonumber \\
    && |\sin\theta_{24,5}|\in\, [2.2\times 10^{-4}, 0.036]\,,\nonumber \\
    && |\sin\theta_{34,5}|\in\, [1.0\times 10^{-3}, 0.13]\,.
\end{eqnarray}
It is worth noticing that these ranges lead to scenarios complying with experimental bounds (see Section~\ref{sec:exp_bounds} and Appendix~\ref{app:constraints}).
In our analysis, we thus (randomly) select $10^4$ points consistent with all experimental data. 
For each tuple of mixing angles we then vary {\it all} CPV phases associated with the sterile states, i.e. $\delta_{\alpha 4, 5}, \varphi_{4, 5}\,\in\,[0, 2\pi]$, drawing 100 values for each of the four from a uniform distribution.
The upper limits on the intervals for the mixing angles are inferred from requiring agreement with the most constraining  current bounds; clearly 
no lower limit for the mixing angles is phenomenologically relevant. 
However, we have limited ourselves to regimes that do not lead to cLFV predictions
excessively far away 
from the corresponding future experimental sensitivity. Thus, it is important to stress that the resulting predictions for the observables could in principle be extended to extremely tiny values of the rates, should we have explored all the allowed ranges for the mixing angles. In summary, no conclusions concerning lower limits for the cLFV observables should be drawn from this analysis.

\begin{figure}[t!]
    \centering
    \mbox{\hspace*{-8.5mm}
    \includegraphics[width = 0.53\textwidth]{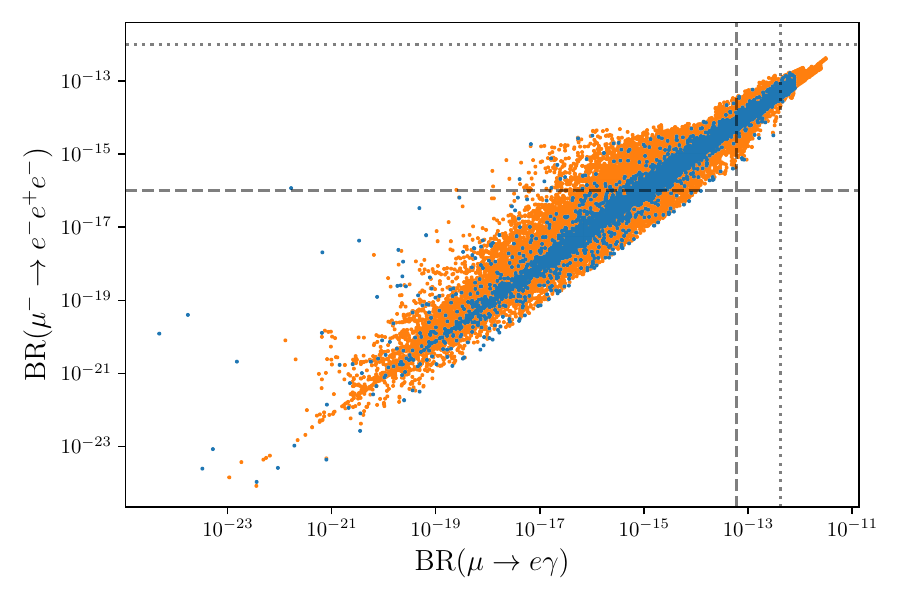}
    \hspace*{2mm} \includegraphics[width = 0.53\textwidth]{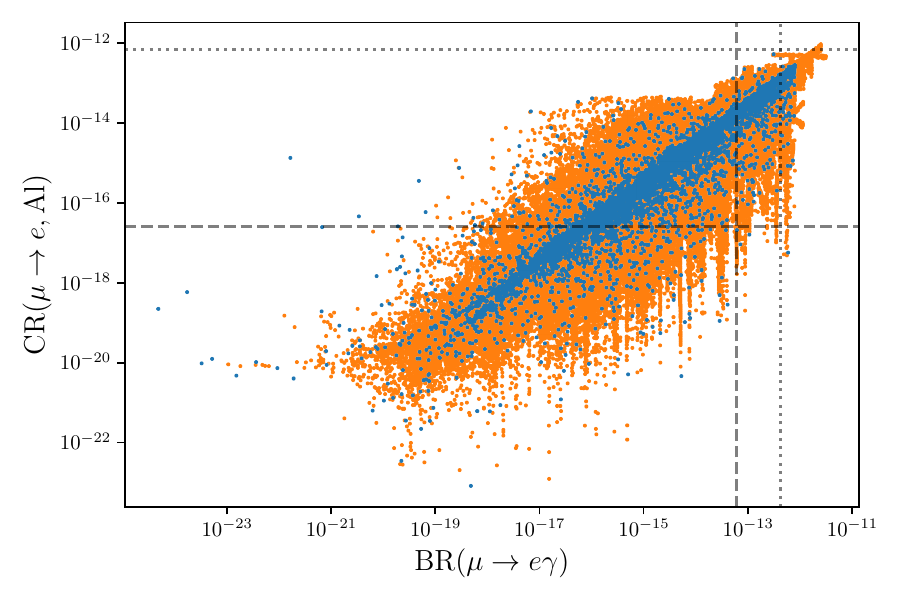}}\\
    \mbox{\hspace*{-8.5mm}\includegraphics[width = 0.53\textwidth]{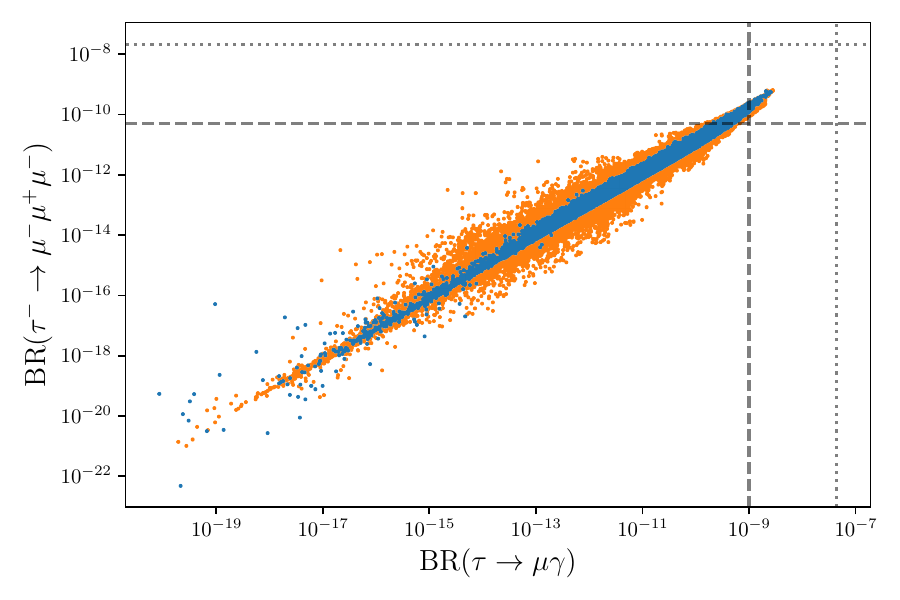}
    \hspace*{2mm} \includegraphics[width = 0.53\textwidth]{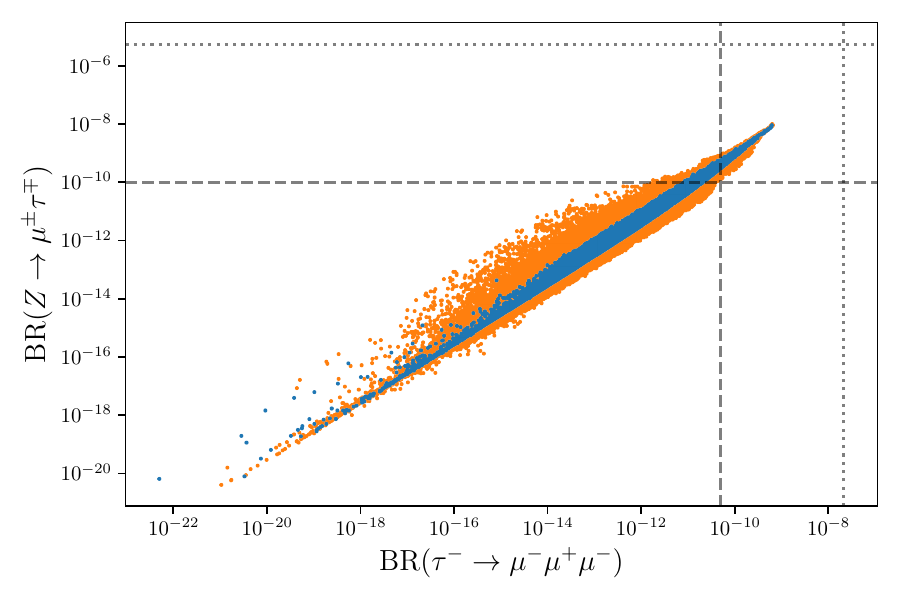}}
    \caption{General overview of cLFV observables (correlations) in the ``3+2 toy model'' parameter space. 
    All active-sterile mixing angles, as well as Dirac and Majorana CP phases, are randomly varied (see detailed description in the text). In all panels, $m_4=1$~TeV, with $m_5-m_4 \in [40~\text{MeV}, 210~\text{GeV}]$. Blue points correspond to vanishing phases, while orange denote random values of all phases ($\delta_{\alpha i}$ and $\varphi_i$, with $\alpha = e,\mu,\tau$ and $i=4,5$). Dotted (dashed) lines denote current bounds (future sensitivity) as given in Table~\ref{tab:cLFVdata}.}
    \label{fig:6dscan}
\end{figure}

The outcome of this comprehensive analysis is shown in Fig.\ref{fig:6dscan}, where
we display the predictions for several cLFV rates (focusing on the $\mu-e$ and $\tau-\mu$ sectors), in particular in what concerns correlations between same-sector observables. As before, we present the predictions obtained in the 
case in which all CPV phases are set to zero (blue points), and then those corresponding to a random scan over {\it all} Dirac and Majorana phases (orange points). 
For a fixed scale of the heavy propagators (with the latter sufficiently close in mass), and as would be expected, one finds correlated
same-sector observables.
In  addition to enlarging the range of the predictions for the different observables (possibly leading to a conflict with current bounds for certain $\mu-e$ observables), non-vanishing CP phases lead to a visible spread of the predicted rates. Should ``special'' values of the phases be imposed on the scan, the cancellation effects (and associated decrease of the rates) would have been even more striking. 

\bigskip
Throughout the discussion we have focused on the $\mu-e$ and $\tau-\mu$ sectors, in view of the most promising prospects (both theoretical and experimental). In general, the associated predictions for $\tau-e$ cLFV transitions typically lie beyond future experimental sensitivity. However, and to fully explore this very minimal
SM extension, one would require a comprehensive probing of the associated cLFV predictions in all 3 flavour sectors (i.e. $\mu-e$, $\tau-\mu$ and $\tau-e$ transitions).

Profiting from the data collected leading to 
the results displayed in Fig.~\ref{fig:6dscan}, we have tried to infer which would be the required future sensitivity for the 
$\tau-e$ channels 
so that the regimes (mixing angles and CP phases)
leading to predictions for $\mu \to e\gamma$, $\mu \to 3e$, $\mu - e$ conversion in  Al, $\tau \to 3\mu$ and $Z \to \mu \tau$, all within future experimental sensitivities, would also be within reach of $\tau\to e\gamma$ and $\tau\to 3e$ dedicated searches.
Requiring that at least 68\% of the previously mentioned subset be within $\tau-e$ future reach would imply the following ideal experimental sensitivities\footnote{We have assumed the same ratio between the envisaged $\tau\to e\gamma$ and $\tau\to 3e$ sensitivities as the one of the future prospects of Belle II~\cite{Kou:2018nap}.}:  
\begin{equation}
    \text{BR}(\tau\to e\gamma) \geq 2\times 10^{-13}\,,
    \quad
    \text{BR}(\tau\to 3e) \geq 3\times 10^{-14}\,.
\end{equation}
In other words, should a signal of cLFV in $\mu-e$ and $\tau-\mu$ transitions be observed at the current and near-future facilities, an improvement of circa
$4$ orders of magnitude in the $\tau-e$ sensitivity is needed in order to obtain competitive constraints on 
these SM extensions via heavy neutral leptons from all flavour sectors.

\subsection{Reconciling cLFV predictions with future observations}
As discussed extensively in the previous (sub)sections, 
CPV phases can impact the predictions for the cLFV observables, 
enhancing or suppressing the distinct rates. 
To conclude the discussion, we have identified a small set of representative (benchmark) points, which reflect not only the effect on the rates, but also the impact that taking into account the CPV phases might have on 
the interpretation of 
experimental data (or negative search results). 

In Table~\ref{table:changingcLFVphases} we present the predictions for several cLFV observables for three configurations of active-sterile mixing angles, in the 
case of vanishing CPV phases (P$_i$), and for non-vanishing values of the phases (P$^\prime_i$):
\begin{eqnarray}
\label{eq:Pi:angles}
& \text{P}_1: &
s_{14} = 0.0023\,, \:s_{15} = -0.0024\,,\:
s_{24} = 0.0035\,, \:s_{25} = 0.0037\,, \:
s_{34} = 0.0670\,, \:s_{35} = -0.0654\,, \nonumber \\
& \text{P}_2: &
s_{14} = 0.0006\,, \: s_{15} = -0.0006\,, \: 
s_{24} = 0.008\,, \: s_{25} = 0.008\,, \: 
s_{34} = 0.038\,, \: s_{35} = 0.038\,, \nonumber \\
& \text{P}_3: &
s_{14} = 0.003\,, \: s_{15} = 0.003\,, \:  
s_{24} = 0.023\,, \:  s_{25} = 0.023\,, \:  
s_{34} = 0.068\,, \:  s_{35} = 0.068\,. 
\end{eqnarray}
The variants P$^\prime_i$ have identical mixing angles, but in association with the following phase configurations:
\begin{eqnarray}\label{eq:Pi:phases}
\text{P}^\prime_1:    
\delta_{14} = \frac{\pi}{2}\,, \:  
\varphi_4 = \frac{3\pi}{4}\,;\quad  
\text{P}^\prime_2:    
\delta_{24}=\frac{3\pi}{4}\,, \:  
\delta_{34} = \frac{\pi}{2}\,, \:  
\varphi_4 = \frac{\pi}{\sqrt{8}}\,; \quad  
\text{P}^\prime_3: 
\delta_{14}\approx \pi\,, \:  
\varphi_4\approx \frac{\pi}{2}\,.
\end{eqnarray}

\noindent We have chosen $m_4=m_5=5$ TeV for all three benchmark points.

\renewcommand{\arraystretch}{1.3}
\begin{table}[h!]
    \centering
    \begin{tabular}{|l|c|c|c|c|c|}
    \hline
 & BR($\mu \to e\gamma$) & BR($\mu \to 3e$) & CR($\mu - e$, Al) & BR($\tau \to 3\mu$)& BR($Z \to \mu \tau$)\\
 \hline\hline
$\text P_1$ & $ 3\times 10^{-16}$ \:\:$\circ$ & 
$ 1\times 10^{-15}$ \:\:$\checkmark$& 
$ 9\times 10^{-15}$ \:\:$\checkmark$& 
$ 2\times 10^{-13}$ \:\:$\circ$&
$ 3\times 10^{-12}$ \:\:$\circ$\\
$\text P_1^\prime$
& $ 1\times 10^{-13}$ \:\:$\checkmark$& 
$2\times 10^{-14}$ \:\:$\checkmark$& 
$1\times 10^{-16}$ \:\:$\checkmark$& 
$1\times 10^{-10}$ \:\:$\checkmark$& 
$2\times 10^{-9}$ \:\:$\checkmark$\\
\hline
\hline
$\text P_2$ 
& $2\times 10^{-23}$ \:\:$\circ$
& $2\times 10^{-20}$ \:\:$\circ$ 
& $2\times 10^{-19}$ \:\:$\circ$ 
&  $1\times 10^{-10}$ \:\:$\checkmark$
& $3\times 10^{-9}$ \:\:$\checkmark$\\
$\text P_2^\prime$
& $6\times 10^{-14}$ \:\:$\checkmark$
& $4\times 10^{-14}$ \:\:$\checkmark$
& $9\times 10^{-14}$ \:\:$\checkmark$
&  $8\times 10^{-11}$ \:\:$\checkmark$
& $1\times 10^{-9}$ \:\:$\checkmark$\\
 \hline
 \hline
 $\text{P}_3$ 
 & $2\times 10^{-11}$ \:\:{\footnotesize \XSolidBrush}
 & $3\times 10^{-10}$ \:\:{\footnotesize \XSolidBrush} 
 & $3\times 10^{-9}$ \:\:{\footnotesize \XSolidBrush}
 & $2\times 10^{-8}$ \:\:$\checkmark$
 & $8\times 10^{-7}$ \:\:$\checkmark$\\
$\text P_3^\prime$
& $8\times 10^{-15}$ \:\:$\circ$
  & $1\times 10^{-14}$ \:\:$\checkmark$
  & $6\times 10^{-14}$ \:\:$\checkmark$
  & $2\times 10^{-9}$ \:\:$\checkmark$
  & $1\times 10^{-8}$ \:\:$\checkmark$\\
 \hline
\end{tabular}
\caption{Predictions for several cLFV observables in association with three 
configurations with vanishing CPV phases,
P$_i$ ($i=1-3$) and associated variants with non-vanishing CP violating phases, P$_i'$, see Eqs.~(\ref{eq:Pi:angles}, \ref{eq:Pi:phases}). We have taken
$m_4 = m_5 = 5~\mathrm{TeV}$. The symbols 
({\small\XSolidBrush}, $\checkmark$, $\circ$) respectively denote 
rates already in conflict with current experimental bounds, predictions within future sensitivity and those beyond future experimental reach.
}\label{table:changingcLFVphases}
\end{table}
\renewcommand{\arraystretch}{1.}

\noindent
The first point (P$_1$) represents a case for which only two 
cLFV observables would be within future experimental reach, 
$\mu \to 3e$ and $\mu-e$ conversion in Aluminium; however, in the presence of CP phases (P$_1^\prime$), the predictions for the different considered observables are now {\it all} within future sensitivity. 

\noindent
The points P$_2$ and P$_2^\prime$ correspond to a similar scenario, but for which only the two considered 
$\mu-\tau$ observables lie within future reach  in the case of vanishing phases.

\noindent
The third and final point (P$_3$) clearly illustrates the importance of taking into account the possibility of  CP violating phases upon interpretation of experimental data.
Negative search results for the different $\mu-e$ flavour violating transitions would lead to the exclusion of the associated mixing angles (for heavy masses $\sim5$~TeV); however, and should CPV phases be present, the considered active-sterile mixing regime can be readily reconciled with current bounds\footnote{A similar approach was pursued in Ref.~\cite{Heeck:2018ntp}, albeit for the $3\times3$ PMNS mixing matrix.} (with 
$\mu \to e \gamma$ now even lying beyond experimental reach). 
A similar exercise could be carried for other heavy mass regimes, leading to analogous conclusions.

\noindent
This demonstrates the crucial role of CPV phases in evaluating the viability of a given scenario in what regards conflict/agreement with the associated cLFV bounds.

\section{Conclusions}
\label{sec:concs}
In this work we have thoroughly addressed the impact of leptonic CP violating phases on the predictions of the rates of several cLFV observables, focusing on minimal SM extensions by
heavy Majorana sterile fermions. 
Despite their minimality and simplicity, these 
extensions can be interpreted as representative  of more complete constructions calling upon the addition  of heavy neutral fermions (as is the case of several low-scale seesaw realisations). 
Via their mixings with the light (mostly active) states, and as a  consequence of a departure from unitarity of the would-be PMNS mixing matrix, the new states open the door to contributions to numerous observables.

Here we have considered a simple case with 2 heavy neutral fermions, taking them close in mass in order to explore the potential impact of the new CPV phases on cLFV observables.  
These states could very well be embedded in a seesaw, and the latter even incorporated in more complete BSM frameworks. The conclusions drawn in this work are thus always valid once one considers that the source of lepton flavour violation stems from the enlarged leptonic mixing. 

Building upon an analytical insight, our numerical study reveals that the CP violating phases can indeed lead to important effects in  cLFV transitions and decays, with an 
impact for the rates (enhancement or suppression of the predictions obtained for vanishing phases).
Moreover, whenever correlations between observables would be typically expected (in association with the dominance of a given topology for certain regimes of the model), one also encounters a potential loss of correlation. 
Furthermore, our analysis suggests that  
the non-observation of a given observable (usually expected to be within experimental reach in view of the measurement of another one) should not be a conclusive reason to disfavour a given regime. 

The conclusions of this work can be generalised for a given BSM construction, provided that all complex degrees of freedom are consistently taken into account, and predictions for cLFV observables re-evaluated in view of the potential presence of new CP violating phases.

In the near future, should potentially new (unexpected) cLFV patterns emerge upon observation of certain processes, this could be interpreted as possibly hinting towards the presence of non-vanishing CP violating phases (under the working hypothesis of SM extensions via heavy neutral fermions).

\section*{Acknowledgements}
The authors are grateful to Cédric Weiland for useful discussions.
This project has received support from the European Union's Horizon 2020 research and innovation programme under the Marie Sk\l{}odowska-Curie grant agreement No.~860881 (HIDDe$\nu$ network) and from the IN2P3 (CNRS) Master Project, ``Flavour probes: lepton sector and beyond'' (16-PH-169).

\appendix

\section{A minimal ``3+2 toy model'' for cLFV studies and constraints}
\label{app:model}
Here we collect the most relevant information concerning this minimal extension of the SM. 
As mentioned before, we will consider an effective toy model, in which 2 sterile (Majorana) states are added to the SM, a so-called ``3+2 toy model''.
The neutral spectrum comprises 5 states, with masses $m_i$ (with $i=1,...,5$): this includes the 3 light (mostly active) neutrinos and two heavier states (with masses $m_{4,5}$). 

\subsection{Modified interaction Lagrangian}
The leptonic mixings 
are parametrised by a $5\times5$ unitary mixing matrix, $\mathcal U$;  its upper left $3\times3$ block corresponds to the left-handed leptonic mixing matrix, the would-be PMNS, $\tilde{U}_\text{PMNS}$. The ensuing non-unitarity of 
$\tilde{U}_\text{PMNS}$ will lead to modified charged and neutral currents, which can be cast in the physical basis as
\begin{align}\label{eq:lagrangian:WGHZ}
& \mathcal{L}_{W^\pm}\, =\, -\frac{g_w}{\sqrt{2}} \, W^-_\mu \,
\sum_{\alpha=1}^{3} \sum_{j=1}^{3 + n_S} \mathcal{U}_{\alpha j} \bar \ell_\alpha 
\gamma^\mu P_L \nu_j \, + \, \text{H.c.}\,, \nonumber \\
& \mathcal{L}_{Z^0}^{\nu}\, = \,-\frac{g_w}{2 \cos \theta_w} \, Z_\mu \,
\sum_{i,j=1}^{3 + n_S} \bar \nu_i \gamma ^\mu \left(
P_L {C}_{ij} - P_R {C}_{ij}^* \right) \nu_j\,, \nonumber \\
& \mathcal{L}_{Z^0}^{\ell}\, = \,-\frac{g_w}{4 \cos \theta_w} \, Z_\mu \,
\sum_{\alpha=1}^{3}  \bar \ell_\alpha \gamma ^\mu \left(
{\bf C}_{V} - {\bf C}_{A} \gamma_5 \right) \ell_\alpha\,, \nonumber \\
& \mathcal{L}_{H^0}\, = \, -\frac{g_w}{2 M_W} \, H  \,
\sum_{i\ne j= 1}^{3 + n_S}  {C}_{ij}  \bar \nu_i\left(
P_R m_i + P_L m_j \right) \nu_j + \, \text{H.c.}\ , \nonumber \\
& \mathcal{L}_{G^0}\, =\,\frac{i g_w}{2 M_W} \, G^0 \,
\sum_{i,j=1}^{3 + n_S} {C}_{ij}  \bar \nu_i  
\left(P_R m_j  - P_L m_i  \right) \nu_j\,+ \, \text{H.c.}, \nonumber  \\
& \mathcal{L}_{G^\pm}\, =\, -\frac{g_w}{\sqrt{2} M_W} \, G^- \,
\sum_{\alpha=1}^{3}\sum_{j=1}^{3 + n_S} \mathcal{U}_{\alpha j} 
\bar \ell_\alpha\left(
m_i P_L - m_j P_R \right) \nu_j\, + \, \text{H.c.}\,, 
\end{align}
with $n_S=2$, and in which we recall that  
\begin{equation}
    {C}_{ij} = \sum_{\rho = 1}^3
  \mathcal{U}_{i\rho}^\dagger \,\mathcal{U}_{\rho j}^{\phantom{\dagger}}\:. 
\end{equation}
In the above, the indices 
$\alpha, \rho = 1, \dots, 3$ denote the flavour of the charged leptons, while $i, j = 1, \dots, 3+n_S$ correspond to the physical (massive) 
neutrino states; $P_{L,R} = (1 \mp \gamma_5)/2$, 
$g_w$ denotes the weak coupling constant, and
$\cos^2 \theta_w =  M_W^2 /M_Z^2$.
The coefficients ${\bf C}_{V}$ and ${\bf C}_{A}$ 
parametrise the SM vector and axial-vector currents 
for the interaction of neutrinos with charged leptons, respectively given by 
${\bf C}_{V} = \frac{1}{2} + 2 \sin^2\theta_w$ and 
${\bf C}_{A} = \frac{1}{2}$.

\subsection{Parametrisation of the leptonic mixings}
Following for instance~\cite{Abada:2015trh}, $\mathcal{U}$
can be parametrised through five subsequent rotations $R_{ij}$ (with $i\neq j$), and a diagonal matrix including the four physical Majorana phases, $\varphi_i$ 
\begin{eqnarray}
    \mathcal{U} \,= \,R_{45}\,R_{35}\,R_{25}\,R_{15}\,
    R_{34}\,R_{24}\,R_{14}\,R_{23}\,R_{13}\,R_{12}\times\mathrm{diag}(1, e^{i\varphi_2}, e^{i\varphi_3}, e^{i\varphi_4}, e^{i\varphi_5})\,.
    \label{eqn:allrot}
\end{eqnarray}
The above rotations are of the form (illustrated by $R_{45}$):
\begin{equation}\label{eq:R45}
    R_{45} = \begin{pmatrix}
                1 & 0 & 0 & 0 & 0\\
                0 & 1 & 0 & 0 & 0\\
                0 & 0 & 1 & 0 & 0\\
                0 & 0 & 0 & \cos\theta_{45} & \sin \theta_{45} e^{-i\delta_{45}}\\
                0 & 0 & 0 & -\sin\theta_{45} e^{i\delta_{45}} & \cos\theta_{45}
            \end{pmatrix}\,.
\end{equation}

As already noticed (and clear from the $\mathcal{L}_{W^\pm}$ term in Eq.~(\ref{eq:lagrangian:WGHZ})), the mixing in charged current interactions is parametrised via a
rectangular $3 \times (3 +n_S)$ (i.e. $3\times 5$) mixing matrix, of which the $3 \times 3$ sub-block encodes the mixing between the left-handed leptons, $\tilde U_\text{PMNS}$. 
The deviations of $\tilde U_\text{PMNS}$ from unitarity~\cite{Schechter:1980gr,Gronau:1984ct} can be conveniently parametrised in terms of a matrix 
$\eta$~\cite{FernandezMartinez:2007ms}
\begin{equation}
\label{eq:defPMNSeta}
U_\text{PMNS} \, \to \, \tilde U_\text{PMNS} \, = \,(\mathbb{1} - \eta)\, 
U_\text{PMNS}\,.
\end{equation}

In our numerical analysis, we fix the mixing parameters of the mostly active neutrinos to the central values of the NuFIT 5.0 results~\cite{Esteban:2020cvm} (without atmospheric data), which we quote for convenience in Table~\ref{tab:nufit}.

\renewcommand{\arraystretch}{1.3}
\begin{table}[h!]
    \centering
    \begin{tabular}{|c|c|c|}
        \hline
        & Normal ordering & Inverted ordering \\
        \hline
        \hline
      $\sin^2\theta_{12}$  & $0.304^{+0.013}_{-0.012}$ & $0.304^{+0.013}_{-0.012}$ \\
      \hline
      $\sin^2\theta_{23}$ & $0.570^{+0.018}_{-0.024}$ & $0.575^{+0.017}_{-0.021}$\\
      \hline
      $\sin^2\theta_{13}$ & $0.02221^{+0.00068}_{-0.00062}$ & $0.02240^{+0.00062}_{-0.00062}$\\
      \hline
      $\Delta m_{21}^2/10^{-5}\,\mathrm{eV}$ & $7.42^{+0.21}_{-0.20}$ & $7.42^{+0.21}_{-0.20}$\\
      \hline
      $\Delta m_{3\ell}^2/10^{-3}\,\mathrm{eV}$ & $2.514^{+0.028}_{-0.027}$ & $-2.497^{+0.028}_{-0.028}$\\
      \hline
    \end{tabular}
    \caption{Global fit results obtained by NuFIT 5.0~\cite{Esteban:2020cvm} for neutrino mixing data, not including experimental data from oscillation experiments measuring atmospheric neutrinos. In our numerical analysis we assume normal ordering of the light neutrino spectrum and fix the neutrino mixing parameters to their central values.}
    \label{tab:nufit}
\end{table}
\renewcommand{\arraystretch}{1.}
The predictions for cLFV transitions are mostly independent of the ordering of the spectrum and of the lightest neutrino mass.
For concreteness (and especially in what regards constraints from neutrinoless double beta decay) throughout our numerical analysis we assume the light spectrum to follow a normal ordering, and vary the lightest neutrino mass in the range $m_0 \in [10^{-5}, 10^{-3}]\:\mathrm{eV}$.

\subsection{Constraints on SM extensions via sterile fermions}\label{app:constraints}
In addition to the cLFV observables, we constrain the mixing angles $\theta_{\alpha 4, 5}$ using experimental data on various precision observables that do not depend on the new CPV phases.
These can be divided into low- and high-energy observables.
At high energies, one has lepton flavour universality ratios of $W$ decays (see e.g.~\cite{Abada:2013aba})
\begin{equation}
    R_W^{\ell_1\ell_2} = \frac{\Gamma(W\to \ell_1 \nu)}{\Gamma(W\to \ell_2 \nu)}\,,
\end{equation}
with $\ell_{1, 2}\in \{e, \mu, \tau\}$.
Of paramount importance is also the invisible decay width of the $Z$ boson, $\Gamma(Z\to\mathrm{inv})$, which is typically reduced in scenarios with heavy sterile states that have sizeable mixings with the mostly active neutrinos. 

At low energy, we take into account (semi-)leptonic decays of the $\tau$-lepton and leptonic decays of light mesons, out of which various ratios can be constructed; these are sensitive to a modified $W\ell\nu$ vertex, and therefore indirectly sensitive to effects of the sterile states.
These are (see for instance~\cite{Abada:2012mc, Abada:2013aba})
\begin{eqnarray}
    R_\tau &\equiv& \frac{\Gamma (\tau\to \mu\nu\nu)}{\Gamma(\tau\to e\nu\nu)}\\
    \Delta r_P &\equiv&\frac{R_P}{R_P^\mathrm{SM}} - 1\\
    \mathrm{where}\:\: R_P &\equiv& \frac{\Gamma(P^+\to e\nu)}{\Gamma(P^+\to \mu\nu)}\:\:\:\:(P = \pi,K)\\
    R_e &\equiv& \frac{\Gamma(\pi^+\to e\nu)}{\Gamma(K^+\to e\nu)}\:\:,\:\: R_\mu \equiv \frac{\Gamma(\pi^+\to \mu\nu)}{\Gamma(K^+\to \mu\nu)}\,.
\end{eqnarray}

Finally, we take into account the upper bounds on the entries of $\eta$ as derived in Ref.~\cite{Fernandez-Martinez:2016lgt}, indirectly taking into account further constraints from modifications of the Fermi constant $G_F$, the weak mixing angle $\sin^2\theta_w$, the mass of the $W$-boson among others.

For HNL with masses around the TeV~scale (as these here considered), constraints from direct searches at colliders and limits from cosmological considerations such as big bang nucleosynthesis (BBN), are typically not competitive and therefore not taken into account.
However, another important constraint of theoretical nature can be derived by imposing that decays of the HNL comply with perturbative unitarity~\cite{Chanowitz:1978mv,Durand:1989zs,Korner:1992an,Bernabeu:1993up,Fajfer:1998px,Ilakovac:1999md}, which gives a direct bound on their decay width $\frac{\Gamma_{\nu_i}}{m_{\nu_i}} < \frac{1}{2} (i \geq 4)$. 
Since the dominant contribution arises from a $W$-boson exchange, one can obtain a bound on the sterile masses and their couplings to active states which can be written as
\begin{equation}
    m_{\nu_i}^2 C_{ii} < 2\ \frac{M_W^2}{\alpha_w}\quad(i\geq4)\,.
\end{equation}

\medskip
Additional sterile states are known to lead to modifications of the predictions for the effective mass to which the amplitude of neutrinoless double beta decay is proportional to, $m_{ee}$. In the presence of $n_S$ heavy states, the contributions to $m_{ee}$ can be written as~\cite{Blennow:2010th,Abada:2014nwa}
\begin{equation}
\label{eq:def:0nubb_nS}
m_{ee} \simeq \sum_{i=1}^{3+n_s} \, \mathcal U_{e i}^2 \, p^2 \, \frac{m_{ i}}{p^2-m_{i}^2} \simeq \sum_{i=1}^3 \, \mathcal U^2_{e i} \, m_i + \sum_{k=4}^{3+n_s} \mathcal U_{e k}^2 \, p^2 \, \frac{m_{k}}{p^2-m_{k}^2} \; ,
\end{equation}
in which $p^2$ corresponds to the virtual momentum, with $p^2 \simeq -(100 \, \mathrm{MeV})^2$. 
In our analysis, we take into account the KamLAND-ZEN~\cite{KamLAND-Zen:2016pfg} upper limit on $m_{ee}^\mathrm{eff} \lesssim  (61 \div 165)\:\mathrm{MeV}$ which is obtained using the isotope $^{136}\mathrm{Xe}$.

\section{Additional information on cLFV transitions and decays}\label{app:cLFV.others}
\subsection{Loop functions}\label{app:loopfunctions}
Below we summarise the loop functions for the cLFV transitions discussed in our study, as well as their relevant limits\footnote{Note that in
  Ref.~\cite{Ilakovac:1994kj} the loop function $F_\text{Xbox}$ is
  named $F_\text{box}$ and has an opposite global sign when compared to
  Ref.~\cite{Alonso:2012ji}, which also reflects in the form factor
  $F_\text{box}^{\beta 3\alpha}$.}, as taken from
Refs.~\cite{Alonso:2012ji,Ilakovac:1994kj}. 
The photon dipole and anapole functions and relevant asymptotic limits, are given by
\begin{eqnarray}
    F_\gamma(x) &=& \frac{7 x^3 - x^2 - 12x}{12(1-x)^3} - \frac{x^4 -
      10x^3 + 12x^2}{6(1-x)^4}\log x\,,\nonumber\\ 
    F_\gamma(x) &\xrightarrow[x\gg1]{}& -\frac{7}{12} -
    \frac{1}{6}\log x\,,\nonumber\\ 
    F_\gamma(0) &=& 0\,,\label{eqn:lfun:fgamma}\\
    G_\gamma(x) &=& -\frac{x(2x^2 + 5x - 1)}{4(1-x)^3} -
    \frac{3x^3}{2(1-x)^4}\log x\,,\nonumber\\ 
    G_\gamma(x) &\xrightarrow[x\gg1]{}& \frac{1}{2}\,,\nonumber\\
    G_\gamma(0) &=& 0\,.\label{eqn:lfun:ggamma}
\end{eqnarray}
The loop functions of the $Z$-penguins are given by a
two-point 
function
\begin{eqnarray}
    F_Z(x) &=& -\frac{5 x}{2(1 - x)} - \frac{5x^2}{2(1-x)^2}\log x\,,\nonumber\\
    F_Z(x) &\xrightarrow[x\gg 1]{}& \frac{5}{2} - \frac{5}{2}\log
    x\,,\nonumber\\ 
    F_Z(0) &=& 0\,,\label{eqn:lfun:fz}
\end{eqnarray}
and two three-point functions which are symmetric under interchange
of the arguments. 
\begin{eqnarray}
    G_Z(x,y) &=& -\frac{1}{2(x-y)}\left[\frac{x^2(1-y)}{1-x}\log x -
      \frac{y^2(1-x)}{1-y}\log y \right]\,,\nonumber\\ 
    G_Z(x, x) &=& -\frac{x}{2} - \frac{x\log x}{1-x}\,,\nonumber\\
    G_Z(0,x) &=& -\frac{x\log x}{2(1-x)}\,,\nonumber\\
    G_Z(0,x)&\xrightarrow[x\gg 1]{}& \frac{1}{2}\log x\,,\nonumber\\
    G_Z(0,0) &=& 0\,,\label{eqn:lfun:gz}\\
    H_Z(x,y) &=& \frac{\sqrt{xy}}{4(x-y)}\left[\frac{x^2 - 4x}{1 -
        x}\log x - \frac{y^2 - 4y}{1 - y}\log
      y\ \right]\,,\nonumber\\ 
    H_Z(x,x) &=& \frac{(3 - x)(1-x) - 3}{4(1-x)} - \frac{x^3 - 2x^2 +
      4x}{4(1-x)^2}\log x\,,\nonumber\\ 
    H_Z(0,x) &=& 0\,.\label{eqn:lfun:hz}
\end{eqnarray}
The (symmetric) box-loop-functions and their limits are given by
\begin{eqnarray}
    F_\text{box}(x,y) &=& \frac{1}{x-y}\left\{\left(4 +
    \frac{xy}{4}\right)\left[\frac{1}{1-x} + \frac{x^2}{(1-x)^2} \log
      x - \frac{1}{1-y} - \frac{y^2}{(1-y)^2}\log
      y\right]\right.\nonumber\\  
    &\phantom{=}& \left. -2xy\left[\frac{1}{1-x} + \frac{x}{(1-x)^2}
      \log x - \frac{1}{1-y} - \frac{y}{(1-y)^2}\log y
      \right]\right\}\,,\nonumber\\ 
    F_\text{box}(x,x) &=& -\frac{1}{4(1-x)^3}\left[x^4 - 16x^3 + 31x^2
      - 16 + 2x\left(3x^2 + 4x - 16\right)\log x\right]\,,\nonumber\\ 
    F_\text{box}(0,x) &=& \frac{4}{1 - x} + \frac{4x}{(1-x)^2}\log
    x\,,\nonumber\\ 
    F_\text{box}(0,x)&\xrightarrow[x\gg 1]{}& 0\,,\nonumber\\
    F_\text{box}(0,0) &=& 4\,,\label{eqn:lfun:fbox}\\
    F_\text{Xbox}(x,y) &=& -\frac{1}{x-y}\left\{\left(1 + \frac{xy}{4}
    \right)\left[\frac{1}{1-x} + \frac{x^2}{(1-x)^2} \log x -
      \frac{1}{1-y} - \frac{y^2}{(1-y)^2}\log
      y\right]\right.\nonumber\\  
    &\phantom{=}& \left. -2xy\left[\frac{1}{1-x} + \frac{x}{(1-x)^2}
      \log x - \frac{1}{1-y} - \frac{y}{(1-y)^2}\log y
      \right]\right\}\,,\nonumber\\ 
    F_\text{Xbox}(x,x) &=& \frac{x^4 - 16x^3 + 19x^2 - 4}{4(1-x)^3} +
    \frac{3x^3 + 4x^2 - 4x}{2(1-x)^3}\log x\,,\nonumber\\ 
    F_\text{Xbox}(0,x) &=& -\frac{1}{1-x} - \frac{x}{(1 - x)^2}\log
    x\,,\nonumber\\ 
    F_\text{Xbox}(0,x)&\xrightarrow[x\gg 1]{}& 0\,,\nonumber\\
    F_\text{Xbox}(0,0) &=& -1\,,\label{eqn:lfun:fxbox}\\
    G_\text{box}(x,y) &=& -\frac{\sqrt{xy}}{x-y}\left\{(4 +
    xy)\left[\frac{1}{1-x} + \frac{x}{(1-x)^2} \log x - \frac{1}{1-y}
      - \frac{y}{(1-y)^2}\log y\right]\right.\nonumber\\  
    &\phantom{=}& \left. -2\left[\frac{1}{1-x} + \frac{x^2}{(1-x)^2}
      \log x - \frac{1}{1-y} - \frac{y^2}{(1-y)^2}\log y
      \right]\right\}\,,\nonumber\\ 
    G_\text{box}(x,x) &=& \frac{2x^4 - 4x^3 + 8x^2 - 6x}{(1-x)^3} -
    \frac{x^4 + x^3 + 4x}{(1-x)^3}\log x\,,\nonumber\\ 
    G_\text{box}(0,x) &=& 0\,.\label{eqn:lfun:gbox}
\end{eqnarray}

\subsection{Muonic atoms}\label{app:cLFV.others.muonic}
Several cLFV observables concern processes occurring in the presence of a (short-lived) muonic atom. These include muonium oscillations and decay ($\text{Mu}-\overline{\text{Mu}}$, $\text{Mu}\to e e $), neutrinoless muon-electron conversion in nuclei ($\mu-e$, N), and the Coulomb enhanced decay
$\mu e \to e e$.

\paragraph{Muonium oscillations and decay}
Muonium (Mu) is a hydrogen-like Coulomb bound state formed by an anti-muon which slows down inside matter and captures an electron, $e^-\mu^+$~\cite{Pontecorvo:1957cp}. 
The bound state is free of hadronic interactions, and its electromagnetic binding is well described by the SM electroweak interactions. 
The spontaneous conversion of a Muonium atom to its anti-atom 
($\overline{\text{Mu}} = e^+\mu^-$)~\cite{Feinberg:1961zza}, as well as its decay to a pair of electrons, have been identified as promising cLFV observables. 

In the presence of $(V -A) \times (V - A)$ interactions, Muonium anti-Muonium oscillations can be described by the following effective
four-fermion interaction, in terms of an effective coupling $G_{M\overline{M}}$, 
\begin{equation}
 \mathcal{L}_\text{eff}^{M\overline{M}} \,= \, \frac{G_{
     M\overline{M}}}{ \sqrt{2} } 
\left[\, {\overline \mu}\,  \gamma^{\alpha} (1 - \gamma_5) \,e
\,\right] \left[ \,{\overline \mu}\, \gamma_{\alpha} (1 - \gamma_5)\,
e \,\right] \, .
\label{eq:def:Leff_anti-muonium}
\end{equation}
In extensions of the SM with sterile 
neutrinos
Mu-$\rm \overline{Mu}$ conversion occurs at the loop level, being exclusively mediated by a set of 4 independent box diagrams; while a first set is common to both Dirac and Majorana
neutrinos 
a second one is only present if neutrinos are Majorana particles,
 effectively amounting to two Majorana mass insertions (see Refs.~\cite{Clark:2003tv,Cvetic:2005gx}). 
Working in the unitary gauge, the computation of the different box diagrams allows to write the effective coupling of Eq.~(\ref{eq:def:Leff_anti-muonium}) as~\cite{Clark:2003tv,Cvetic:2005gx}:
\begin{equation}
\frac{G_{M\overline{M}}}{\sqrt{2}}\,=\,
-\frac{G_F^2 M_W^2}{16\pi^2}\left[\sum_{i,j =1}^{3+n_s}
2\, \mathcal U_{\mu i}^\ast\,\mathcal U_{\mu j}^\ast\, \mathcal U_{e i}\, \mathcal U_{e j} F_\text{Xbox}(x_i, x_j)+ (\mathcal U_{e i})^2\,(\mathcal U_{\mu j}^\ast)^2 G_\text{box}(x_i, x_j)
\right]
\label{eq:Gmumu}\,,
\end{equation}
where $F_\text{Xbox}(x_i, x_j)$ and $G_\text{box}(x_i, x_j)$ are the relevant loop functions\footnote{We note a sign difference between the function $F_\text{box}$ in Ref.~\cite{Cvetic:2005gx} and the function $F_\text{Xbox}$ in our convention.} given in Appendix~\ref{app:loopfunctions}, with $x_i =\frac{m_{\nu_i}^2}{M_W^2}, i=1,...,3+n_s$
(further details can be found in \cite{Abada:2015oba}).

\bigskip
In the presence  of new physics, Muonium can also undergo the cLFV decay  $\text{Mu} \, \to \, e^+ \, e^-$. In the SM extended by $n_s$ heavy neutral leptons,
the cLFV Muonium decay rate is given by~\cite{Cvetic:2006yg}  
\begin{equation}\label{eq:Mudecay:BR}
\text{BR}(\text{Mu} \to e^+ e^-) \, =\, 
\frac{\alpha_\text{em}^3}{\Gamma_\mu \, 32 \pi^2}\, 
\frac{m_e^2 m_\mu^2}{(m_e + m_\mu)^3}\,
\sqrt{1 -4\, \frac{m_e^2}{(m_e + m_\mu)^2}}\,
|\mathcal{M}_\text{tot}|^2\ , 
\end{equation}
in which $\Gamma_\mu = G_F^2 m_\mu^5 /(192 \pi^3)$ denotes the muon decay width and $\alpha_{\rm em} = e^2/4\pi$ the electromagnetic coupling constant, with 
$|\mathcal{M}_\text{tot}|$ the full amplitude (summed
(averaged) over final (initial) spins)~\cite{Cvetic:2006yg}, 
\begin{align}
|\mathcal{M}_\text{tot}|^2\,  =\, & 
\frac{\alpha_w^4}{16 M_W^4}
\bigg\{
\left( m_e\,m_\mu^{3}+2\,{m_e}^{2}m_\mu^{2}
+{m_e}^{3}m_\mu \right) {\left| 2F_Z^{\mu e} + F_{\rm Box}^{\mu e e e}\right|}^{2}
\nonumber\\
&+ 4 \sin^2 \theta_w\left( 2\,m_e\,m_\mu^{3}
+3\,{m_e}^{2}m_\mu^{2}+3\,{m_e}^{3}m_\mu \right)\
{\rm Re} \left[(2F_Z^{\mu e}+F_{\rm Box}^{\mu e e e})
(F_\gamma^{\mu e} -F_Z^{\mu e})^*\right]\nonumber\\
&+ 12 \sin^2 \theta_w
\left( m_e\,m_\mu^{3}+ 2\,{m_e}^{2}m_\mu^{2}+{m_e}^{3}m_\mu \right)
\ {\rm Re} \left[(2F_Z^{\mu e}+F_{\rm Box}^{\mu e e e})
G_\gamma^{\mu e *}\right] \nonumber\\
 &+ 4 \sin^4\theta_w \left( 7\,m_e\,{m_\mu}^{3}+12\,{m_e}^{2}m_\mu^{2}
 +9\,{m_e}^{3}m_\mu \right) {\left|F_\gamma^{\mu e}-F_Z^{\mu
     e}\right|}^{2}\nonumber\\ 
&+ 4 \sin^4\theta_w \left( -2\,m_\mu^{4}
+12\,m_e\,m_\mu^{3}+36\,{m_e}^{2}m_\mu^{2}+18\,{m_e}^{3}m_\mu \right)
\ {\rm Re} \left[(F_\gamma^{\mu e} - F_Z^{\mu e}) G_\gamma^{\mu e*}
\right]\nonumber\\
&+ 4 \sin^4\theta_w \left( {\frac {m_\mu^{5}}{m_e}}+2\,m_\mu^{4}
+8\,m_e\,{m_\mu}^{3}+24\,{m_e}^{2}m_\mu^{2}
+9\,{m_e}^{3}m_\mu \right) \left|{G_\gamma^{\mu e}}\right|^{2}\bigg\}
\ ,
\label{eq:amplMudecay}
\end{align}
where the corresponding form factors have been previously introduced (see Eqs.~(\ref{eq:cLFV:FF:Ggamma} - \ref{eq:cLFV:FF:Fbox}), setting $\beta=\mu$ and $\alpha=e$). 

\section{Phase dependence of cLFV observables - analytical expressions}
\label{app:analytic.phase.observables}
In order to analytically study the phase dependence of the cLFV form factors, we work under the approximation $\sin\theta_{\alpha 4}\approx\sin\theta_{\alpha5} \ll 1$. Let us further assume that the masses of the heavy states are close to each other 
and of the order of a few $\mathrm{TeV}$, $m_4\approx m_5 \gtrsim \Lambda_\text{EW}$.
\subsection{Photon penguins}
The photon penguin form factors exhibit a similar structure. From inspection of the relevant loop functions (cf. Appendix~\ref{app:loopfunctions}), it can be seen that contributions from the light neutrino mass eigenstates are negligible. Furthermore, since the heavy states are assumed to be close in mass, the loop functions are approximately equal and can be thus factored out of the sum.
The form factor is then given by 
\begin{eqnarray}
    G_\gamma^{\beta \alpha} &=& \sum_{i =1}^{5}
    \mathcal{U}_{\alpha i}\,\mathcal{U}_{\beta i}^\ast\, G_\gamma(x_i)\:,\nonumber \\ 
    &\approx& (\mathcal U_{\alpha 4}\,\mathcal U_{\beta 4}^\ast + \mathcal U_{\alpha 5}\,\mathcal U_{\beta 5}^\ast) \,G_\gamma(x_{4,5}) \ ,\label{eq:cLFV:FF:Ggamma:approx}
\end{eqnarray}
and inserting now the entries of the mixing matrix of Eq.~\eqref{eqn:allrot} 
(in the limit in which $\varphi_{i}=0$) 
yields
\begin{equation}
    G_\gamma^{\beta \alpha} \approx e^{-\frac{i}{2}(\Delta_4^{\alpha\beta}+\Delta_5^{\alpha\beta})}\left( s_{\alpha 4} s_{\beta_4} e^{-\frac{i}{2}(\Delta_4^{\alpha\beta}-\Delta_5^{\alpha\beta})}  + s_{\alpha 5} s_{\beta 5}e^{\frac{i}{2}(\Delta_4^{\alpha\beta}-\Delta_5^{\alpha\beta})}\right) G_\gamma(x_{4,5})\,,
    \label{eqn:Ggaphase}
\end{equation}
with $s_{ij} = \sin\theta_{ij}$ and $x_{4,5} = m_4^2/M_W^2 = m_5^2/M_W^2$, and where we have defined
\begin{equation}
    \Delta_i^{\alpha\beta} = \delta_{\alpha i} - \delta_{\beta i}\,,
\end{equation}
with the properties $\Delta_i^{\alpha\beta} = - \Delta_i^{\beta\alpha}$, $\Delta_i^{\alpha\rho} + \Delta_i^{\rho\beta} = \Delta_i^{\alpha\beta}$ and $\Delta_i^{\alpha\alpha} = 0$.
Further assuming that the mixing between the active neutrinos and $\nu_4$ and $\nu_5$ is approximately equal, i.e. $\sin\theta_{\alpha 4}\approx \sin\theta_{\alpha 5}$, we can simplify $G_\gamma^{\beta\alpha}$ to 
\begin{equation}
    G_\gamma^{\beta \alpha} \approx s_{\alpha 4}s_{\beta 4} e^{-\frac{i}{2}(\Delta_4^{\alpha\beta}+\Delta_5^{\alpha\beta})} 2 \cos\left(\frac{\Delta_4^{\alpha\beta}-\Delta_5^{\alpha\beta}}{2}\right)  G_\gamma(x_{4,5})\,,
\end{equation}
such that the branching fraction for the radiative decays is given by
\begin{equation}
    \mathrm{BR}(\ell_\beta\to \ell_\alpha\gamma)\propto |G_\gamma^{\beta\alpha}|^2 \approx 4 s_{\alpha 4}^2 s_{\beta 4}^2 \cos^2\left(\frac{\Delta_4^{\alpha\beta}-\Delta_5^{\alpha\beta}}{2}\right) \, G^2_\gamma(x_{4,5})\,.
\end{equation}
Similar results can be obtained for $F_\gamma^{\beta\alpha}$.

\subsection{$Z$ penguins}
The form factor generated by $Z$ penguin diagrams can be split into three different parts. It consists of bubble diagrams and triangle diagrams with or without Majorana mass insertions, and is given in the ``$3+2$ toy model'' by
\begin{eqnarray}
    F_Z^{\beta \alpha} &=& F_Z^{(1)} + F_Z^{(2)} + F_Z^{(3)}\\
    &=&\sum_{i,j =1}^{5}
    \mathcal{U}_{\alpha i}\,\mathcal{U}_{\beta j}^\ast \left[\delta_{ij} \,F_Z(x_j) +
    C_{ij}\, G_Z(x_i, x_j) + C_{ij}^\ast \,H_Z(x_i, x_j)\right]\:.
    \label{eq:fz3}
\end{eqnarray}
The first term which is proportional to the function $F_Z$ can be rewritten in the same way as the photon penguins, and is thus given by
\begin{eqnarray}
    F_Z^{(1)} &=& \sum_{i=1}^{5}\, \mathcal U_{\alpha i}\, \mathcal U_{\beta i}^\ast \, F_Z(x_i)\nonumber\\
    &\approx& (\mathcal U_{\alpha 4}\, \mathcal U_{\beta 4}^\ast + \mathcal U_{\alpha 5}\, \mathcal U_{\beta 5}^\ast) \widetilde F_Z(x_{4,5})\nonumber\\
    &\approx& 2 s_{\alpha 4}s_{\beta 4}\,e^{-\frac{i}{2}(\Delta_4^{\alpha\beta}+\Delta_5^{\alpha\beta})}\: \cos\left(\frac{\Delta_4^{\alpha\beta}-\Delta_5^{\alpha\beta}}{2}\right)  F_Z(x_{4,5})\,,
\end{eqnarray}
in analogy to $G_\gamma^{\beta\alpha}$, see Eq.~\eqref{eqn:Ggaphase}.
The second and the third terms are more involved due the presence of $C_{ij}$.
The second term can be written as
\begin{eqnarray}
    F_Z^{(2)} &\approx& \sum_{\rho \in \{e, \mu, \tau\}} \left[(\mathcal U_{\alpha 4} U_{\rho 4}^\ast +\mathcal U_{\alpha 5} \,\mathcal U_{\rho 5}^\ast)(\mathcal U_{\beta 4}^\ast \,\mathcal U_{\rho 4} +\mathcal U_{\beta 5}^\ast \,\mathcal U_{\rho 5})\right]\widetilde G_Z(x_{4,5})\nonumber\\
    &=& \sum_{\rho \in \{e, \mu, \tau\}} \left[(s_{\alpha 4}s_{\rho 4} e^{-i \Delta_4^{\alpha\rho}} + s_{\alpha 5}s_{\rho 5} e^{-i \Delta_5^{\alpha\rho}})(s_{\beta 4}s_{\rho 4} e^{i \Delta_4^{\beta\rho}} + s_{\beta 5}s_{\rho 5} e^{i \Delta_5^{\beta\rho}})\right] \widetilde G_Z\nonumber\\
    &\approx& \sum_{\rho \in \{e, \mu, \tau\}} 4s_{\alpha 4}s_{\beta 4} s_{\rho 4}^2\,e^{-\frac{i}{2}(\Delta_4^{\alpha \beta} + \Delta_5^{\alpha\beta})}\:\cos\left(\frac{\Delta_4^{\alpha \rho} - \Delta_5^{\alpha\rho}}{2}\right) \cos\left(\frac{\Delta_4^{\beta \rho} - \Delta_5^{\beta\rho}}{2}\right)\widetilde G_Z\,,
\end{eqnarray}
where we introduced $\widetilde G_Z = \widetilde G_Z(x_{4,5}) \equiv  G_Z(x_{4,5}, x_{4,5})$, which is also used in the following for loop functions that depend on 2 parameters, in the limit of degenerate masses (cf. App.~\ref{app:loopfunctions}).
Due to the appearance of $C_{ij}^\ast$, in the third  term of Eq.~(\ref{eq:fz3}), the Majorana phases will also be present. This corresponds to a Majorana mass insertion in the corresponding triangle diagram.
The last term can be cast as
\begin{eqnarray}
    F_Z^{(3)} &\approx& \sum_{\rho \in \{e, \mu, \tau\}}\left[(\mathcal U_{\alpha 4} \,\mathcal U_{\rho 4} +\mathcal U_{\alpha 5} \,\mathcal U_{\rho 5})(\mathcal U_{\beta 4}^\ast \,\mathcal U_{\rho 4}^\ast +\mathcal U_{\beta 5}^\ast \,\mathcal U_{\rho 5}^\ast)\right]\widetilde H_Z(x_{4,5})\nonumber\\
    &=&e^{-\frac{i}{2}(\Delta_4^{\alpha\beta} + \Delta_5^{\alpha\beta})}\sum_{\rho \in \{e, \mu, \tau\}}\left[\left(s_{\alpha 4}s_{\rho 4}e^{-\frac{i}{2}(\Delta_\alpha^{45} + \Delta_\rho^{45} - 2(\varphi_4 - \varphi_5))} + s_{\alpha 5}s_{\rho 5}e^{\frac{i}{2}(\Delta_\alpha^{45} + \Delta_\rho^{45} - 2(\varphi_4 - \varphi_5))}\right)\right.\nonumber\\
    &\phantom{=}&\hspace{2cm}
    \times\left.\left(s_{\beta 4}s_{\rho 4}e^{\frac{i}{2}(\Delta_\beta^{45} + \Delta_\rho^{45} - 2(\varphi_4 - \varphi_5))} + s_{\beta 5}s_{\rho 5}e^{-\frac{i}{2}(\Delta_\beta^{45} + \Delta_\rho^{45} - 2(\varphi_4 - \varphi_5))}\right)\right]\widetilde H_Z\nonumber
\end{eqnarray}
\begin{eqnarray}
    &\approx& 4e^{-\frac{i}{2}(\Delta_4^{\alpha\beta} + \Delta_5^{\alpha\beta})} \widetilde H_Z\,s_{\alpha 4}s_{\beta 4}\times\nonumber\\
    &\phantom{=}&\times\,\sum_{\rho \in \{e, \mu, \tau\}}\left[s_{\rho 4}^2\cos\left(\frac{\Delta_\alpha^{45} + \Delta_\rho^{45}}{2} - (\varphi_4 - \varphi_5)\right)\cos\left(\frac{\Delta_\beta^{45} + \Delta_\rho^{45}}{2} - (\varphi_4 - \varphi_5)\right)\right]\,.
\end{eqnarray}
In the case of vanishing Dirac phases this can be further simplified to 
\begin{eqnarray}
    F_Z^{(3)} \approx 4s_{\alpha 4}s_{\beta 4} \widetilde H_Z(x_{4,5}, x_{4,5})\,\sum_\rho \left[s_{\rho 4}^2 \cos^2(\varphi_4 - \varphi_5)\right]\,.
\end{eqnarray}

Moreover, and unique to the $Z$-penguin, there are non-negligible contributions to the form factor stemming from light and heavy virtual neutrinos in the loop.
We write the corresponding limit of the loop function as $G_Z(0, x_{4,5}) = \overline{G}_Z$.
As usual, using the above approximations, the corresponding part of the form factor can be cast as
\begin{eqnarray}
    F_Z^{(2)}(0,x_{4,5}) &\approx&\left[ \sum_{i = 1}^3\sum_{j=4,5}\sum_{\rho=e,\mu,\tau} \mathcal U_{\alpha i}\,\mathcal U_{\beta j}^\ast\,\mathcal U_{\rho i}^\ast\,\mathcal U_{\rho j} + \sum_{i = 4,5}\sum_{j=1}^3\sum_{\rho=e,\mu,\tau} \mathcal U_{\alpha i}\,\mathcal U_{\beta j}^\ast\,\mathcal U_{\rho i}^\ast\,\mathcal U_{\rho j}\right] \overline{G}_Z\nonumber\\
    &=& \sum_\rho \left[\left(\mathcal U_{\alpha 1}\,\mathcal U_{\rho 1}^\ast + \mathcal U_{\alpha 2}\,\mathcal U_{\rho 2}^\ast + \mathcal U_{\alpha 3}\,\mathcal U_{\rho 3}^\ast \right) \left( \mathcal U_{\beta 4}^\ast\,\mathcal U_{\rho 4} + \mathcal U_{\beta 5}^\ast\,\mathcal U_{\rho 5}\right) \right. + \nonumber\\
    &\phantom{=}&\hspace{0.5cm} +\left. \left(\mathcal U_{\alpha 4}\,\mathcal U_{\rho 4}^\ast + \mathcal U_{\alpha 5}\,\mathcal U_{\rho 5}^\ast \right) \left(\mathcal U_{\beta 1}^\ast\,\mathcal U_{\rho 1} + \mathcal U_{\beta 2}^\ast\,\mathcal U_{\rho 2} + \mathcal U_{\beta 3}^\ast\,\mathcal U_{\rho 3}\right)\right]\,\overline{G}_Z\,.
\end{eqnarray}
Making use of the unitarity of the full $5\times5$ mixing matrix, i.e. 
\begin{equation}
    \sum_{i = 1}^{5} \mathcal U_{\alpha i}\, \mathcal U_{\rho i}^\ast = \delta_{\alpha\rho}\:\:\Rightarrow\:\: \sum_{i = 1}^{3} \mathcal U_{\alpha i}\, \mathcal U_{\rho i}^\ast = \delta_{\alpha\rho} - \mathcal U_{\alpha 4}\, \mathcal U_{\rho 4}^\ast - \mathcal U_{\alpha 5}\, \mathcal U_{\rho 5}^\ast\,,
\end{equation}
one finally has
\begin{eqnarray}
    F_Z^{(2)}(0,x_{4,5}) &\approx& \sum_\rho \left[\left(\delta_{\alpha\rho} - \mathcal U_{\alpha 4}\, \mathcal U_{\rho 4}^\ast - \mathcal U_{\alpha 5}\, \mathcal U_{\rho 5}^\ast\,\right)\left( \mathcal U_{\beta 4}^\ast\,\mathcal U_{\rho 4} + \mathcal U_{\beta 5}^\ast\,\mathcal U_{\rho 5} \right)\right.\nonumber\\
    &\phantom{=}&\left. + \left(\mathcal U_{\alpha 4}\,\mathcal U_{\rho 4}^\ast + \mathcal U_{\alpha 5}\,\mathcal U_{\rho 5}^\ast \right)\left(\delta_{\beta\rho} -  \mathcal U_{\beta 4}^\ast\,\mathcal U_{\rho 4} - \mathcal U_{\beta 5}^\ast\,\mathcal U_{\rho 5} \right)\right]\overline{G}_Z\nonumber\\
    &=& 2\,\overline{G}_Z \sum_\rho\left[\delta_{\alpha\rho} s_{\beta 4}s_{\rho 4} e^{-\frac{i}{2}(\Delta_4^{\rho\beta} + \Delta_5^{\rho\beta})}\cos\left(\frac{\Delta_4^{\rho\beta} - \Delta_5^{\rho\beta}}{2}\right) + \right.\nonumber\\
    &\phantom{=}& \left.+\delta_{\beta\rho} s_{\alpha 4}s_{\rho 4} e^{-\frac{i}{2}(\Delta_4^{\alpha\rho} + \Delta_5^{\alpha\rho})}\cos\left(\frac{\Delta_4^{\alpha\rho} - \Delta_5^{\alpha\rho}}{2}\right)\right.\nonumber\\
    &\phantom{=}& \left. - 4s_{\alpha 4}s_{\beta 4}s_{\rho 4}^2 e^{-\frac{i}{2}(\Delta_4^{\alpha\beta} + \Delta_5^{\alpha\beta})}\cos\left(\frac{\Delta_4^{\alpha\rho} - \Delta_5^{\alpha\rho}}{2}\right)\cos\left(\frac{\Delta_4^{\beta\rho} - \Delta_5^{\beta\rho}}{2}\right)\right]\,.
\end{eqnarray}

\subsection{Box diagrams}
The form factor generated by box diagrams is given by
\begin{eqnarray}
    F_\text{box}^{\beta 3 \alpha} &=& F_\text{box}^{(1)} + F_\text{box}^{(2)}\nonumber\\
    &=&\sum_{i,j = 1}^{3+k} \mathcal{U}_{\alpha i}\,\mathcal{U}_{\beta j}^\ast\left[\mathcal{U}_{\alpha i} \,\mathcal{U}_{\alpha j}^\ast\, G_\text{box}(x_i, x_j) - 2 \,\mathcal{U}_{\alpha i}^\ast \,\mathcal{U}_{\alpha j}\, F_\text{Xbox}(x_i, x_j) \right]\:,
\end{eqnarray}
where the first term corresponds to a diagram with a possible Majorana mass insertion, thus depending on the Majorana phases.
The first term $F_\text{box}^{(1)}$ can then be written as
\begin{eqnarray}
    F_\text{box}^{(1)} &\approx& (\mathcal U_{\alpha 4}^2 + \mathcal U_{\alpha 5}^2)(\mathcal U_{\beta 4}^\ast \,\mathcal U_{\alpha 4}^\ast + \mathcal U_{\beta 5}^\ast \,\mathcal U_{\alpha 5}^\ast )\, \widetilde G_\text{box}(x_{4,5})\nonumber\\
    &=& \left(s_{\alpha 4}^2 e^{-2i(\delta_{\alpha 4} - \varphi_4)} + s_{\alpha 5}^2 e^{-2i(\delta_{\alpha 5} - \varphi_5)}\right)\left(s_{\alpha 4}s_{\beta 4} e^{i(\delta_{\alpha 4} + \delta_{\beta 4} - 2\varphi_4)} + s_{\alpha 5}s_{\beta 5} e^{i(\delta_{\alpha 5} + \delta_{\beta 5} - 2\varphi_5)}\right) \widetilde G_\text{box}\nonumber\\
    &=& e^{-\frac{i}{2}(\Delta_4^{\alpha\beta} + \Delta_5^{\alpha \beta})}\left(s_{\alpha 4}^2e^{-i(\Delta_\alpha^{45} - (\varphi_4 - \varphi_5))} + s_{\alpha 5}^2e^{i(\Delta_\alpha^{45} - (\varphi_4 - \varphi_5)}\right)\nonumber\\
    &\phantom{=}&\hspace{2cm} \times \left(s_{\alpha 4}s_{\beta 4}e^{\frac{i}{2}(\Delta_\alpha^{45} + \Delta_{\beta}^{45} - 2(\varphi_4 - \varphi_5))} + s_{\alpha 5}s_{\beta 5}e^{-\frac{i}{2}(\Delta_\alpha^{45} + \Delta_{\beta}^{45} - 2(\varphi_4 - \varphi_5))}\right)\widetilde G_\text{box}\nonumber\\
    &\approx& 4 e^{-\frac{i}{2}(\Delta_4^{\alpha\beta} + \Delta_5^{\alpha \beta})} s_{\alpha 4}^3s_{\beta 4}\cos\left(\Delta_{\alpha}^{45} - (\varphi_4 - \varphi_5)\right)\cos\left(\frac{\Delta_{\alpha}^{45} + \Delta_{\beta}^{45}}{2} - (\varphi_4 - \varphi_5)\right)\widetilde G_\text{box}\,,
\end{eqnarray}
which again can be further simplified in the case of vanishing Dirac phases to
\begin{equation}
    F_\text{box}^{(1)} \approx 4 s_{\alpha 4}^3 s_{\beta 4}\cos^2(\varphi_4 - \varphi_5)\,\widetilde G_\text{box}\,.
\end{equation}
The second term is independent of the Majorana phases and can be written as
\begin{eqnarray}
    F_\text{box}^{(2)} &\approx& -2 (|\mathcal U_{\alpha 4}|^2 + |\mathcal U_{\alpha 5}^2|)(\mathcal U_{\beta 4}^\ast \,\mathcal U_{\alpha 4} + \mathcal U_{\beta 5}^\ast \,\mathcal U_{\alpha 5})\,\widetilde F_\text{Xbox}(x_{4,5})\nonumber\\
    &=& -2 (s_{\alpha 4}^2 + s_{\alpha 5}^2)\left(s_{\alpha 4}s_{\beta 4} e^{-i(\delta_{\alpha 4} - \delta_{\beta 4})} + s_{\alpha 5}s_{\beta 5} e^{-i(\delta_{\alpha 5} - \delta_{\beta 5})}\right) \widetilde F_\text{Xbox}\nonumber\\
    &=& -2(s_{\alpha 4}^2 + s_{\alpha 5}^2)e^{-\frac{i}{2}(\Delta_4^{\alpha \beta} + \Delta_5^{\alpha\beta})}\left(s_{\alpha 4}s_{\beta 4}e^{-\frac{i}{2}(\Delta_4^{\alpha\beta} -\Delta_5^{\alpha\beta})} + s_{\alpha 5}s_{\beta 5}e^{\frac{i}{2}(\Delta_4^{\alpha\beta} -\Delta_5^{\alpha\beta})}\right)\widetilde F_\text{Xbox}\nonumber\\
    &\approx& - 8 e^{-\frac{i}{2}(\Delta_4^{\alpha \beta} + \Delta_5^{\alpha\beta})} s_{\alpha 4}^3s_{\beta 4}\cos\left(\frac{\Delta_4^{\alpha\beta} - \Delta_5^{\alpha\beta}}{2}\right)\widetilde F_\text{Xbox}\,.
\end{eqnarray}

\medskip
The box diagrams contributing to neutrinoless muon-electron conversion show a similar behaviour as that of photon- and $Z$-penguin diagrams with one neutrino in the loop.

\end{document}